\documentclass[english,aps,a4paper,prd,groupedaddress,preprintnumbers,nofootinbib,preprint,floatfix,showpacs]{revtex4}

\usepackage{hyperref}%link to eqs, figs and references
\usepackage{amsmath}
\usepackage{amssymb}
\usepackage{graphicx}
\usepackage{rotating}
\usepackage{color} %This is essential for placing graphics onto
\usepackage{multirow} % To use multirow command in tables
\usepackage[T1]{fontenc} 
\usepackage[utf8]{inputenc}
\usepackage{lmodern}
\makeatletter

\usepackage{booktabs}
\usepackage{mathrsfs}   %\mathscr{L}
\usepackage{slashed}     %\slashed{p}
\usepackage{url}
\usepackage{multirow}
\usepackage{units}
\usepackage{wasysym }
\usepackage{comment}

\usepackage{float}
\newcommand{\bea}{\begin{eqnarray}}
\newcommand{\eea}{\end{eqnarray}}

\def\ba{$$\begin{array}}
\def\ea{\end{array}$$}
\def\bra{$\begin{array}}
 \def\era{\end{array}$}

\makeatother

\usepackage{babel}

\newcommand {\be} {\begin{equation}}
\newcommand {\ee} {\end{equation}}

\def\vb#1{\vbox to #1 pt{}}

\newcommand{\AddrCFTP}{%
 Departamento de F\'\i sica and CFTP, Instituto Superior T\'ecnico\\
 Universidade de Lisboa, 
          Av. Rovisco Pais 1, 1049-001 Lisboa, Portugal }

\def\gsim{\raise0.3ex\hbox{$\;>$\kern-0.75em\raise-1.1ex\hbox{$\sim\;$}}}
\def\lsim{\raise0.3ex\hbox{$\;<$\kern-0.75em\raise-1.1ex\hbox{$\sim\;$}}}

\allowdisplaybreaks

\begin{document}

\preprint{CFTP/21-009}

\title{Off diagonal charged scalar couplings with the Z boson:
  the Zee model as an example}
%%
% \author{Duarte Fontes} \email{duartefontes@tecnico.ulisboa.pt}
% \affiliation{\AddrCFTP}
 \author{Ricardo R. Florentino} \email{ricardomflorentino@tecnico.ulisboa.pt}
 \affiliation{\AddrCFTP}
\author{Jorge C.~Rom\~ao}\email{jorge.romao@tecnico.ulisboa.pt}
\affiliation{\AddrCFTP} 
 \author{João  P.~Silva} \email{jpsilva@cftp.ist.utl.pt}
 \affiliation{\AddrCFTP}

%\today

%\pacs{14.60.Pq 12.60.Fr 14.60.St }
\begin{abstract}
Models with scalar doublets and charged scalar singlets
have the interesting property that they have couplings
between one $Z$ boson and two charged scalars of different
masses.
This property is often ignored in phenomenological analysis,
as it is absent from models with only extra scalar doublets.
We explore this issue in detail, considering $h \rightarrow Z \gamma$,
$B \to X_s \gamma$, and the decay of a heavy charged scalar into a lighter one and
a $Z$ boson.
We propose that the latter be actively searched for at the LHC,
using the scalar sector of the Zee model as a prototype and proposing benchmark points
which obey all current experimental data, and could be within reach of the LHC.
\end{abstract}

\maketitle

\section{Introduction}

Over many decades, the Standard Model (SM)
\cite{Glashow:1961tr,Weinberg:1967tq,Salam:1968rm}
has been confirmed to unprecedented precision.
This culminated with the 2012 experimental detection of a
fundamental scalar particle with mass 125GeV (the Higgs Boson $h_{125}$) 
\cite{Aad:2012tfa,Chatrchyan:2012ufa},
which had been proposed in the early 1960's \cite{Higgs:1964ia,Englert:1964et}.
Still, the SM leaves unanswered questions, from the nature of neutrino masses,
to the origin of Dark Matter (DM).
Having found one fundamental scalar, the most pressing question is:
are there more fundamental scalars in Nature?
There is a large international effort to answer this question,
both from the theoretical point of view, and from the robust experimental physics
programs currently pursued at CERN's LHC.

Thus, one is lead to study and search for signals of extra scalars.
It is known experimentally that the masses of the $W$ and $Z$ bosons
bear a relation very close to that predicted in the SM:
$M_Z \cos{\theta_W} / M_W \sim 1$, where $\theta_W$ is the
Weinberg angle.
This holds automatically if the extra scalars are in doublets or singlets
of the electroweak gauge group.
Thus, we are lead to study theories with any number of scalar doublets
and/or singlets; the latter neutral and/or charged.

A case of particular interest is the Zee model \cite{zee:1980ai}
with two Higgs doublets and one charged singlet,
originally proposed to explain naturally small neutrino masses, and
later adapted to explain also DM \cite{Smirnov:1996bv,Krauss:2002px}.
The Zee model with an extra $Z_2$ symmetry proposed by Wolfenstein
\cite{Wolfenstein:1980sy} is not consistent with current
data from neutrino oscillations \cite{Koide:2001xy,He:2003ih},
but the original proposal is still consistent with all leptonic experimental
results \cite{Herrero-Garcia:2017xdu,Babu:2019mfe}.
But the scalar sector of the Zee model also has another striking feature which is
mostly ignored; it is the minimal model predicting the
existence of couplings $Z H_1^\pm H_2^\mp$ between the $Z$ gauge boson
and two charged scalars ($H_1^+$ and $H_2^+$) of different mass.
This is the feature highlighted in this article.

Even before direct detection of the extra charged scalar particles,
$Z H_1^\pm H_2^\mp$ couplings could potentially have a virtual effect on
current measurements,
such as $h_{125} \rightarrow Z \gamma$.
We discuss this example in detail. In fact, the contribution of the charged
scalars to the branching ratio can even vanish,
but that is not because the $Z$ couples to two different charged
scalars, but rather because there are two charged scalars contributing in the
loop.
Indeed, this feature is already present for instance in the 3HDM, where
there are two charged scalars but the coupling of the $Z$ to them is
diagonal. Although there is a modulation of the result with the mixing
angle between the two charged Higgs, this is hidden when the sum over all
diagrams is performed.

To study this model we took into account all the theoretical and
experimental constraints coming from the scalar and quark sectors.
In particular, we considered in detail the
influence of the bounds coming from
BR$(B \to X_s \gamma)$ \cite{Borzumati:1998tg}.
This is especially important because,
as there are two charged Higgs, one can evade the $580$ GeV limit for the
2HDM \cite{Misiak:2017bgg}. We will discuss the implications of this
for the Zee model.

A distinctive signal for this model with its $Z H_1^\pm H_2^\mp$ couplings
is the decay of the heavier charged Higgs into the lightest one and one $Z$.
We performed an analysis of the parameter space to look for regions
where this decay can be large.
This lead us to identify examples of benchmark points
where the decay $H^+_2\to H^+_1 Z$ can be large as well as the decay
$H_1^+ \to t \bar{b}$, leading to a clear signature that should
be searched for at the LHC.

The paper is organized as follows. In section~\ref{sec:generic}, we review the
formalism for models with an arbitrary number of doublets and
singlets, and in section~\ref{sec:Zee} we apply it for the case of the
scalar and quark sectors of the Zee model.
In section~\ref{sec:constraints} we discuss the constraints,
both theoretical and
experimental on the model. Our results are presented in section~\ref{sec:loopdecays}
where we discuss the impact on $h\to Z\gamma$ and in section~\ref{sec:newdecay}
where we study the novel decay $H^+_{2} \to H^+_{1} Z$. For this decay we
propose benchmark points with noteworthy features in section~\ref{sec:benchmark}.
After the conclusions in section~\ref{sec:conclusions}, some appendices are included.
In appendix~\ref{app:A} we collect the relevant couplings of the charged Higgs.
The detailed formulas for the loop decays are presented in appendix~\ref{sec:hdecays}
and for perturbative unitarity in appendix~\ref{sec:PertUni}.
As far as we know, the latter are presented here for the first time.

\section{\label{sec:generic}Models with an arbitrary number of doublets and singlets}
We consider the models studied in \cite{Grimus:2007if} and use a
similar notation to the one presented there. The scalar part of the
model consists of $n_d$ doublets of $SU(2)$, $n_c$ singly charged
singlets and $n_n$ real neutral singlets. The fermionic and vector
fields are identical to the SM content. 

The scalars are denoted by
\begin{equation}
    \phi_a=
    \begin{pmatrix}
    \varphi_a^+\\
    \varphi_a^0
    \end{pmatrix},
    \;\;\;a=1,2,...,n_d\nonumber\, ,
\end{equation}
\begin{equation}
    \chi_i^+,
    \;\;\;i=1,2,...,n_c\, ,
\end{equation}
\begin{equation}
    \chi_r^0,
    \;\;\;r=1,2,...,n_n\nonumber\, ,
\end{equation}
and the neutral fields can be expanded around their vevs as
\begin{equation}
    \varphi_a^0=\frac{1}{\sqrt{2}}(v_a+\varphi_a^0{}')
    \label{eq:vevs}
\end{equation}
\begin{equation}
    \chi_r^0=u_r+\chi_r^0{}'\, ,
    \nonumber
\end{equation}
with complex $v_a$ and real $u_r$, where the former satisfy $v=\left(\sum|v_a|^2\right)\simeq246\text{GeV}$.
With a total of $n=n_d+n_c$ complex singly charged scalar fields and $m=2n_d+n_n$ real neutral scalar fields, we can define the change to the physical basis $S_\alpha^+$ $(\alpha=1,2,...,n)$ and $S_\beta^0$ $(\beta=1,2,...,m)$ with masses $m_{\pm\alpha}$ and $m_{0\beta}$ respectively, throughout the unitary transformations
\begin{align}
    \varphi_a^+=U_a^\alpha S_\alpha^+\, ,
    \nonumber\\
    \chi_i^+=T_i^\alpha S_\alpha^+\, ,
    \nonumber\\
    \varphi_a^0{}'=V_a^\beta S_\beta^0\, ,
    \label{eq:basis_change}\\
    \chi_r^0{}'=R_r^\beta S_\beta^0\, ,
    \nonumber
\end{align}
where the last matrix is real and the others are complex. In this text, every index appearing up and down in the same expression is assumed to be summed over. The matrices
\begin{equation}
    \tilde{U}_{\alpha'}^{\alpha}=
    \begin{pmatrix}
    U^\alpha_a\\
    T^\alpha_i
    \end{pmatrix}\, ,
    \nonumber
\end{equation}
\begin{equation}
    \tilde{V}_{\beta'}^{\beta}=
    \begin{pmatrix}
    \text{Re }V^\beta_a\\
    \text{Im }V^\beta_a\\
    R^\beta_r
    \end{pmatrix}\, ,
    \label{eq:matrices}
\end{equation}
are, respectively, the unitary and orthogonal matrices that diagonalize
the charged and neutral mass matrices. The physical fields with
indices $\alpha=1$ and $\beta=1$ are assigned to the unphysical
Goldstone bosons, and the neutral $S_2^0$ field is assigned to the
Higgs  particle measured at the LHC with mass
$m_h\simeq125\text{GeV}$. We note that even though the matrices
defined in eq.~\eqref{eq:matrices} are unitary, the matrices in
eq.~\eqref{eq:basis_change} do not need to be. In fact, only if there
is no mixing between the doublet fields and the charged singlets, can the
matrices be brought to a basis where they become composed of zeros
surrounding a unitary square matrix. This characteristic is of
significant importance as we will show later.

\subsection{\label{subsec:V}Scalar potential}

For simplicity, we assume a discrete symmetry under which all fields transform trivially,
except the neutral singlet scalars, for which $\chi^0_r\rightarrow-\chi^0_r$.
The scalar potential may then be conveniently written as
\begin{align}
    V=&\;\mu_1^{ab}\phi^\dagger_a\phi_b+\mu_2^{ij}\chi^+_i\chi^-_j+
    \mu_3^{rs}\chi^0_r\chi^0_s+(\mu_4^{abi}\phi_ai\sigma_2\phi_b\chi_i^-+h.c.)\nonumber\\
    &+\lambda_{1}^{abcd}\phi^\dagger_a\phi_b\phi^\dagger_c\phi_d+
    \lambda_{2}^{ijkl}\chi^+_i\chi^-_j\chi^+_k\chi^-_l+
    \lambda_{3}^{rstu}\chi^0_r\chi^0_s\chi^0_t\chi^0_u
    \label{eq:pot}\\
    &+\lambda_{4}^{abij}\phi^\dagger_a\phi_b\chi^+_i\chi^-_j+
    \lambda_{5}^{abrs}\phi^\dagger_a\phi_b\chi^0_r\chi^0_s+
    \lambda_{6}^{ijrs}\chi^+_i\chi^-_j\chi^0_r\chi^0_s\, ,\nonumber
\end{align}
where $\sigma_2$ is the second Pauli matrix, $\mu_3$ and $\lambda_3$ are real and the rest complex,
while $h.c.$ stands for hermitian conjugate.
The parameters are subject to the relations
\begin{equation}
\mu_1^{ab}=\mu_1^{ba*}\, ,
\hspace{5ex} \mu_2^{ij}=\mu_2^{ji*}\, ,
\hspace{5ex} \mu_3^{rs}=\mu_3^{sr}\, ,
\hspace{5ex}\mu_4^{abi}=-\mu_4^{bai}\, ,
\nonumber
\end{equation}
and
\begin{align}
    \lambda_1^{abcd}&=\lambda_1^{cdab}=\lambda_1^{badc*},
    \,\;\;\;\lambda_2^{ijkl}=\lambda_2^{klij}=\lambda_2^{jilk*},
    \;\;\;\;\;\lambda_3^{rstu}=\lambda_3^{(rstu)},\\
    \lambda_4^{abij}&=\lambda_4^{baji*},
    \;\;\;\;\;\;\;\;\;\;\;\;\;\;\;\;\;\lambda_5^{abrs}=\lambda_5^{bars*}=\lambda_5^{absr},
    \;\;\;\lambda_6^{ijrs}=\lambda_6^{jirs*}=\lambda_6^{ijsr}\, ,\nonumber
\end{align}
where $(rstu)$ stands for any permutation of the indices $rstu$.
After expanding around the vevs with eq.~\eqref{eq:vevs} and using
eqs.~\eqref{eq:basis_change}-\eqref{eq:matrices}
we are interested in the cubic terms
\begin{align}
    V\supset&\lambda_1^{abcd}(\varphi_a^{0}{}'^*v_b+v_a^*\varphi_b^{0}{}')
    \varphi_c^{-}\varphi_d^{+}+
    \frac{1}{2}\lambda_4^{abij}(\varphi_a^{0}{}'^*v_b+v_a^*\varphi_b^{0}{}')
    \chi_i^{+}\chi_j^{-}
    \nonumber\\
    \label{eq:h}
    &+2u_s\lambda_5^{abrs}\varphi_a^{-}
    \varphi_b^{+}\chi_r^{0}{}'+2u_s\lambda_6^{ijrs}\chi_i^{+}\chi_j^{-}\chi_r^{0}{}'
    \\
    &\frac{\mu_4^{abi}}{\sqrt{2}}(\varphi_a^{+}\varphi_b^{0}{}'-\varphi_a^
    {0}{}'\varphi_b^{+})\chi_i^{-}+
    \frac{\mu_4^{abi*}}{\sqrt{2}}(\varphi_a^{-}\varphi_b^{0}{}'^*
    -\varphi_a^{0}{}'^*\varphi_b^{-})\chi_i^{+}
    \nonumber\\
    =&\Big[\lambda_1^{abcd}(V_a^{\beta*}v_b+v_a^*V_b^\beta)
    U_c^{\alpha_1*}U_d^{\alpha_2}+
    \frac{1}{2}\lambda_4^{abij}(V_a^{\beta*}v_b+v_a^*V_b^\beta)
    T_i^{\alpha_1}T_j^{\alpha_2*}
    \nonumber\\
    \label{eq:h2}
    &+2u_s\lambda_5^{abrs}U_a^{\alpha_1*}
    U_b^{\alpha_2}R_r^\beta+
    2u_s\lambda_6^{ijrs}T_i^{\alpha_1}T_j^{\alpha_2*}R_r^\beta
    \\
    &\frac{\mu_4^{abi}}{\sqrt{2}}(U_a^{\alpha_1}V_b^\beta-V_a^
    {\beta}U_b^{\alpha_1})T_i^{\alpha_2*}+
    \frac{\mu_4^{abi*}}{\sqrt{2}}(U_a^{\alpha_2*}V_b^{\beta*}
    -V_a^{\beta*}U_b^{\alpha_2*})T_i^{\alpha_1}
    \Big]
    S_{\alpha_1}^+S_{\alpha_2}^-S_\beta^0
    \nonumber\\
    \equiv&g^{\beta\alpha_1\alpha_2}\, v\,
    S_{\alpha_1}^+S_{\alpha_2}^-S_\beta^0\, ,\nonumber
\end{align}
and in the quadratic terms with charged scalars, given by
\begin{align}
    V\supset(&\mu_1^{ab}+\lambda_1^{abcd}v_dv_c^*+
    \lambda_5^{abrs}u_ru_s)\varphi_b^+\varphi_a^-+(\mu_2^{ij}+
    \frac{1}{2}\lambda_4^{abij}v_bv_a^*+
    \lambda_6^{ijrs}u_ru_s)\chi_i^+\chi_j^-
    \nonumber\\\label{eq:seemu4}
    &+\frac{\mu_4^{abi}}{\sqrt{2}}(v_a\varphi_b^+ - v_b\varphi_a^+)\chi_i^-
    +\frac{\mu_4^{abi*}}{\sqrt{2}}(v_a^*\varphi_b^--v_b^*\varphi_a^-)\chi_i^+\\
	=
	&
	\Big[
	\left( \mu_1^{ab} + \lambda_1^{abcd} v_d v_c^* +
	\lambda_5^{abrs} u_r u_s \right) U_b^{\alpha_1}	U_a^{\alpha_2*}
	+ \left( \mu_2^{ij} + \frac{1}{2} \lambda_4^{abij} v_b v_a^*
	+ \lambda_6^{ijrs} u_r u_s \right) T_i^{\alpha_1} T_j^{\alpha_2*}
	\nonumber\\
	& + \frac{\mu_4^{abi}}{\sqrt{2}}
	\left( v_a U_b^{\alpha_1} - v_b U_a^{\alpha_1} \right) T_i^{\alpha_2*}
	+ \frac{\mu_4^{abi*}}{\sqrt{2}}
	\left( v_a^* U_b^{\alpha_2*} - v_b^* U_a^{\alpha_2*} \right) T_i^{\alpha_1}
	\Big] S_{\alpha_1}^ + S_{\alpha_2}^-\, .
\end{align}
We see from eq.~\eqref{eq:seemu4} that there is no mixing between the charged
fields originating from doublets with the charged
fields originating from singlets, unless $\mu_4^{abi} \neq 0$ for some
combination of indices.
Thus, the cubic terms in the potential \eqref{eq:pot} are essential
for the non-unitary behaviour of the matrix $U^\alpha_a$ that will be
shown to be mandatory for the appearance of $Z H_1^+ H_2^-$
couplings that change the ``flavour'' of the charged scalars. Also,
eq.~\eqref{eq:h} tells us that the coupling $h^0 H_1^+ H_2^-$ exists with
$\mu_4^{abi} \neq 0$ only for $H_1^+$ and $H_2^-$ belonging both to
the doublet sector or both to the singlet sector. Since $\mu_4^{abi}$
is anti-symmetric in $(a,b)$, the minimal scalar sector containing
such a coupling is a model with two doublets and one charged
singlet. This corresponds to the Zee model, which we study in the next section.

\subsection{\label{subsec:K}Gauge-scalar couplings}
% Tirar do Luis\\
% https://arxiv.org/pdf/0711.4022.pdf\\
% informado pelas adaptações que fizemos em\\
% https://arxiv.org/pdf/1708.09408.pdf\\
% e em\\
% https://arxiv.org/pdf/1808.07123.pdf\\

The part of the Lagrangian regarding the covariant derivative of the scalars, was derived in equation (29) of \cite{Grimus:2007if}. The relevant terms for our purposes are
\begin{align}
    \mathcal{L}\supset&\;ieA_\mu\delta^{\alpha\alpha'}(S_\alpha^+\partial^\mu
    S_{\alpha'}^--S_{\alpha'}^-\partial^\mu S_\alpha^+)+e^2A_\mu
    A^\mu\delta^{\alpha\alpha'}S_{\alpha'}^-S_\alpha^+\nonumber\\[+2mm]
&    +g\left(M_WW_\mu^+W^{-\mu}+\frac{M_Z}{2c_W}Z_\mu
  Z^\mu\right)\text{Re}(\omega^\dagger V)^\beta S_\beta^0 
-i\frac{g}{2c_W}Z_\mu(2s_W^2\delta^{\alpha\alpha'} \nonumber\\[+2mm]
&-(U^\dagger U)^{\alpha'\alpha})(S_\alpha^+\partial^\mu S_{\alpha'}^--S_{\alpha'}^-\partial^\mu S_\alpha^+)-\frac{eg}{c_W}A_\mu Z^\mu(2s_W^2\delta^{\alpha\alpha'}-(U^\dagger U)^{\alpha'\alpha})S_{\alpha'}^-S_\alpha^+\, .
\end{align}
where $\omega_a=v_a/v$.
Here we finally check the appearance of the expression $(U^\dagger
U)^{\alpha'\alpha}$ that is diagonal if $U^\alpha_a$ is unitary.
In models without a $\mu_4^{abi}$ coupling, this expression
will then be diagonal and there will be no ``flavour'' changing $Z
H_1^+ H_2^-$ coupling. The exploration of this under-appreciated point
is one of the distinguishing features of this work.
\subsection{\label{subsec:f}Fermion-scalar couplings}
% Notas manuscritas que enviei. Ver
% https://arxiv.org/pdf/1808.07123.pdf\\
% e tb
% https://arxiv.org/pdf/1507.07941.pdf\\
%
The Yukawa Lagrangian is the same as for the NHDM for $N=n_d$, and the fermion-scalar couplings were calculated for that model in \cite{Bento:2018fmy}.  The calculation for our model proceeds in a similar fashion, leaving us with the relevant Lagrangian term
\begin{align}
    \mathcal{L}&\supset-\frac{1}{v}\Bar{d}_L\left(N_d^{\alpha}B_{\alpha}^\beta S_\beta^0\right)d_R-\frac{1}{v}\Bar{u}_L\left(N_u^{\alpha}B_{\alpha}^{\beta*}S_\beta^0\right)u_R-\frac{1}{v}\Bar{e}_L\left(N_e^{\alpha}B_{\alpha}^\beta S_\beta^0\right)e_R\\
    &-\Bar{u}_LV\left(N_d^{\alpha}S_\alpha^+\right)d_R+\Bar{d}_LV^\dagger\left(N_u^{\alpha}S_\alpha^-\right)u_R+\text{h.c.}\, ,\nonumber
\end{align}
where
\begin{align}
    B_\alpha^\beta&=U^{\dagger a}_\alpha V_a^\beta,\;\;\;\;\;\;\;\;\;\;\;\;\;\;\;\;\;\;\;\;\;\,N_d^{\alpha}=\frac{v}{\sqrt{2}}U^\dagger_{dL}\Gamma^aU_{dR}U^\alpha_{a}\, ,\\
    N_u^{\alpha}&=\frac{v}{\sqrt{2}}U^\dagger_{uL}\Delta^aU_{uR}U^{\alpha*}_{a},\;\;\;N_e^{\alpha}=\frac{v}{\sqrt{2}}U^\dagger_{eL}\Gamma^a_eU_{eR}U^\alpha_{a}\, .\nonumber\nonumber
\end{align}
$V$ %and $V_l$ are 
is the CKM 
%and PMNS matrices, 
matrix, $\Gamma^a$, $\Delta^a$ and $\Gamma^a_e$ are the Yukawa coupling matrices, and $U_{fL/R}$ are the rotation matrices to the physical basis. We ignore neutrino masses for simplicity.
To calculate the Higgs decays, the relevant terms may be written as
%
% OLD form
% \begin{align}
%     \mathcal{L}\supset-\sum_f\Bar{\psi}_f^i(A^{f\beta}_{ij}+
%     i\gamma_5B^{f\beta}_{ij})\psi_f^jS_\beta^0
%     \supset-\sum_{fi}\left(\sqrt{2}G_\mu\right)^{\frac{1}{2}}
%     m_f\Bar{\psi}_f^i(a^{\beta}_{fi}+i\gamma_5b^{\beta}_{fi})\psi_f^iS_\beta^0
% \end{align}
%
\begin{align}\label{eq:f}
    \mathcal{L}\supset-\sum_{f}\left(\sqrt{2}G_\mu\right)^{\frac{1}{2}}
    m_f\Bar{\psi}_f(a^{\beta}_{f}+i\gamma_5b^{\beta}_{f})\psi_fS_\beta^0\, ,
\end{align}
where $m_f$ are the fermion masses, $G_\mu$ is the Fermi constant,
satisfying $\left(\sqrt{2}G_\mu\right)^{-\frac{1}{2}}= v$, and
\begin{align}
    a^{\beta}_f&=\frac{v}{2 m_f}
    (R^{f\beta}+L^{f\beta})\, ,\qquad
    b^{\beta}_f=-i\frac{v}{2 m_f}
    (R^{f\beta}+L^{f\beta})\, ,\nonumber\\
    R^{f\beta}&=\frac{1}{v}N_f^\alpha B_\alpha^\beta\, ,\;\;\;\;\;\;\;
    L^{f\beta}=\frac{1}{v}N_f^{\dagger\alpha} B_\alpha^{\beta*}\, ,\;\;\;\;\; f=d,e\, ,\\
    R^{u\beta}&=\frac{1}{v}N_u^\alpha B_\alpha^{\beta*}\, ,\;\;\;\;\;
    L^{u\beta}=\frac{1}{v}N_u^{\dagger\alpha} B_\alpha^{\beta}\, .\nonumber
\end{align}

\section{\label{sec:Zee}The scalar sector of the Zee model}

As an example, we look at a particular case of the Zee
model \cite{zee:1980ai} consisting of a type II 2HDM with a complex
singly charged singlet scalar. In a type II 2HDM, the fields satisfy a
$Z_2$ symmetry where $\phi_2$ and $u_R$ transform as $\psi\to-\psi$,
while the other fields do not transform under the symmetry. This means
that $\phi_2$ will only couple to the up type quarks while $\phi_1$
will only couple to the rest of the fermions.

Our purpose is not to make a global fit to the quark, scalar and
also the lepton sectors of any specific Zee model,
but rather to highlight those features of such types of model that
could be probed at LHC.
As a result, we do not explore the bounds coming
from the lepton sector, including neutrino oscillations;
an analysis which can be found, for example,
in Refs.~\cite{Herrero-Garcia:2017xdu,Babu:2019mfe}.
These references simplify the analysis by effectively using the $Z_2$ symmetry in the
quark sector, which is helpful to fix the production and some
branching ratios at LHC.
Those simulations also assume some scalar couplings to vanish,
effectively bringing the result close to that in the $Z_2$ scalar sector
used here.
For simplicity, we take couplings consistent with
$Z_2$ in the quark-scalar sectors,
reducing the number of parameters to scan,
and simplifying the analysis of some theoretical constraints,
such as bounded from below (BFB) and absence of charge breaking (CB) vacua.
Our main result,
the importance of searching for the decay $H_2^+ \to H_1^+ Z$,
is not affected by this simplification.

\subsection{The Higgs potential and rotation matrices}

The Higgs potential can in general be written as a particular case of
eq.~(\ref{eq:pot}), 
\begin{align}
  \label{eq:ZeePot}
    V=\;&m_C^2\chi^+\chi^-+\lambda_C(\chi^+\chi^-)^2+
    \left[\mu_4\; \phi_1 i \sigma_2 \phi_2 \chi^-
    +h.c.\right]
  +m_1^2\phi_1^\dagger\phi_1+m_2^2\phi_2^\dagger\phi_2\nonumber\\
 &  -m_{12}^2\left(\phi_1^\dagger\phi_2+\phi_2^\dagger\phi_1\right)
    +\left[k_{1}\phi_1^\dagger\phi_1+k_{2}\phi_2^\dagger\phi_2
      -k_{12}\left(\phi_1^\dagger\phi_2
        +\phi_2^\dagger\phi_1\right)\right]\chi^+\chi^-\\ 
    &+\frac{\lambda_1}{2}\left(\phi_1^\dagger\phi_1\right)^2
    +\frac{\lambda_2}{2}\left(\phi_2^\dagger\phi_2\right)^2 
    +\lambda_3\phi_1^\dagger\phi_1\phi_2^\dagger\phi_2
    +\lambda_4\phi_1^\dagger\phi_2\phi_2^\dagger\phi_1+\frac{\lambda_5}{2} 
    \left[\left(\phi_1^\dagger\phi_2\right)^2
      +\left(\phi_2^\dagger\phi_1\right)^2\right]\,  ,   
    \nonumber
\end{align}
where we generalized the 2HDM potential with a $Z_2$ symmetry
in \cite{Branco:2011iw}.
For simplicity, we consider all parameters and vevs real,
corresponding to CP conservation.

Allowing the doublets to develop vevs, the minimum conditions read
\begin{align}
    m_1^2 &=
    \frac{2m_{12}^2v_2-v_1^3\lambda_1-v_1v_2^2(\lambda_3+\lambda_4+\lambda_5)}{2v_1}\, ,\label{eq:26}\\
    m_2^2 &=
    \frac{2m_{12}^2v_1-v_2^3\lambda_1-v_2v_1^2(\lambda_3+\lambda_4+\lambda_5)}{2v_2}\, .
    \nonumber
\end{align}

The analytic expressions for the mass matrices have no inherent interest,
so we will just state some of their properties, while defining the
rotation to the physical basis.
First, we note that CP-odd fields do not mix with the CP-even fields.
The mass matrix for the CP-odd fields has the eigenvectors
$(v_1,v_2)$ and $(v_2,-v_1)$,
with the first corresponding to a null eigenvalue, which
is the Goldstone boson. We can transform the fields
into the physical mass basis through\footnote{For convenience, we place
  the pseudoscalar as the last of the neutral scalars. So $ S^0_4$ is
  the CP odd scalar and $S^0_2,S^0_3$ are the two CP even
  eigenstates. Notice that this choice affects the order of the columns
  in the matrix $V$ in eq.~(\ref{eq:matV}).}
\begin{align}
    \begin{pmatrix}
    S^0_1\equiv G^0\\
    S^0_4\equiv A
    \end{pmatrix}
    =
    \begin{pmatrix}
    \cos{\beta} & \sin{\beta}\\
    -\sin{\beta} & \cos{\beta}
    \end{pmatrix}
    \begin{pmatrix}
    \text{Im}{\varphi_1^0{}'}\\
    \text{Im}{\varphi_2^0{}'}
    \end{pmatrix}
    \equiv
    \mathcal{O}_{\beta}
    \begin{pmatrix}
    \text{Im}{\varphi_1^0{}'}\\
    \text{Im}{\varphi_2^0{}'}
    \end{pmatrix}\, ,
\end{align}
where
\begin{align}
    \cos{\beta}=v_1/v,\;\;\;\sin{\beta}=v_2/v,\;\;\;v=\sqrt{v_1^2+v_2^2}\, .
\end{align}
By applying the same rotation to the doublets' charged scalars
\begin{align}
    \begin{pmatrix}
    S^+_1\equiv G^+\\
    H^+
    \end{pmatrix}
    =
    \begin{pmatrix}
    \cos{\beta} & \sin{\beta}\\
    -\sin{\beta} & \cos{\beta}
    \end{pmatrix}
    \begin{pmatrix}
    \varphi_1^+\\
    \varphi_2^+
    \end{pmatrix}\, ,
\end{align}
we find the charged Goldstone boson $G^+$,
and the intermediate field $H^+$, not yet a mass eigenstate.
Finally, the remaining charged and neutral scalars
do not follow such simple relations. So we need to diagonalize,
in the general case, with two new independent angles
\begin{align}
    \begin{pmatrix}
    S^0_2\\
    S^0_3
    \end{pmatrix}
    =
    \begin{pmatrix}
    \cos{\alpha} & \sin{\alpha}\\
    -\sin{\alpha} & \cos{\alpha}
    \end{pmatrix}
    \begin{pmatrix}
    \text{Re}{\varphi^0_1{}'}\\
    \text{Re}{\varphi^0_2{}'}
    \end{pmatrix}
    \equiv
    \mathcal{O}_{\alpha}
    \begin{pmatrix}
    \text{Re}{\varphi^0_1{}'}\\
    \text{Re}{\varphi^0_2{}'}
    \end{pmatrix}\, ,
\end{align}
\begin{align}
    \begin{pmatrix}
    S^+_2\\
    S^+_3
    \end{pmatrix}
    =
    \begin{pmatrix}
    \cos{\gamma} & \sin{\gamma}\\
    -\sin{\gamma} & \cos{\gamma}
    \end{pmatrix}
    \begin{pmatrix}
    H^+\\
    \chi^+
    \end{pmatrix}
    \equiv
    \mathcal{O}_{\gamma}
    \begin{pmatrix}
    H^+\\
    \chi^+
    \end{pmatrix}\, .
\end{align}
Note that, if we had applied the rotation by $\beta$ initially to
the doublets themselves,
we would get to the so-called Higgs basis \cite{Botella:1994cs}.

Inverting all the transformations and joining the two charged
transformation above, we find that the matrices defined
in eqs.~(\ref{eq:basis_change})-(\ref{eq:matrices}) are
\begin{align}
  \label{eq:matV}
    V=&
    \begin{pmatrix}
    i\cos{\beta}  & \cos{\alpha} & -\sin{\alpha}& -i\sin{\beta}\\
    i\sin{\beta}  & \sin{\alpha} & \cos{\alpha}& i\cos{\beta}
  \end{pmatrix}\, ,\\
  \label{eq:matU}
    U=&
    \begin{pmatrix}
    \cos\beta & -\sin\beta\cos\gamma & \sin\beta\sin\gamma\\
    \sin\beta & \cos\beta\cos\gamma & -\cos\beta\sin\gamma
  \end{pmatrix}\, ,\\
  \label{eq:matT}
    T=&
    \begin{pmatrix}
    0 & \sin\gamma & \cos\gamma
    \end{pmatrix}\, .
\end{align}
Some of the relevant combinations of these matrices that
appear in the Lagrangian terms calculated in the previous section are
\begin{align}
  \label{eq:UdagU}
    U^\dagger U=&
    \begin{pmatrix}
    1 & 0 & 0\\
    0 & \cos^2{\gamma} & -\sin{\gamma}\cos{\gamma}\\
    0 & -\sin{\gamma}\cos{\gamma} & \sin^2{\gamma}
  \end{pmatrix}\, ,\\
    \label{eq:UdagV}
    B=U^\dagger V=&
    \begin{pmatrix}
    i  & \cos(\beta-\alpha) & \sin(\beta-\alpha)& 0\\
    0  & -\cos\gamma\sin(\beta-\alpha) &
    \cos\gamma\cos(\beta-\alpha)& i\cos\gamma\\ 
    0  & \sin\gamma\sin(\beta-\alpha) &
    -\sin\gamma\cos(\beta-\alpha) & -i\sin\gamma 
    \end{pmatrix}\, ,\\
    \text{Re}{\omega^\dagger V}=&
    \begin{pmatrix}
    0  & \cos(\beta-\alpha) & \sin(\beta-\alpha) &0 
    \end{pmatrix}\, .
\end{align}
Note that,
if we had started by bringing the doublets to the Higgs basis,
and then defined $\alpha$ as the rotation of the neutral CP-even fields
from that basis to the physical one,
then $\alpha$ would transform as $\alpha\to\alpha+\beta$,
and these matrices would become independent of $\beta$.

The non diagonal nature of $U^\dagger U$ is what gives rise
to the flavour changing coupling of the charged scalars with
the $Z$ boson, adding a new type of diagrams to
the process $h\to Z\gamma$ when compared to the general NHDM.
In that same sense,
the non mixture of the first component of that matrix with
the rest ensures that the Goldstone bosons do
not take part on those flavour changing couplings,
so that the diagrams involving the $W$ bosons remain safely of
the same nature.

\subsection{The choice of independent parameters}

The Higgs potential of eq.~(\ref{eq:pot}), after using the minimization
eqs.~(\ref{eq:26}) has twelve real independent parameters,
\begin{equation}
  \label{eq:27}
  m_C^2,\lambda_C,\mu_4,m_{12}^2,k_{1},k_{2},k_{12},
  \lambda_1,\lambda_2,\lambda_3,\lambda_4,\lambda_5 \, .
\end{equation}
For phenomenological studies it is convenient to trade some of these
parameters for the physical masses of the neutral and charged
scalars: $m_{H_1^0},m_{H_2^0},m_{A^0},m_{H_1^+}$, and $m_{H_2^+}$. This
follows a standard procedure. We just give the example of the mass
matrix for the pseudo-scalars. We have
\begin{equation}
  \label{eq:28}
  \mathcal{L}\supset -\frac{1}{2} \left[\text{Im}{\varphi_1^0{}'},
    \text{Im}{\varphi_2^0{}'}\right] M^2_P
    \begin{bmatrix}
      \text{Im}{\varphi_1^0{}'}\\
    \text{Im}{\varphi_2^0{}'} 
    \end{bmatrix}+ \cdots \ ,
\end{equation}
where
\begin{equation}
  \label{eq:29}
   M^2_P =
   \begin{bmatrix}\displaystyle
     \frac{v_2}{v_1}  \left(m^2_{12} - \lambda_5 v_1 v_2\right)&
     -m^2_{12} + \lambda_5 v_1 v_2\\
     -m^2_{12} + \lambda_5 v_1 v_2 &
     \displaystyle
     \frac{v_1}{v_2}\left(m^2_{12} - \lambda_5 v_1 v_2\right)
   \end{bmatrix}\, .
\end{equation}
Now using
\begin{equation}
  \label{eq:30}
  \begin{bmatrix}
      \text{Im}{\varphi_1^0{}'}\\
    \text{Im}{\varphi_2^0{}'} 
    \end{bmatrix} =\mathcal{O}_\beta^T
    \begin{bmatrix}
      G^0\\
      A^0
    \end{bmatrix}\, ,
  \end{equation}
we obtain
\begin{equation}
  \label{eq:31}
  \mathcal{O}_{\beta} M^2_P \mathcal{O}_\beta^T =
  \begin{bmatrix}
    0 &0\\
    0& m^2_{A^0}
  \end{bmatrix}\, .
\end{equation}
From here we can get $\lambda_5$ as a function of the mass $m_{A^0}$ and other
independent parameters,
\begin{equation}
  \label{eq:32}
  \lambda_5 = \frac{1}{v^2}\left(-m^2_{A^0}
    +\frac{m^2_{12}}{\sin\beta \cos\beta}\right)\, .
\end{equation}
Following this procedure for the other mass matrices we can solve for
the other $\lambda$'s as well as for $\mu_4, m_C^2$. We find
\begin{subequations}
\begin{align}
  \lambda_1 =&\frac{1}{v^2 \cos^2\beta}
  \left(m^2_{H_1^0} \cos^2\alpha + m^2_{H_2^0} \sin^2\alpha - 
     m^2_{12} \tan\beta\right),\\[+2mm]
   \lambda_2 =& \frac{1}{v^2\sin^2\beta}
   \left(m^2_{H_2^0} \cos^2\alpha^2 - m^2_{12} \cot\beta + 
  m^2_{H_1^0} \sin^2\alpha\right),\\[+2mm]
\lambda_3 =& \frac{1}{v^2}\left(2m_{H_1^+}^2\cos^2\gamma+2m_{H_1^+}^2\sin^2\gamma
-\frac{m_{12}^2+(m_{H_2^0}^2-m_{H_1^0}^2)
\cos\alpha\sin\alpha}{\sin\beta\cos\beta}\right),\\[+2mm] 
\lambda_4 =& -\frac{1}{v^2}
\left(\lambda_5 v^2 + 2 m^2_{H_1^+} \cos^2\gamma - \frac{2
    m^2_{12}}{\sin\beta \cos\beta}
 +   2 m^2_{H_2^+} \sin^2\gamma\right), \\[+2mm]
\mu_4 =& -\frac{\sqrt{2}}{v} (m^2_{H_1^+} - m^2_{H_2^+}) \cos\gamma
 \sin\gamma,\\[+2mm] 
 m^2_C =& -\frac{1}{2} k_{1} v^2 \cos^2\beta  + 
   k_{12} v^2 \cos\beta \sin\beta\ - \frac{1}{2} k_{2} v^2
   \sin^2\beta \nonumber\\[1mm]
   &+  m^2_{H_1^+} \sin^2\gamma  + m^2_{H_2^+} \cos^2\gamma\, .
\end{align}
\end{subequations}
This choice is, of course, not unique but it is a convenient one. In
the end,  our
set of twelve independent parameters is
\begin{equation}
  \label{eq:33}
  m_{H_1^0}, m_{H_2^0}, m_{A^0}, m_{H_1^+}, m_{H_2^+},
  \alpha, \beta, \gamma, \lambda_C, k_{1}, k_{2}, k_{12}\, .
\end{equation}

\subsection{Fermion couplings to scalars}

The Yukawa couplings to the quarks can be written as
\begin{equation}
    -\mathcal{L}_Y=
    \Bar{Q}_L\; \Tilde{\phi}_2Y_uu_R
    + \Bar{Q}_L\; \phi_1Y_dd_R+h.c.
    %+\Bar{u}_RY^\dagger_u\Tilde{\phi}^\dagger_2Q_L+\Bar{d}_RY_d^\dagger\phi^\dagger_1Q_L
\end{equation}
Going to the charged physical basis, we find the couplings
\begin{equation}
    -\mathcal{L}_Y\supset\frac{\sqrt{2}V_{ud}}{v}\;
    \Bar{u}\; (m_u\xi^u_AP_L+m_d\xi^d_AP_R)\; d\;
    (\cos{\gamma}S_1^+ -\sin{\gamma}S_2^+) + h.c.\, ,
\end{equation}
where,
with these definitions,
\begin{equation}
  \label{eq:51}
    \xi^u_A=\cot{\beta}\, ,\ \ \ \  \xi^d_A=\tan{\beta}\, .
\end{equation}
These are exactly the 2HDM couplings of fermions to
the only charged scalar existent in that case:
$H_\textrm{2HDM}^+$ \cite{Branco:2011iw}.
We re-obtain them with the substitution
$(\cos{\gamma}S_1^+ -\sin{\gamma}S_2^+) \rightarrow H_\textrm{2HDM}^+$.
Said otherwise, the vertices $udS_1^+$ and $udS_2^+$ are the same
as the 2HDM vertex $udH_\textrm{2HDM}^+$, but with the factors
$\cos{\gamma}$ and $-\sin{\gamma}$, respectively.
This is not surprising.
Indeed, the combination of scalars appearing above corresponds to the
$H^+$ field. This field is the one we find in the doublets when
in the Higgs basis and so the result is the same as treating the
model as we would treat the 2HDM, and then replace the charged scalar
by this combination.

\section{\label{sec:constraints}Constraints on the Model}

\subsection{Theoretical Constraints}

\subsubsection{Bounded from Below}

The necessary and sufficient conditions for the potential to be
bounded from below (BFB) are know \cite{Kanemura:1993hm,Ferreira:2004yd} for the
neutral part of the potential, that coincides with the 2HDM. They are
\begin{equation}
  \label{eq:39}
     \lambda_1 \geq  0,\quad \lambda_2 \geq  0,\quad
      \lambda_3+\sqrt{\lambda_1 \lambda_2} \geq 0,\quad
      \lambda_3+\lambda_4- |\lambda_5| +\sqrt{\lambda_1 \lambda_2}\geq 0\, .
\end{equation}
For the Zee model they were studied in ref.~\cite{Barroso:2005hc}. They
extended the conditions in eq.~(\ref{eq:39}) but were not able to find
necessary and sufficient conditions, only necessary conditions. To
explain these conditions it is better to use their notation and
indicate the correspondence with ours. They write the quartic part of
the potential as
\begin{align}
  \label{eq:41}
  V_{\rm Q}=& b_{00} x_0^2 +b_{11} x_1^2 +b_{22} x_2^2 +b_{33} x_3^2+b_{44} x_4^2
  \nonumber \\[+1mm]
 &+ b_{01} x_0 x_1 + b_{02} x_0 x_2 + b_{03} x_0 x_3 + b_{12} x_1 x_2
 + b_{13} x_1 x_3 + b_{23} x_2 x_3\, ,  
\end{align}
where
\begin{align}
  \label{eq:42}
  x_0=|\chi^+|^2,\ x_1=|\phi_1|^2,\ x_2=|\phi_2|^2,\
  x_3=\text{Re}(\phi_1^\dagger \phi_2),\
  x_4=\text{Im}(\phi_1^\dagger \phi_2)\ .
\end{align}
Comparing with the potential in eq.~(\ref{eq:ZeePot}) we obtain
\begin{align}
  \label{eq:43}
  &b_{00}= \lambda_C, \ b_{11}=\frac{1}{2} \lambda_1, \
  b_{22}=\frac{1}{2} \lambda_2, \ b_{33}=\lambda_4+\lambda_5, \
  b_{44}=\lambda_4-\lambda_5, \nonumber\\
  &b_{01}=k_1,\ b_{02}=k_2,\ b_{03}=-2 k_{12}, \
  b_{12}=\lambda_3,\ b_{13}=0,\ b_{23}=0\, .  
\end{align}
They found the following necessary conditions for the potential to be
BFB,
\begin{subequations}
\begin{align}
  &b_{11}\geq 0,\ b_{22}\geq 0,\ b_{12}\geq -2\sqrt{b_{11} b_{22}},
  b_{12}+b_{44} \geq  -2 \sqrt{b_{11} b_{22}},\
  b_{12}+b_{33} \geq  -2 \sqrt{b_{11} b_{22}}, \label{eq:a}\\[+2mm]
  & b_{01}\geq  -2 \sqrt{b_{00} b_{11}},\ 
  b_{02}\geq  -2 \sqrt{b_{00} b_{22}},\ f(\alpha,\theta) \geq 0,\
  \forall_{\alpha,\theta}. 
\end{align}
\end{subequations}
where
\begin{align}
  \label{eq:44}
  f(\alpha,\theta)=&\ \frac{1}{8} b_{03} \sin 2\theta \sin^2 2\alpha +
  \frac{1}{4} \left( b_{01} \cos^2\theta + b_{02} \sin^2\theta \right)
  \sin^2 2\alpha\nonumber\\[+2mm]
  &+ \left[b_{11} \cos^4\theta+ b_{22}\sin^4\theta
    + \frac{1}{4}\left( b_{12}+b_{33}\right)
    \sin^22\theta\right] \sin^4\alpha\, .
\end{align}
It is easy to verify that the conditions in eq.~(\ref{eq:a})
correspond to the usual conditions for the 2HDM in
eq.~(\ref{eq:39}).
The others are new for the Zee model.
The condition in eq.~(\ref{eq:44}) cannot
be solved analytically for the $b_{ij}$. Therefore we took a large random
sample of $\theta$ and $\alpha$ and excluded points that have
$f(\alpha,\theta)<0$. As explained in ref.~\cite{Barroso:2005hc}, even
after applying these constraints there are a few points that are still
not BFB. We have verified this fact when considering the analysis of the
charged breaking minima in the following section, and he have also discarded
those points.

\subsubsection{Charged Breaking Minima}

The analysis of the charged breaking (CB) minima it is much more
complicated that in 2HDM \cite{Ferreira:2004yd} because of cubic term
in potential. Indeed, contrary to
the 2HDM, the condition
\begin{equation}
  \label{eq:3}
  V_{CB} > V_N
\end{equation}
is not guaranteed to be verified even when we are at the normal neutral
minimum, $V_N$.
As it is very complicated (if not impossible) to solve a set of nonlinear 
equations for the stationary points of $V_{CB}$,
we took a different approach,
based on ref.~\cite{Barroso:2005hc}.
We parameterize the possible charged break minima as
\begin{equation}
  \label{eq:45}
  \phi_1=
  \begin{bmatrix}
   y_1\\y_2 
 \end{bmatrix},\
  \phi_2=
  \begin{bmatrix}
   y_3\\y_4 
 \end{bmatrix},\
   \chi^+=y_5 \, .
 \end{equation}
Then, for the parameters for which we have a normal minimum $V_N$,
 \begin{equation}
   \label{eq:46}
   \text{Set}_{\rm min}=m_1^2, m_2^2,
   m_C^2,\lambda_C,\mu_4,m_{12}^2,k_{1},k_{2},k_{12},
  \lambda_1,\lambda_2,\lambda_3,\lambda_4,\lambda_5 \, ,
 \end{equation}
we consider the function $V_\textrm{other}(\text{Set}_{\rm min}, y_i)$. We start
by taking a large set of random values for $y_i$
\begin{equation}
  \label{eq:47}
  y_i \in [-1000,1000]\, \text{GeV}\, ,
\end{equation}
and then for each of these initial values we apply the method of gradient
descent to obtain the lowest possible value for $V_\textrm{other}(y_i)$ and compare
it with $V_N$. If $V_N < V_\textrm{other}$ we keep the point. In doing this we
also verified the claim \cite{Barroso:2005hc} that the BFB conditions are
not sufficient, as we found a small amount of points corresponding
to potentials unbounded from below.

There is a final point deserving a comment. When doing the
procedure described above, in many cases we got to a point where $y_5=0$
(of course numerically there is no such thing as zero and we have
considered $|y_5| < 10^{-6}$). As $y_1$ and $y_3$ are non-zero, the
question is if this is really a charged breaking minimum or not. We can make
an SU(2) rotation to bring to zero the upper component of the first
doublet
\begin{equation}
  \label{eq:48}
  \begin{bmatrix}
    \cos\theta & \sin\theta\\
    -\sin\theta & \cos\theta
  \end{bmatrix}
  \begin{bmatrix}
    y_1\\y_2
  \end{bmatrix}
  =
  \begin{bmatrix}
    0\\
    y_2'
  \end{bmatrix},\quad \tan\theta = -\frac{y_1}{y_2}\, .
\end{equation}
Now, if the \textit{same} rotation on the second doublet also gives
\begin{equation}
  \label{eq:49}
  \begin{bmatrix}
    \cos\theta & \sin\theta\\
    -\sin\theta & \cos\theta
  \end{bmatrix}
  \begin{bmatrix}
    y_3\\y_4
  \end{bmatrix}
  =
  \begin{bmatrix}
    0\\
    y'_4
  \end{bmatrix}\, ,
\end{equation}  
and
\begin{equation}
  \label{eq:50}
  \sqrt{y'_2{}^2+y'_4{}^2} = \frac{v}{\sqrt{2}}\, ,
\end{equation}
then this is just a normal minimum.
In all occasions we found,
this was precisely the same normal minimum $V_N$ in a different
guise.\footnote{This explains why we used $V_\textrm{other}$ above,
and not $V_{CB}$.}
We have looked at these situations and
kept the points if these conditions were verified.

\subsubsection{Perturbative Unitarity}

To ensure perturbative unitarity of the quartic couplings we
implemented the general algorithm presented in
ref.~\cite{Bento:2017eti}.  As we are interested in the high energy
limit, one just needs to evaluate the scattering S-matrix for the two
body scalar bosons, and these arise exclusively from the quartic part
of the potential.  Since the electric charge and the hypercharge are
conserved in this high energy scattering, we can separate the states
according to these quantum numbers. In the notation of ref.~\cite{Bento:2017eti},
\begin{equation}
  \label{eq:35}
  \phi_i=
  \begin{bmatrix}
    w_i^+\\
    n_i
  \end{bmatrix},\quad
  \phi_i^\dagger=
  \begin{bmatrix}
    w_i^-\\
    n_i^*
  \end{bmatrix}^T,\quad
  \chi =\chi^+,\quad \chi^* = \chi^-\, .
\end{equation}
This corresponds to the following possibilities, 
\begin{subequations}
\begin{align}
Q=2,Y=1&&S_{\alpha}^{++}=&\{w_1^+ w_1^+,w_1^+ w_2^+,w_1^+\chi^+,w_2^+
w_2^+,w_2^+ \chi^+,\chi^+ \chi^+\}, \label{equnit:8a}\\
Q=1,Y=1&&S_{\alpha}^{+}=&\{w_1^+ n_1,w_1^+ n_2,w_2^+ n_1,w_2^+ n_2,
\chi^+ n_1, \chi^+ n_2\}, \label{equnit:8b}\\
Q=1,Y=0&&T_{\alpha}^{+}=&
\{w_1^+ n^*_1,w_1^+ n^*_2,w_2^+ n^*_1,w_2^+ n^*_2,\chi^+ n^*_1,\chi^+n^*_2\},
\label{equnit:8c}\\
Q=0,Y=1&&S_{\alpha}^{0}=&\{n_1 n_1,n_1 n_2,n_2 n_2\}, \label{equnit:8d}\\
Q=0,Y=0&&T_{\alpha}^{0}=&
\{w_1^- w_1^+,w_1^- w_2^+,w_1^- \chi^+,
w_2^- w_1^+,w_2^- w_2^+,w_2^- \chi^+,\nonumber\\
&& &\hskip 2mm
\chi^- w_1^+,\chi^- \chi^+,\chi^- \chi^+,
n_1n^*_1,n_1 n^*_2,n_2 n^*_1,n_2 n^*_2\}. \label{equnit:8e}
\end{align}
\end{subequations}
With this setup we have to find the scattering matrices for each
$(Q,Y)$ combination and their eigenvalues. Let us call this set
$\Lambda_i$. Then the perturbative unitarity constraints are
\begin{equation}
  \label{eq:34}
  \text{max}(\Lambda_i) < 8 \pi, \quad i=1,\ldots,19  .
\end{equation}
In appendix~\ref{sec:PertUni} we write explicitly the various
scattering matrices and their eigenvalues. In total we have 19
different eigenvalues, as we already anticipated in
eq.~(\ref{eq:34}). 

\subsubsection{The oblique parameters $S,T,U$}

All the points in parameter space have to satisfy the electroweak
precision measurements, using the oblique parameters S, T and
U. We demand that S, T and U are within 2$\sigma$ of the fit
given in \cite{Baak:2014ora}.
For general models with an arbitrary number of doublets and
singlets the expressions for the oblique parameters were
given in Refs.~\cite{Grimus:2007if,Grimus:2008nb}. They depend on
combinations of the matrices $V$ and $U$ defined in
eqs.~(\ref{eq:matV})-(\ref{eq:matU}). The needed matrices are $U^\dagger
U$ in eq.~(\ref{eq:UdagU}) , $U^\dagger V$ in eq.~(\ref{eq:UdagV}), and
\begin{equation}
  \label{eq:ImVdagV}
  \text{Im}V^\dagger V=
  \begin{pmatrix}
    0& -\cos(\beta -\alpha)& -\sin(\beta -\alpha)& 0\\
    \cos(\beta-\alpha)& 0& 0& -\sin(\beta-\alpha)\\
  \sin(\beta-\alpha)& 0& 0& \cos(\beta -\alpha)\\
  0& \sin(\beta-\alpha)& -\cos(\beta-\alpha)& 0
  \end{pmatrix}\, .
\end{equation}

\subsection{Constraints from the LHC}

From the LHC data we have two types of constraints. First we consider
the constraints on the $h_{\rm 125}$ Higgs boson. These are normally
enforced through the  signals strengths for each production mode
$i=\texttt{ggF},\texttt{VBF},\texttt{VH},\texttt{ttH}$
and final state $j=H\to \gamma\gamma,H\to ZZ,H\to ZZ,H\to \tau\tau,H\to
bb$, and are defined by 
\begin{equation}
  \label{eq:1}
  \mu_{ij}= \frac{\sigma_i(pp\to H)}{\sigma^{\rm SM}_i(pp\to H)}
            \frac{\text{BR}(H \to j)}{\text{BR}^{\rm SM}(H \to j)}
\end{equation}
\noindent
The values for the signals strengths are given in
Table~\ref{tab:multicol} and were taken from fig.~5 of
ref.~\cite{Aad:2019mbh}.
\begin{table}[htb]
\begin{center}
\begin{tabular}{|c|c|c|c|c|}
    \hline
  Decay &\multicolumn{4}{c|}{Production Processes}\\[+2mm]
  Mode & \texttt{ggF}&\texttt{VBF}&\texttt{VH}&\texttt{ttH}\\[+2mm]
  \hline
  \vb{18}   $H\to \gamma\gamma$
         &$0.96^{+0.14}_{-0.14}$
         &$1.39^{+0.40}_{-0.35}$ 
         &$1.09^{+0.58}_{-0.54}$
         &$1.10^{+0.41}_{-0.35}$ \\[+2mm]
   \vb{18}   $H\to ZZ$
         &$1.04^{+0.16}_{-0.15}$
         &$2.68^{+0.98}_{-0.83}$
         &$0.68^{+1.20}_{-0.78}$
         &$1.50^{+0.59}_{-0.57}$ \\[+2mm]
  \vb{18}  $H\to WW$
         &$1.08^{+0.19}_{-0.19}$
         &$0.59^{+0.36}_{-0.35}$
         &$-$
         &$1.50^{+0.59}_{-0.57}$ \\[+2mm] 
 \vb{18}   $H\to \tau\tau$
         &$0.96^{+0.59}_{-0.52}$
         &$1.16^{+0.58}_{-0.53}$
         &$-$
         &$1.38^{+1.13}_{-0.96}$ \\[+2mm]
   \vb{18}   $H\to bb$
          &$-$
          &$3.01^{+1.67}_{-1.61}$ 
          &$1.19^{+0.27}_{-0.25}$
          &$0.79^{+0.60}_{-0.59}$ \\[+2mm]
  \hline
\end{tabular}
\end{center}
\caption{Values for $\mu_{ij}$ taken from \cite{Aad:2019mbh}}
\label{tab:multicol}
\end{table}
The other type of constraints from the LHC data are the bounds on
other neutral and charged scalars. This we implemented using the most
recent version of \texttt{HiggBounds5} \cite{Bechtle:2020pkv}.

\subsection{Constraints from BR($B\to X_s \gamma$)}
\label{sec:bsgamma}

In models with charged scalar bosons it is well
known \cite{Borzumati:1998nx,Borzumati:1998tg,
  Misiak:2017bgg,Misiak:2018cec,Akeroyd:2020nfj} 
that the experimental limits on the BR($B\to X_s \gamma$) can put
important constraints in the parameter space of these models. For
instance, in ref.~\cite{Misiak:2017bgg} the bound
\begin{equation}
  \label{eq:16}
  m_{H^+}> 580\, \text{GeV}\, ,
\end{equation}
is derived for the type 2 2HDM at 95\% CL
(2$\sigma$). In fact the exact number depends on the errors both in
the theoretical calculation \cite{Bernlochner:2020jlt} as well in the
experimental errors. For instance, the result for the SM at NNLO
is \cite{Misiak:2020vlo,Akeroyd:2020nfj} 
\begin{equation}
  \label{eq:23}
  \text{BR}^{\rm SM}(B\to X_s \gamma) = (3.40 \pm 0.17) \times
  10^{-4}\, ,
\end{equation}
which shows an error of 5\%, to be compared with the world
average \cite{Amhis:2019ckw}
\begin{equation}
  \label{eq:25}
  \text{BR}^{\rm exp}(B\to X_s \gamma) = (3.32 \pm 0.15) \times
  10^{-4}\, .
\end{equation}
Here we take the approach of considering for the theoretical error
a band around the central value of the
calculation with an error of 2.5\%,  and
following \cite{Akeroyd:2020nfj}, for the experimental error,
we consider 99\%CL (3$\sigma$), that is,
\begin{equation}
  \label{eq:17}
  2.78 \times 10^{-4} < \text{BR}(B\to X_s \gamma) < 3.77 \times
  10^{-4}\, .
\end{equation}

\subsubsection{The calculation}

Our calculation follows closely the original calculation of
ref.~\cite{Borzumati:1998tg}. The central point in that calculation is
that the new contributions from the charged scalar bosons are encoded
in the Wilson coefficients,
\begin{subequations}   
\begin{align}
  \label{eq:18}
  C^{0,{\rm eff}}_7(\mu_W) =&
  C^{0,{\rm eff}}_{7,\rm SM}(\mu_W) +|Y|^2 C^{0,{\rm eff}}_{7,\rm YY}(\mu_W)
  +(X Y^*)C^{0,{\rm eff}}_{7,\rm XY}(\mu_W)\, , \\[+2mm]
  C^{0,{\rm eff}}_8(\mu_W) =&
  C^{0,{\rm eff}}_{8,\rm SM}(\mu_W) +|Y|^2 C^{0,{\rm eff}}_{8,\rm YY}(\mu_W)
  +(X Y^*)C^{0,{\rm eff}}_{8,\rm XY}(\mu_W) \, ,\\[+2mm]
  C^{1,{\rm eff}}_4(\mu_W) =&
  E_0(x) + \frac{2}{3} \log(\frac{\mu_W^2}{M_W^2} +|Y|^2 E_H(y)\, ,\\[+2mm]  
  C^{1,{\rm eff}}_7(\mu_W) =&
  C^{1,{\rm eff}}_{7,\rm SM}(\mu_W) +|Y|^2 C^{1,{\rm eff}}_{7,\rm YY}(\mu_W)
  +(X Y^*)C^{1,{\rm eff}}_{7,\rm XY}(\mu_W)\, , \\[+2mm]
  C^{1,{\rm eff}}_8(\mu_W) =&
  C^{1,{\rm eff}}_{8,\rm SM}(\mu_W) +|Y|^2 C^{1,{\rm eff}}_{8,\rm YY}(\mu_W)
  +(X Y^*)C^{1,{\rm eff}}_{8,\rm XY}(\mu_W) \, .
\end{align}
\end{subequations}   

All the expressions needed are given in
ref.~\cite{Borzumati:1998tg}. Also there one finds the way to evolve
these coefficients to the scale $\mu_b=m_b$.
The dependence on the charged scalar mass
appears because the functions, $C^{0,{\rm eff}}_{i,\rm YY},
C^{0,{\rm eff}}_{i,\rm XY}, C^{1,{\rm eff}}_{i,\rm YY},
C^{1,{\rm eff}}_{i,\rm XY}$, depend on $y=m_t^2/m_{H^+}^2$ while the SM
coefficients depend on $x= m_t^2/M_W^2$.

The generalization for models with more charged scalars is
straightforward. The case of two charged scalar bosons was considered in
ref.~\cite{Akeroyd:2020nfj}. We just give the example of
$C^{1,{\rm eff}}_7(\mu_W)$, all the other having similar expressions.
\begin{align}
  \label{eq:19}
  C^{1,{\rm eff}}_7(\mu_W) =&
  C^{1,{\rm eff}}_{7,\rm SM}(\mu_W) +|Y_1|^2 C^{1,{\rm eff}}_{7,\rm  YY}(\mu_W,y_1)
  +|Y_2|^2 C^{1,{\rm eff}}_{7,\rm YY}(\mu_W,y_2)\nonumber\\
& +(X_1 Y_1^*)C^{1,{\rm eff}}_{7,\rm XY}(\mu_W,y_1) 
  +(X_2 Y_2^*)C^{1,{\rm eff}}_{7,\rm XY}(\mu_W,y_2) \, ,
\end{align}
where $X_i,Y_i$ are defined in eq.~(\ref{eq:6}), taking the values
in eq.~(\ref{eq:7}) for the Zee model,
and we wrote explicitly the dependence on the
charged scalar masses,
\begin{equation}
  \label{eq:20}
  y_1= \frac{m_t^2}{m_{H_1^+}^2}, \quad
    y_2= \frac{m_t^2}{m_{H_2^+}^2}\, .
\end{equation}

An important point in the calculation is the value of the input
parameters. We took those of ref.~\cite{Borzumati:1998tg} except for
$\alpha_s(M_Z),m_t,M_Z,M_W$ that were updated to the values of the
PDG \cite{Zyla:2020zbs}. The values are
\begin{subequations}   
\begin{align}
  \label{eq:21}
 & \alpha_s(M_Z)= 0.1179\pm 0.0010, && m_t=172.76 \pm 0.3 \, \text{GeV}\, ,\\
  &m_c/m_b= 0.29 \pm 0.02 && m_b-m_c = 3.39 \pm 0.04\, \text{GeV}\, ,\\
  &\alpha_{em}^{-1}=137.036\pm && |V_{ts}^*V_{tb}/V_{cb}|^2=0.95\pm
  0.03\, ,\\
  &\text{BR}_{SL}= 0.1049\pm 0.0046\, . &&
\end{align}
\end{subequations}   
We should emphasize that, using the input values of
ref.~\cite{Borzumati:1998tg},
we were able to reproduce their results\footnote{We are
indebted to C. Greub for discussions and for having shared with us the
original code for cross checking our independent calculation. One
important point was that the parameter $\lambda_1=0.12\ \text{GeV}^2$ defined in ref.~\cite{Borzumati:1998tg}
should be positive.}  for the SM.

\subsubsection{The result for the 2HDM type 2}

\begin{figure}[htb]
  \centering
  \begin{tabular}{cc}
    \includegraphics[width=0.48\textwidth]{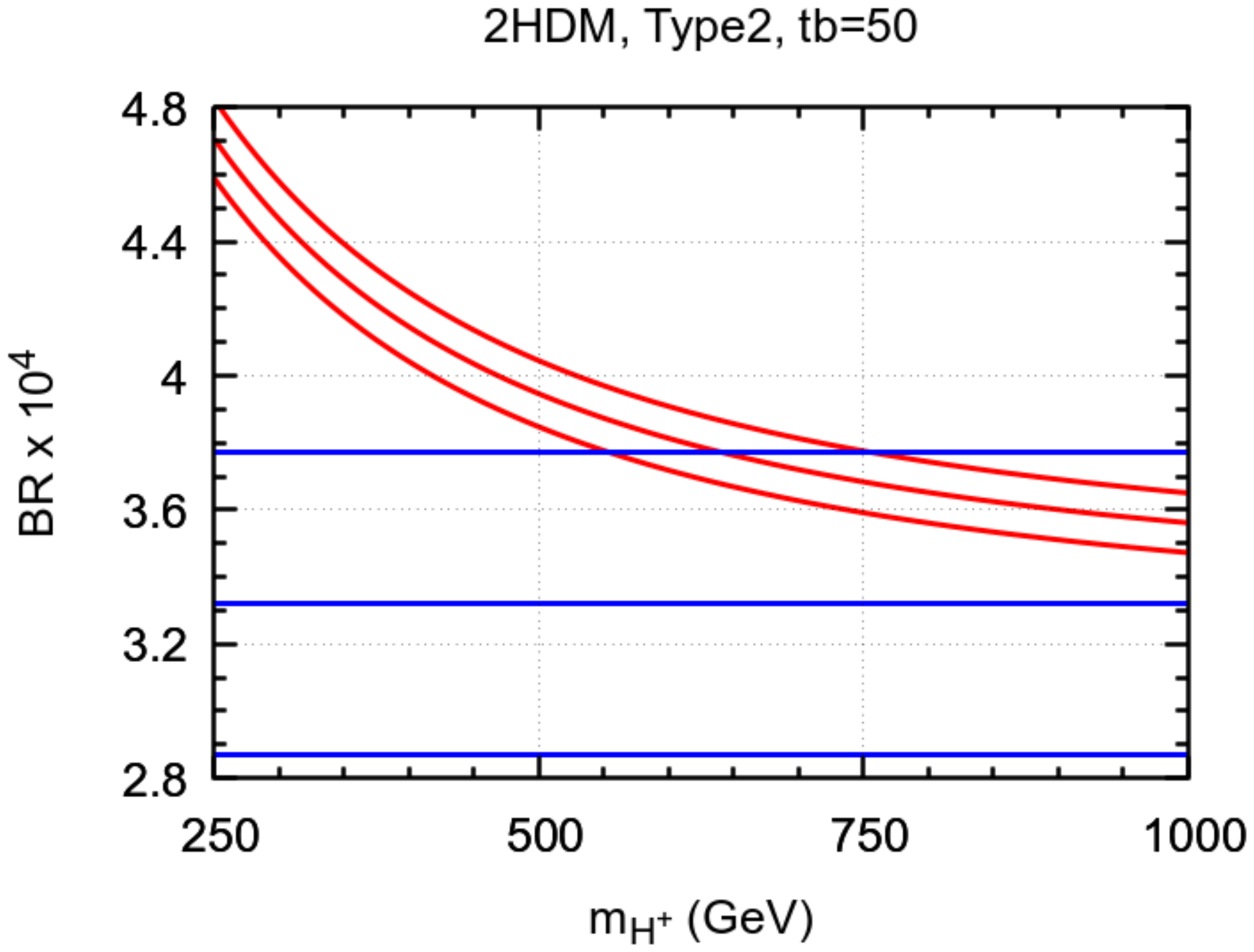} &
    \includegraphics[width=0.48\textwidth]{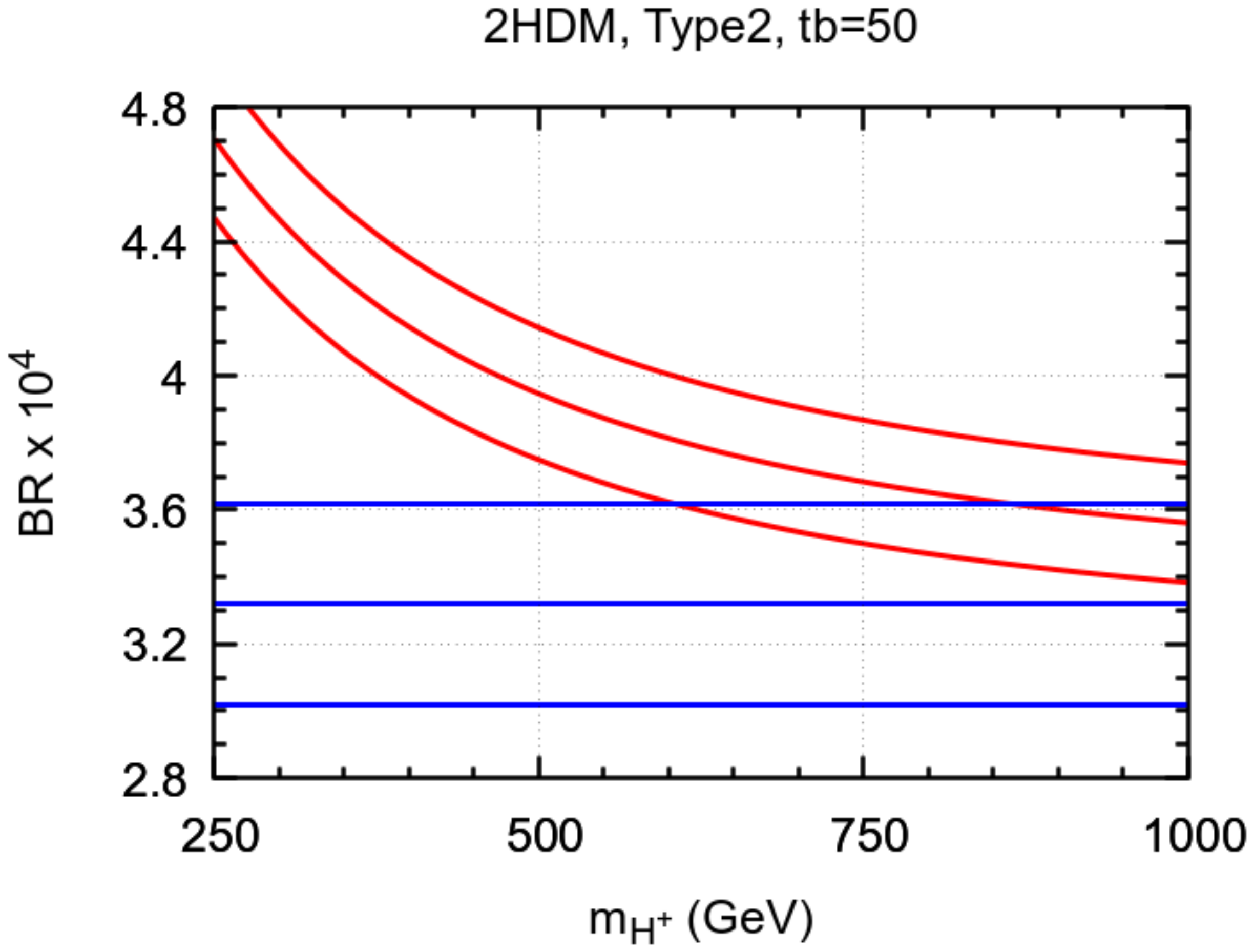}     
  \end{tabular}
   \caption{BR($B\to X_s \gamma$) as a function of the charged scalar
    mass. Left panel: The lines in blue represent the 3$\sigma$
    experimental limits, and those in red to 2.5\% error in the
    calculation. Right panel: The lines in blue represent the 2$\sigma$
    experimental limits, and those in red to 5\% error in the
    calculation.}
  \label{fig:1}
\end{figure}

First we considered the particular case of the 2HDM with type 2
couplings to fermions. In our model this is accomplished by setting
$\gamma=0$. Then the second Higgs decouples completely ($X_2=Y_2=0$)
and we have an effective 2HDM. The results are shown in fig.~\ref{fig:1}.
On the left panel we considered a band corresponding to 2.5\% in the
calculation and a 3$\sigma$ band for the experimental result.  On the
right panel we considered a band corresponding to 5\% in the
calculation and a 2$\sigma$ band for the experimental result.  We see
that the limit for the mass of the charged scalar that we get is
similar in both cases and also similar to what was obtained in
ref.~\cite{Misiak:2017bgg}

As we are not doing a NNLO calculation, our goal here is not to improve
the limit for the 2HDM with type 2 couplings. We just want to show
that in models with more charged scalars, as was addressed in
ref.~\cite{Akeroyd:2020nfj}, the limit in eq.~(\ref{eq:16})
can be relaxed for one of them and this will have implications for the
Zee model. We discuss this in the next section for the
case of the Zee model. For definiteness we take the choice on the
left panel of fig.~\ref{fig:1}.

\subsubsection{Implications for the Zee Model}

\begin{figure}[htb]
  \centering
  \begin{tabular}{cc}
    \includegraphics[width=0.48\textwidth]{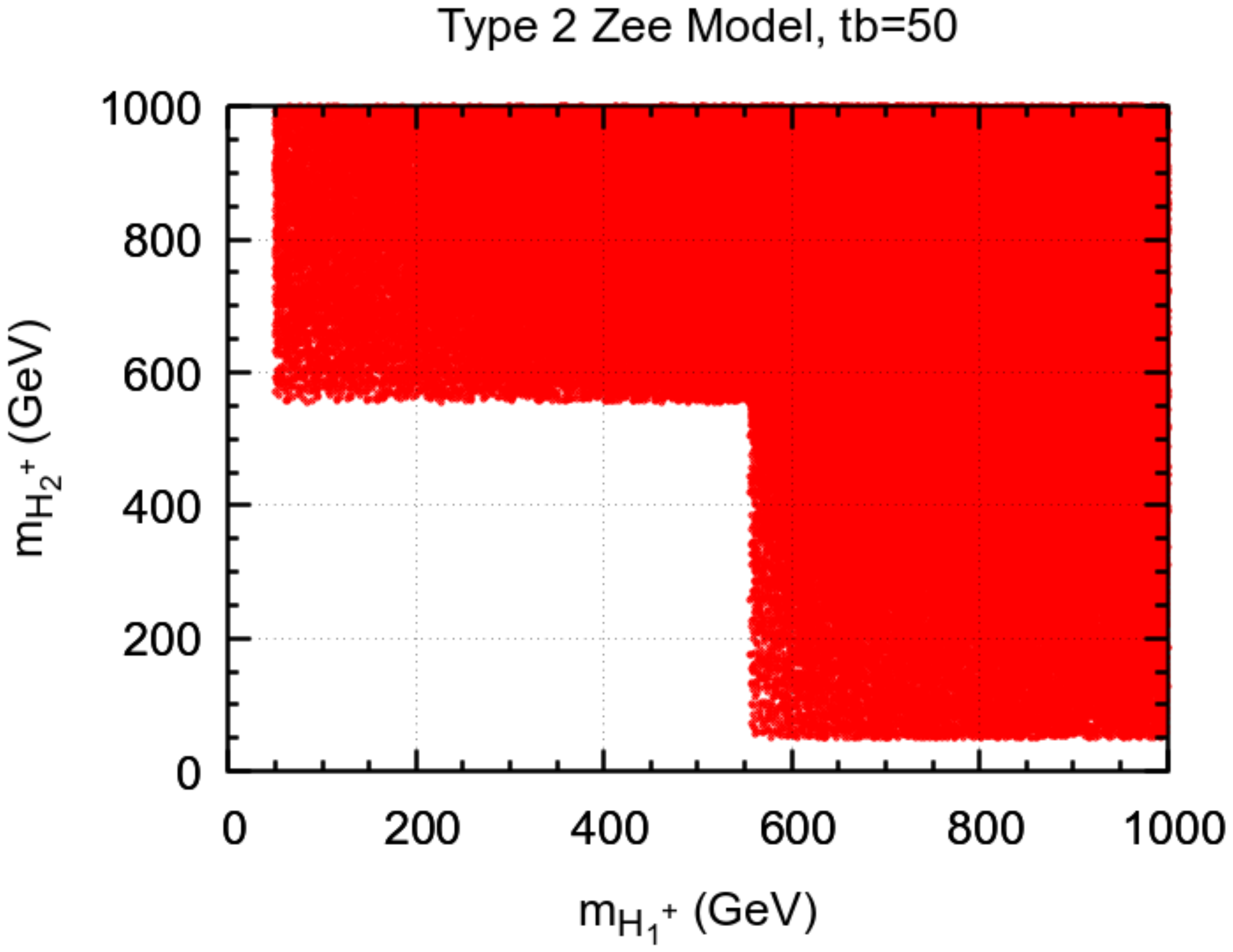} 
    &
    \includegraphics[width=0.48\textwidth]{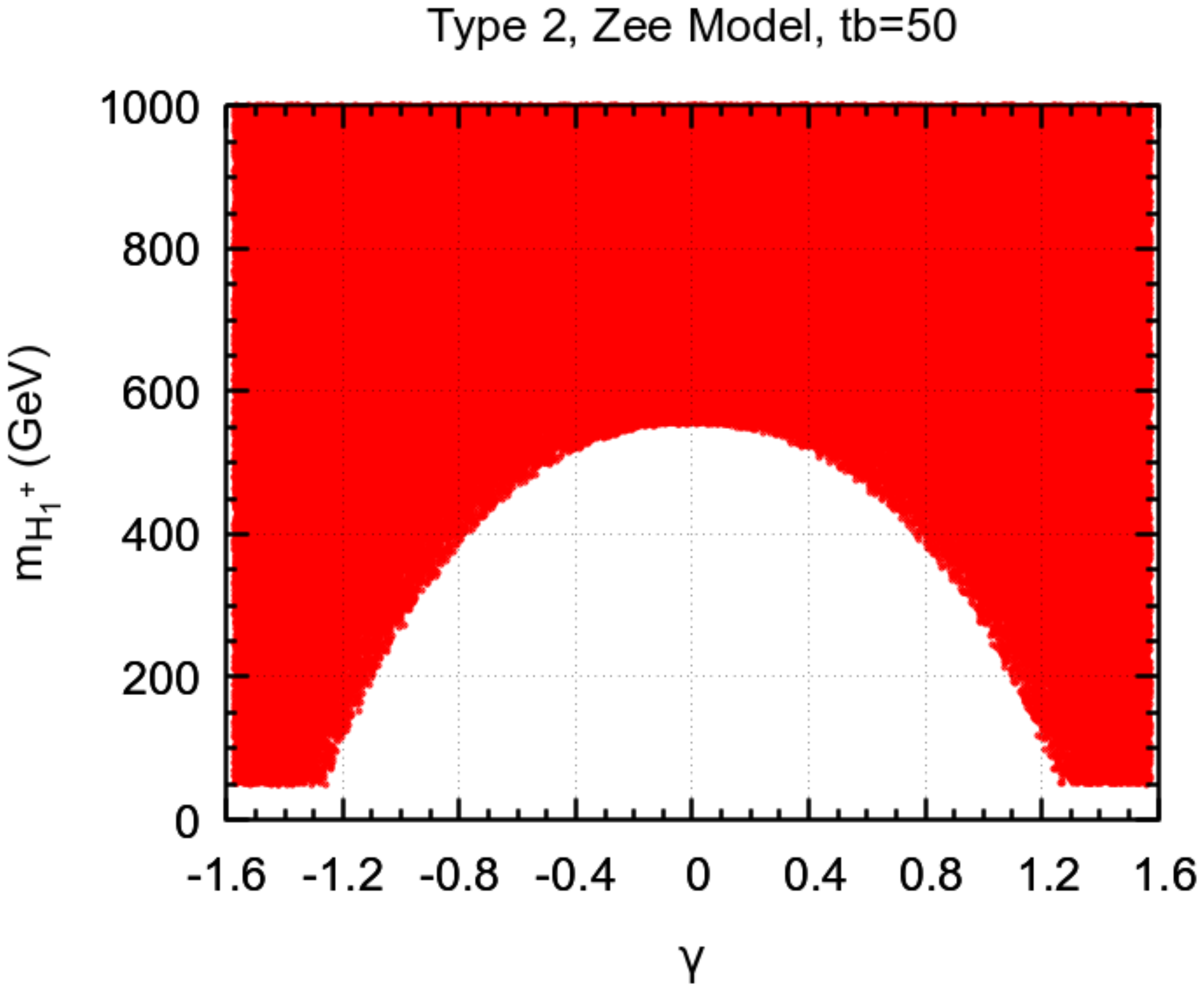} 
  \end{tabular}
  \caption{Left panel: points satisfying eq.~(\ref{eq:17}) for the
    Zee Model. Right panel: mass of the lightest charged scalar
    boson as a function 
    of the mixing angle $\gamma$.}
  \label{fig:2}
\end{figure}

We have just seen that in the case of having just one charged scalar
boson we have a limit for its mass coming from the
BR($B\to X_s\gamma$) for the case of 2HDM with type 2 fermion
couplings. Now we consider the case of the Zee model also with type 2
fermion couplings. We start by just considering the variation of the
masses and of the mixing angle $\gamma$ without imposing all the
theoretical and experimental constraints on the model. That will be
done below when we consider the discussion of benchmark points. Our
purpose here is just to show how the constraints from BR($B\to X_s\gamma$)
can be satisfied in the model. 
Although we can always choose $m_{H_1^+}< m_{H_2^+}$, we start by not
imposing that constraint. All points satisfying eq.~(\ref{eq:17})
are shown on the left panel of fig.~\ref{fig:2}.
We see that we have an exclusion for both masses to be below
the value found (in the 2HDM) with one single charged scalar,
but it is possible that one of the masses is lower
than 580 GeV if the other is above. This is a function of the mixing
angle $\gamma$ as shown on the right panel of fig.~\ref{fig:2}.
We see that $m_{H_1^+}$ can be as low as 50 GeV if the mixing angle is
close to $\pm \pi/2$. Notice that for $\gamma=0$ we recover the
previous result. As can be seen from fig.~\ref{fig:2}, when $m_{H_1^+}$
is low, the other mass has always to be above the 580 GeV limit.

This result means that for each point in parameter space we have to
evaluate the BR($B\to X_s\gamma$) to see if it passes the bounds in
eq.~(\ref{eq:17}), instead of using just one fixed limit for all
points, like in the 2HDM.

Notice that in the Zee model there is no exact cancelation of
the two charge scalar contributions. This is due to the fact
that there is only one charge scalar component coming from a
doublet, as we have explained after eq.~(\ref{eq:51}). In contrast,
in a 3HDM the charged scalars originate in two components of
 doublets and such a cancelation is indeed possible \cite{Boto:2021}.

There is a final comment. The charged Higgs contribute to $\Delta
M_{B_{s,d}}$, coming from the B meson oscillations.  We have not
considered this contribution from flavour data because, as shown
in \cite{Chakraborti:2021bpy}, they are important only for very low
$\tan\beta$, above what we get from the other constraints; see
fig.~\ref{fig:alfa-beta} below.

\subsection{Scanning strategy}

\begin{figure}[htb]
  \centering
  \begin{tabular}{cc}
\includegraphics[width=0.48\textwidth]{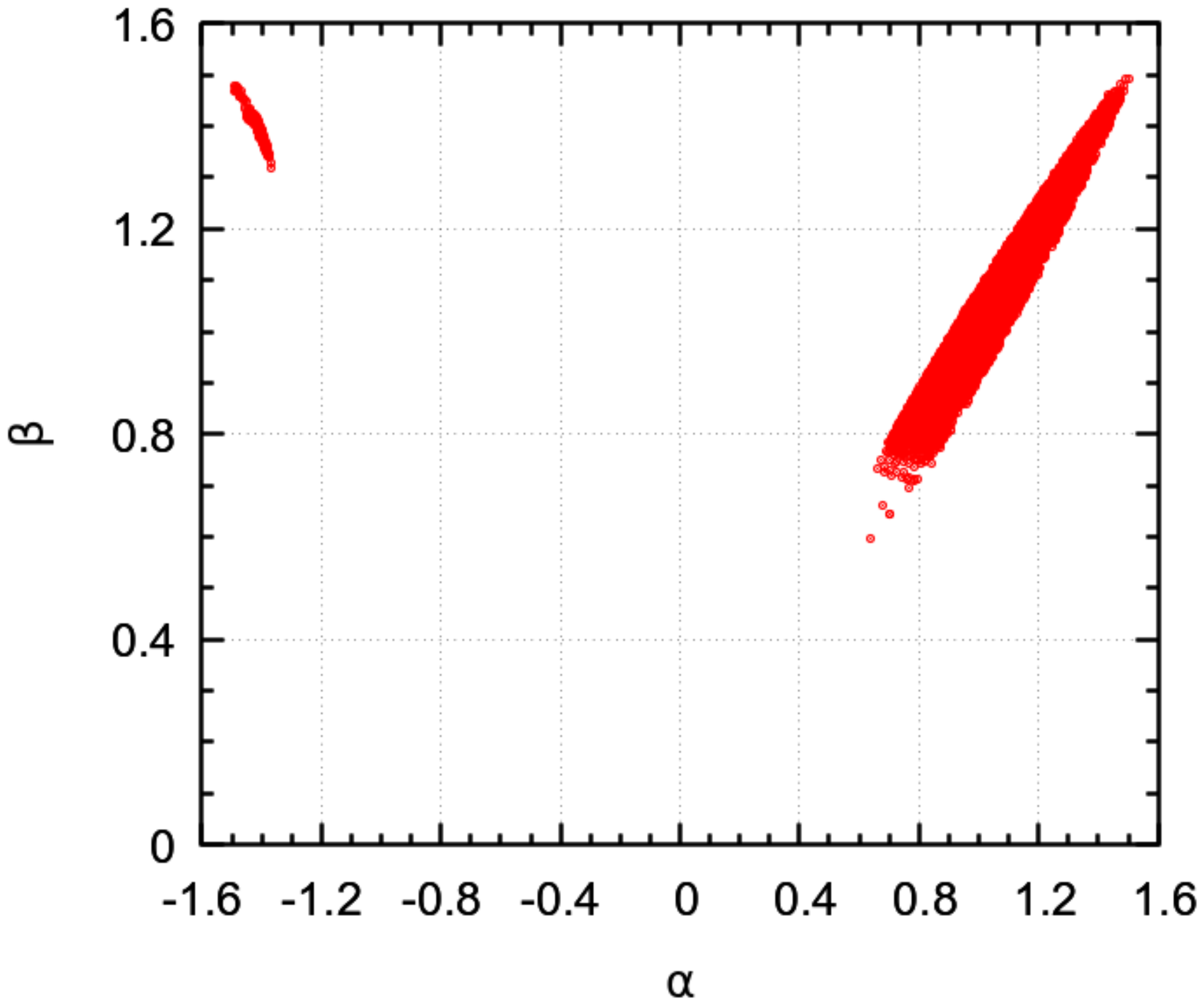}&
\includegraphics[width=0.48\textwidth]{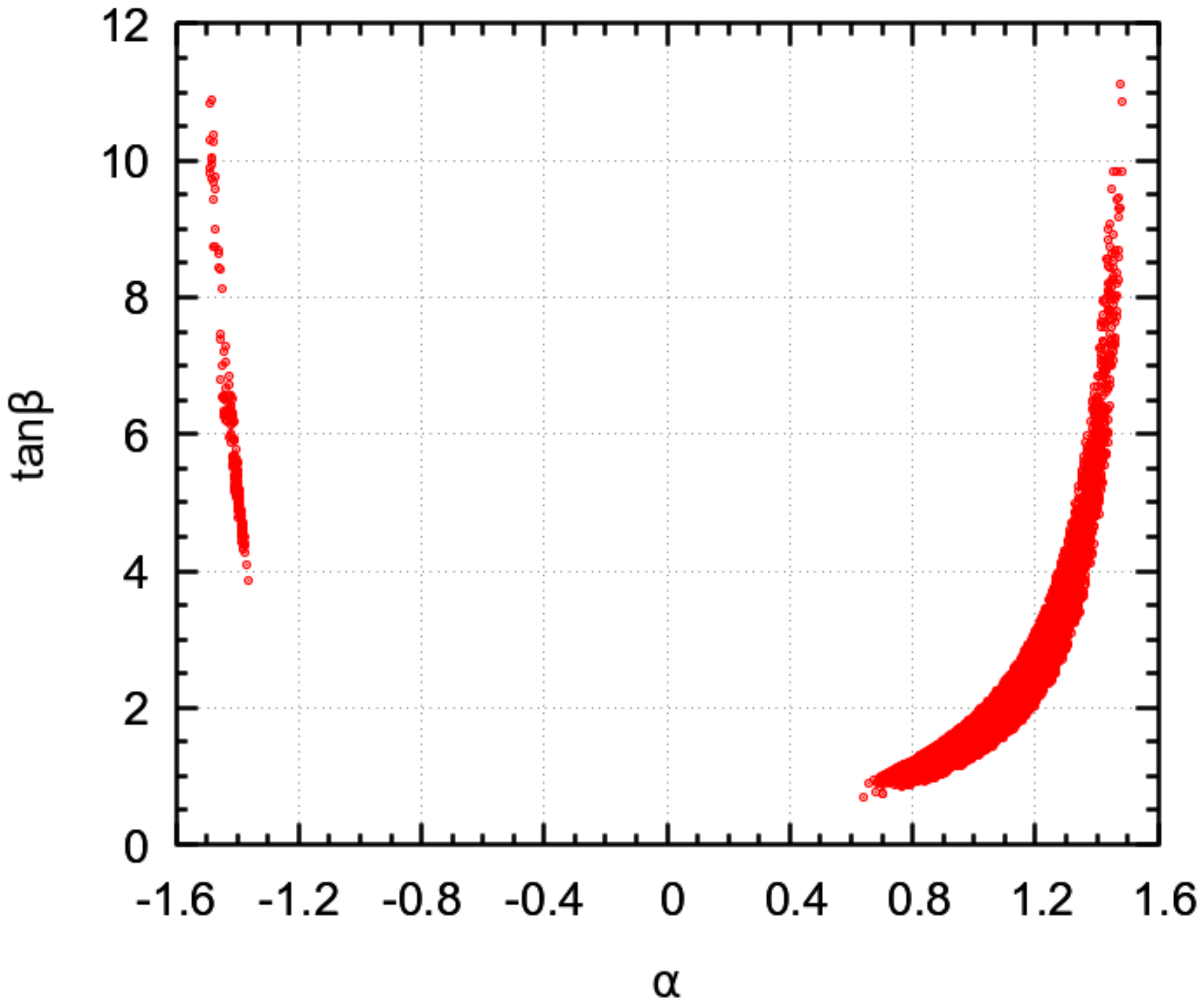}      
  \end{tabular}
  \caption{Left panel: correlation between $\alpha$ and $\beta$; Right
    panel: correlation between $\alpha$ and $\tan\beta$.}
  \label{fig:alfa-beta}
\end{figure}

We made our scans varying the parameters in the following ranges,
\begin{align}\label{eq:40}
  &m_{h_1}=125\, \text{GeV},
  && m_{h_2},m_{h_3},m_{H_1^+}\in [100,1000] \text{GeV},
  &&m_{H_2^+}\in [500,1000]\text{GeV},\\
  &\alpha \in [-\frac{\pi}{2},\frac{\pi}{2}],
  && \tan\beta \in [0,60],
  && \gamma \in [-\frac{\pi}{2},\frac{\pi}{2}], \\
&m_{12}^2 \in  [10^{-1},10^{6}] \text{GeV}^2,
&& \lambda_c\in [10^{-3},10^{2}], &&
k_{_1}\in [10^{-3},10^{2}],\\
&k_{_2}\in [10^{-3},10^{2}],&&  k_{_{12}}\in [10^{-3},10^{2}],&&
\end{align}
and take randomly $m^2_{12},k_{_{12}}$ with both signs. Despite this
flat scan, there are large correlations in the points that satisfy all
the constraints. For instance, we show in fig.~\ref{fig:alfa-beta}
the correlation between $\alpha$ and $\beta$. We see that all the
points satisfy $|\cos(\beta-\alpha)| \lesssim 1$, that is they are
close to the alignment limit,
where the 125GeV neutral scalar has couplings equal to their SM values.
The points with negative $\alpha$
correspond to the wrong sign of the fermion
couplings \cite{Ferreira:2014naa,Fontes:2014tga}. We also see that despite having
varied $\tan\beta$ in a larger interval, the good points have 
$\tan\beta \in [1,10]$.

\section{\label{sec:loopdecays}Impact of the charged scalars on the  decays
$h\to \gamma\gamma$ and $h\to Z \gamma$}

\subsection{The diagrams of the charged scalars}

\begin{figure}[htb]
  \centering
  \includegraphics[scale=0.8]{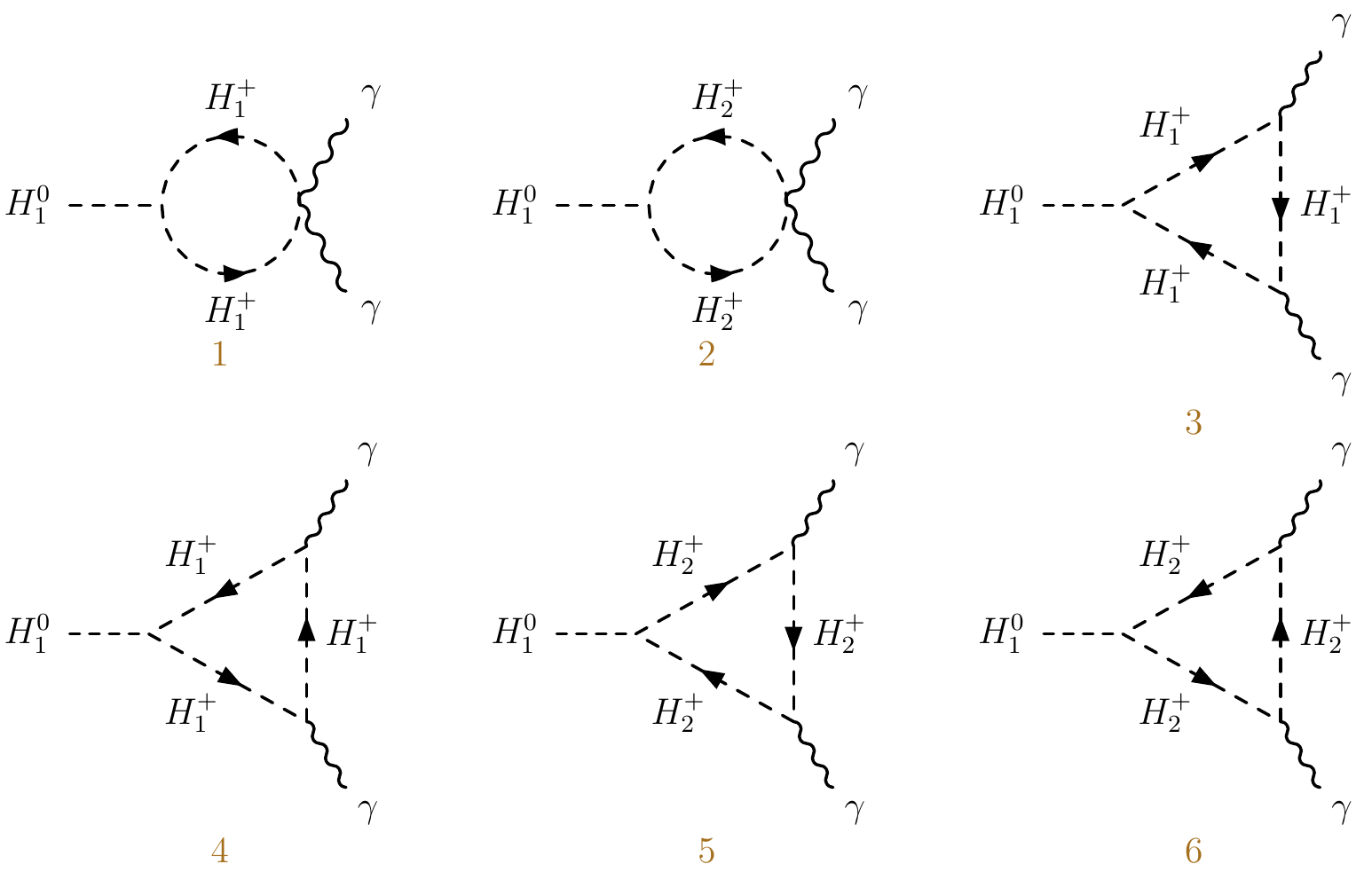}
  \caption{Charged scalars contributions to $h\to\gamma\gamma$}
  \label{fig:HAA}
\end{figure}

\begin{figure}[htb]
  \centering
  \includegraphics[scale=0.8]{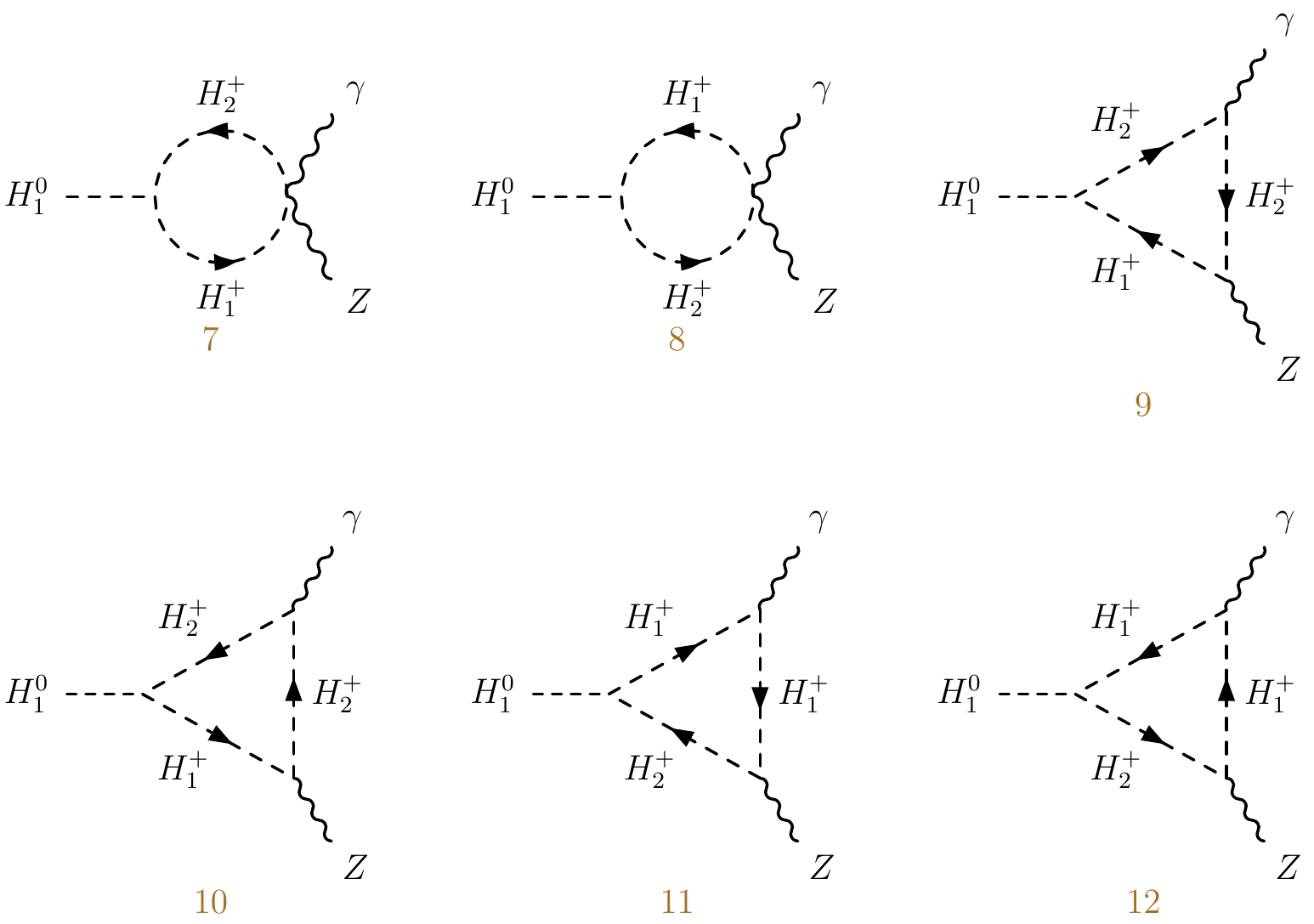}
  \caption{Extra charged scalars contributions to $h\to Z\gamma$}
  \label{fig:HZA}
\end{figure}

As we discussed before, the distinctive feature of our implementation
of the Zee model is the appearance of the off-diagonal coupling
$Z H_1^\pm H_2^\mp$.
This contributes to the loop decay $h\to Z\gamma$ and,
in principle, could lead to some new feature. For the decay $h\to
\gamma\gamma$, on the contrary, because of the photon coupling being
always diagonal, the contribution of the charged scalars will not
depend on the off-diagonal $Z H_1^\pm H_2^\mp$ coupling.
In fact, the diagrams coming from
the charged scalars and contributing in this model for $h\to
\gamma\gamma$ are shown in fig.~\ref{fig:HAA} 
while for the case of the decay $h\to Z\gamma$, besides those
equivalent to fig.~\ref{fig:HAA} (with one $\gamma$ exchanged with a
$Z$) we also have those with the off-diagonal coupling,
as shown in fig.~\ref{fig:HZA}.
The formulas for these loop decays in the absence of couplings of the type $ZH_1^+H_2^+$ are well known. They were
explicitly written for the C2HDM in ref.~\cite{Fontes:2014xva} and,
for $h\to\gamma\gamma$, they 
can be easily adapted for the case of the Zee model. We generalize the
formulas for $h\to Z\gamma$ to include the new couplings, and write
the full expressions in appendix~\ref{sec:hdecays}. The new couplings
needed are given in appendix~\ref{app:A} and were obtained with the
help of the software \texttt{FeynMaster}\cite{Fontes:2019wqh}, that
uses \texttt{QGRAF}\cite{nogueira:1991ex},
\texttt{FeynRules}\cite{christensen:2008py,Alloul:2013bka} and
\texttt{FeynCalc}\cite{Mertig:1990an,Shtabovenko:2016sxi}  in an
integrated way.

\subsection{Discussion of the impact of the charged scalars on the  loop decays}

\subsubsection{Couplings $h_1 H^+_j H^-_k$}

The couplings $h_1 H^+_1 H^-_1$ and $h_1 H^+_2 H^-_2$
do not have a strong dependence on $\gamma$. On the contrary the
couplings $h_1 H^+_1 H^-_2$ and $h_1 H^+_2 H^-_1$ are proportional to
$\sin\gamma$.

\subsubsection{Couplings $Z H^+_j H^-_k$}

The couplings $Z H^+_1 H^-_2$ and $Z H^+_2 H^-_1$ are given in
eqs.~(\ref{eq:3b})-(\ref{eq:3c}). They are proportional to $\sin(2\gamma)$
and vanish for $\gamma=0, \pm \pi/2$, while the couplings $Z H^+_1
H^-_1$ and $Z H^+_2 H^-_2$ vary with $\gamma$ like
\begin{equation}
  \label{eq:1}
  g_{ZH^+_iH^-_i} \propto (-1 + 4 s_W^2 + \cos 2\gamma)\, .
\end{equation}
It is interesting to note that because $-1 + 4 s_W^2 \simeq 0$ they
behave approximately like $\cos 2\gamma$ that vanishes at $\pm \pi/4$.

\subsubsection{Results and Conclusions}

Because of the dependence of the couplings on the mixing angle
$\gamma$, we looked at the contributions of the charged scalars as a
function of this angle. If the loop integral did not vary much with
the masses, the results would be proportional to the products of the
$h_1 H_j^+ H_k^-$ and $Z H_j^+ H_k^-$ couplings, as the photon
coupling is universal. In the following figures all points passed all
the constraints, including HiggsBounds 5.9.0 and  those coming from
BR($B\to X_s \gamma$), as discussed in section~\ref{sec:bsgamma}.

\begin{figure}[!htb]
  \centering
  \begin{tabular}{cc}
    \includegraphics[width=0.48\textwidth]{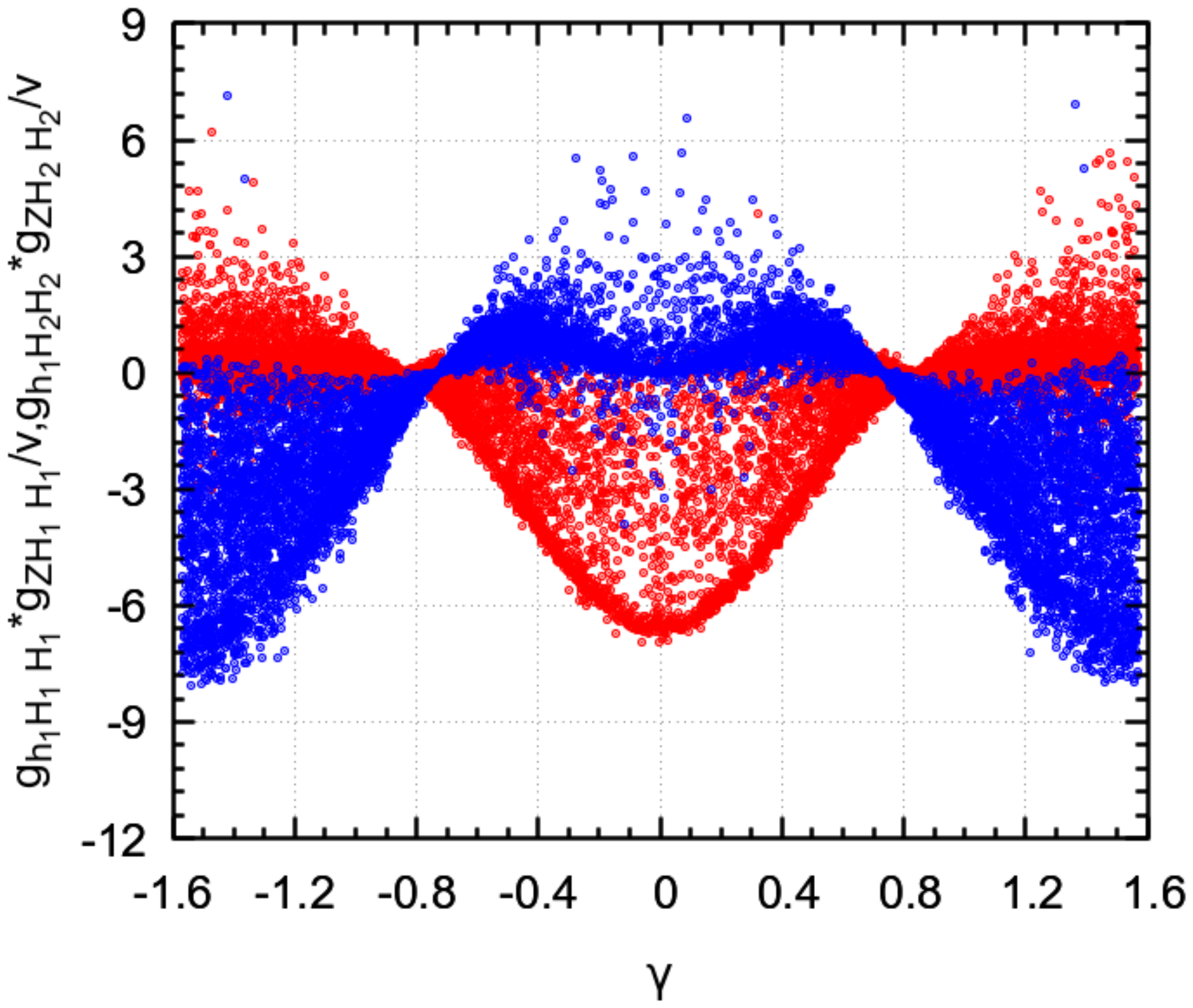}&
    \includegraphics[width=0.48\textwidth]{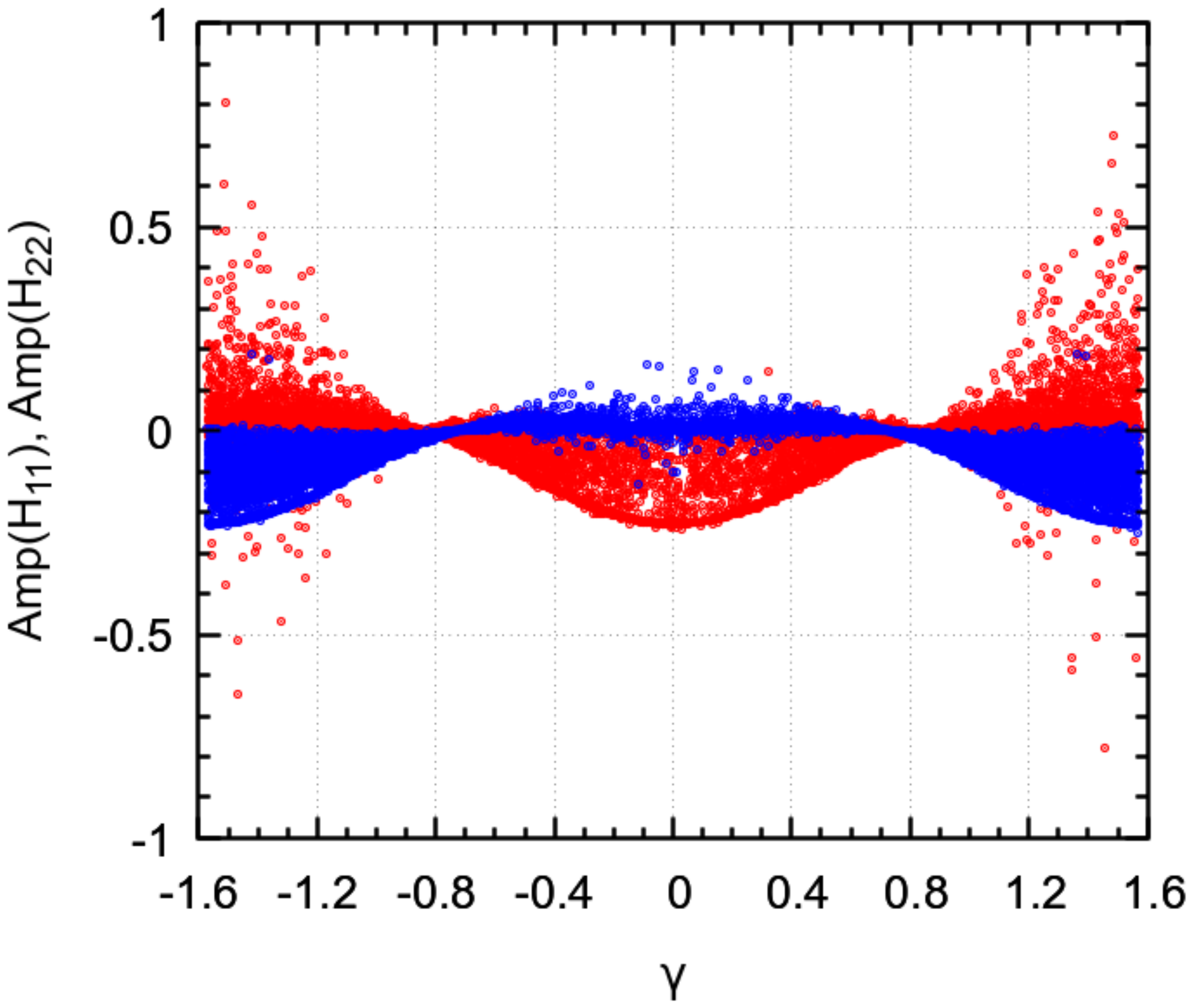}    
  \end{tabular}
  \caption{Results for the charged scalars amplitudes contribution to
    $h\to Z \gamma$. On the left panel the coupling products and
    on the right panel the actual amplitudes.}
  \label{fig:3}
\end{figure}

In fig.~\ref{fig:3} we show on the left panel
the result of the product of the couplings (we divide by $v$ because
the coupling $h_1 H_j^+ H_k^-$ has dimensions of mass), first
for the the case of $H_1^+$ running in the loops of
fig.~\ref{fig:HAA} in red and then for the case of $H_2^+$ in blue. From
the above discussion we expect the result to vary like $\cos 2\gamma$,
and that is indeed the case. Our assumptions that the loop integrals
do not depend much on the masses can be verified in the right panel of
fig.~\ref{fig:3}  where we show the actual plot for the loop
amplitudes. The behaviour as $\cos 2\gamma$ is clear in both cases.

\begin{figure}[htb]
  \centering
  \begin{tabular}{cc}
    \includegraphics[width=0.48\textwidth]{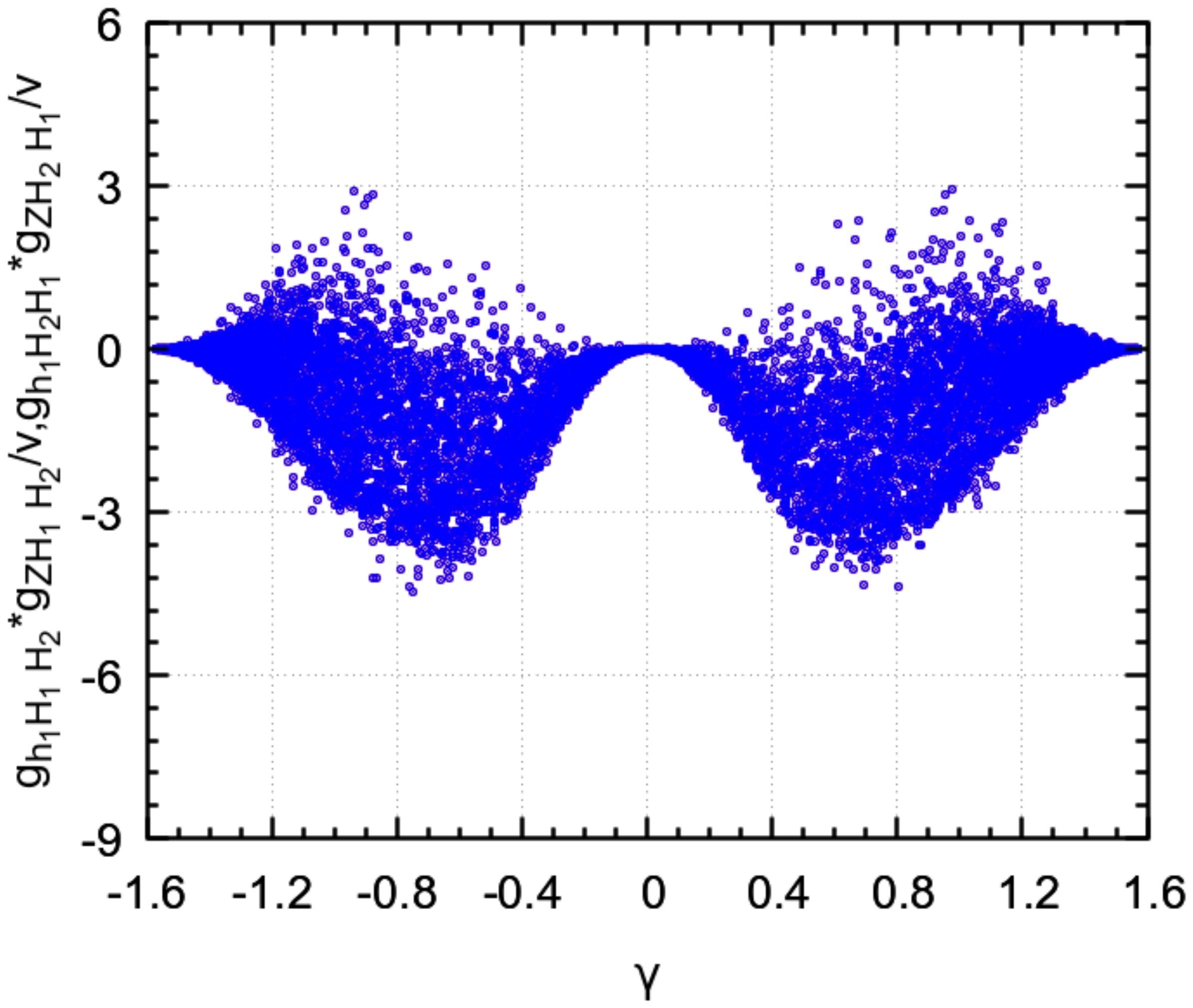}&
    \includegraphics[width=0.48\textwidth]{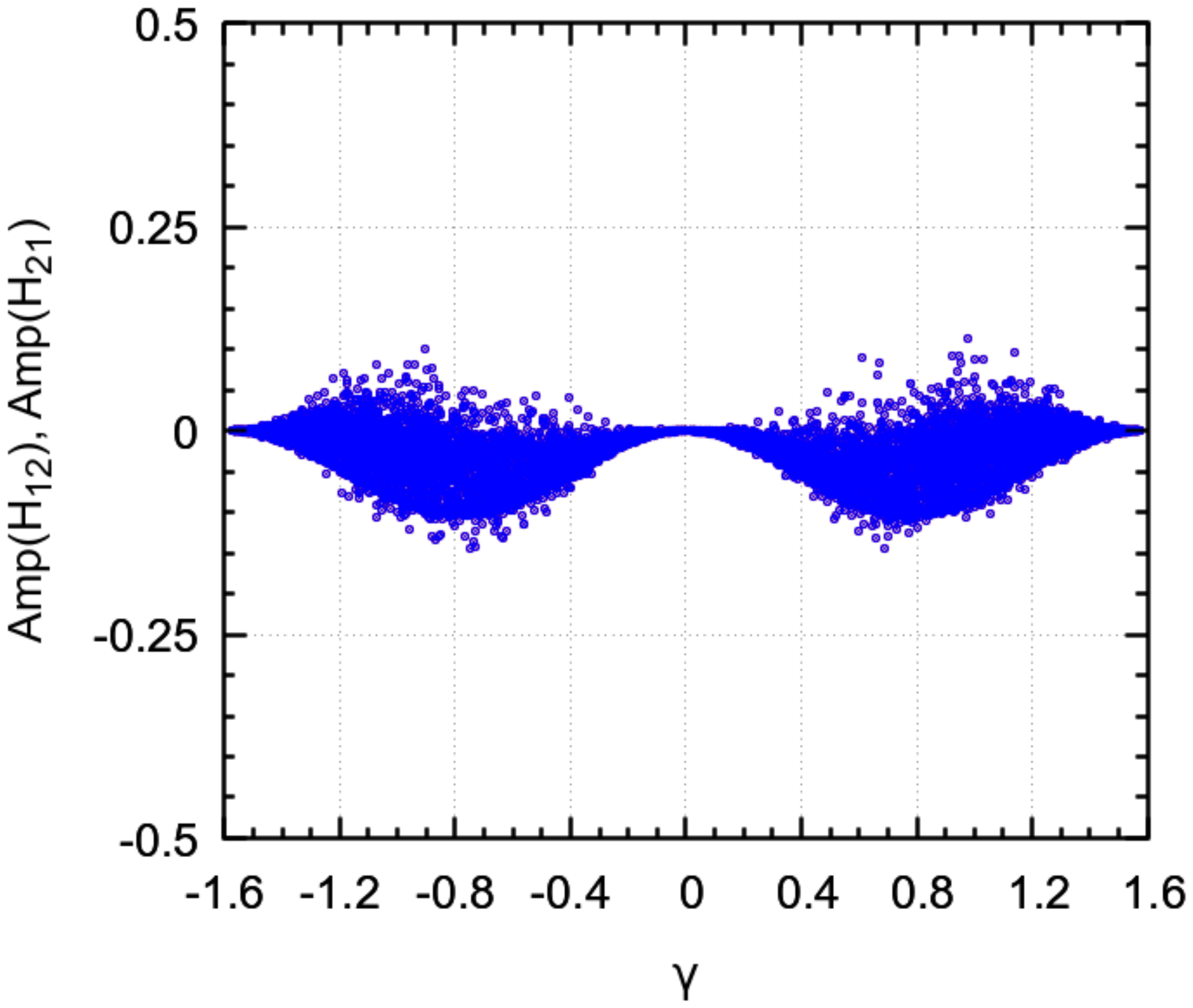}    
  \end{tabular}
  \caption{Results for the charged scalars amplitudes contribution to
    $h\to Z \gamma$. On the left panel the coupling products and
    on the right panel the actual amplitudes.}
  \label{fig:4}
\end{figure}

Now we can study the case where there are two different charged scalars,
$H_1^+$ and $H_2^+$,
running in the loops of fig.~\ref{fig:HZA}. This is shown in
fig.~\ref{fig:4}.
Again on the left panel we plot the product of the couplings, and on
the right panel the loop amplitudes. In this case Amp($H_1^+,H_2^+$),
corresponding to diagrams 7, 10 and 11 of fig.~\ref{fig:HZA}
in red, coincides with Amp($H_2^+,H_1^+$) corresponding
to diagrams 8, 9 and 12. As expected we see clearly a dependence on
$\sin 2 \gamma$, confirming our expectations.

\begin{figure}[htb]
  \centering
  \begin{tabular}{cc}
    \includegraphics[width=0.48\textwidth]{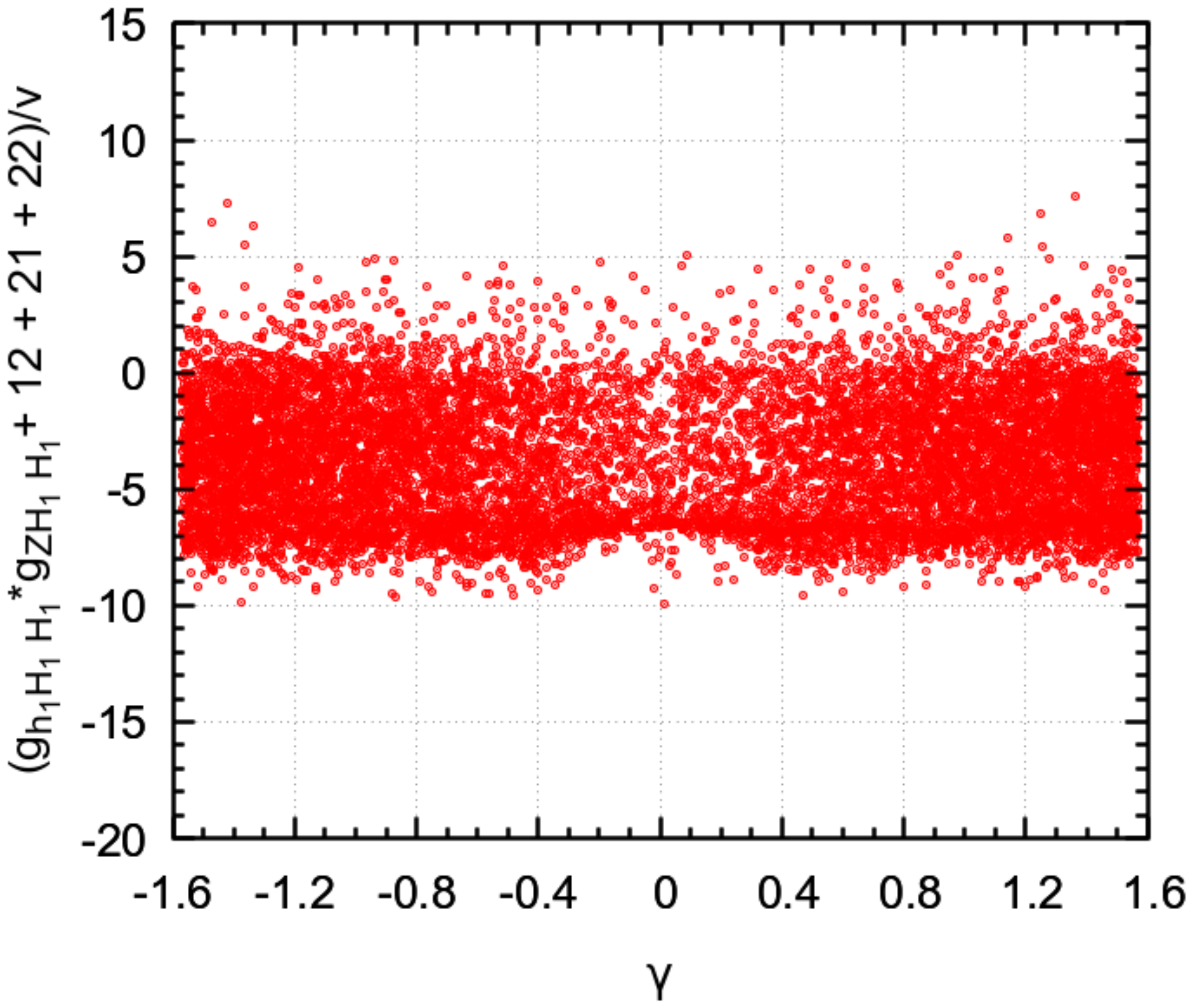}
    &
    \includegraphics[width=0.48\textwidth]{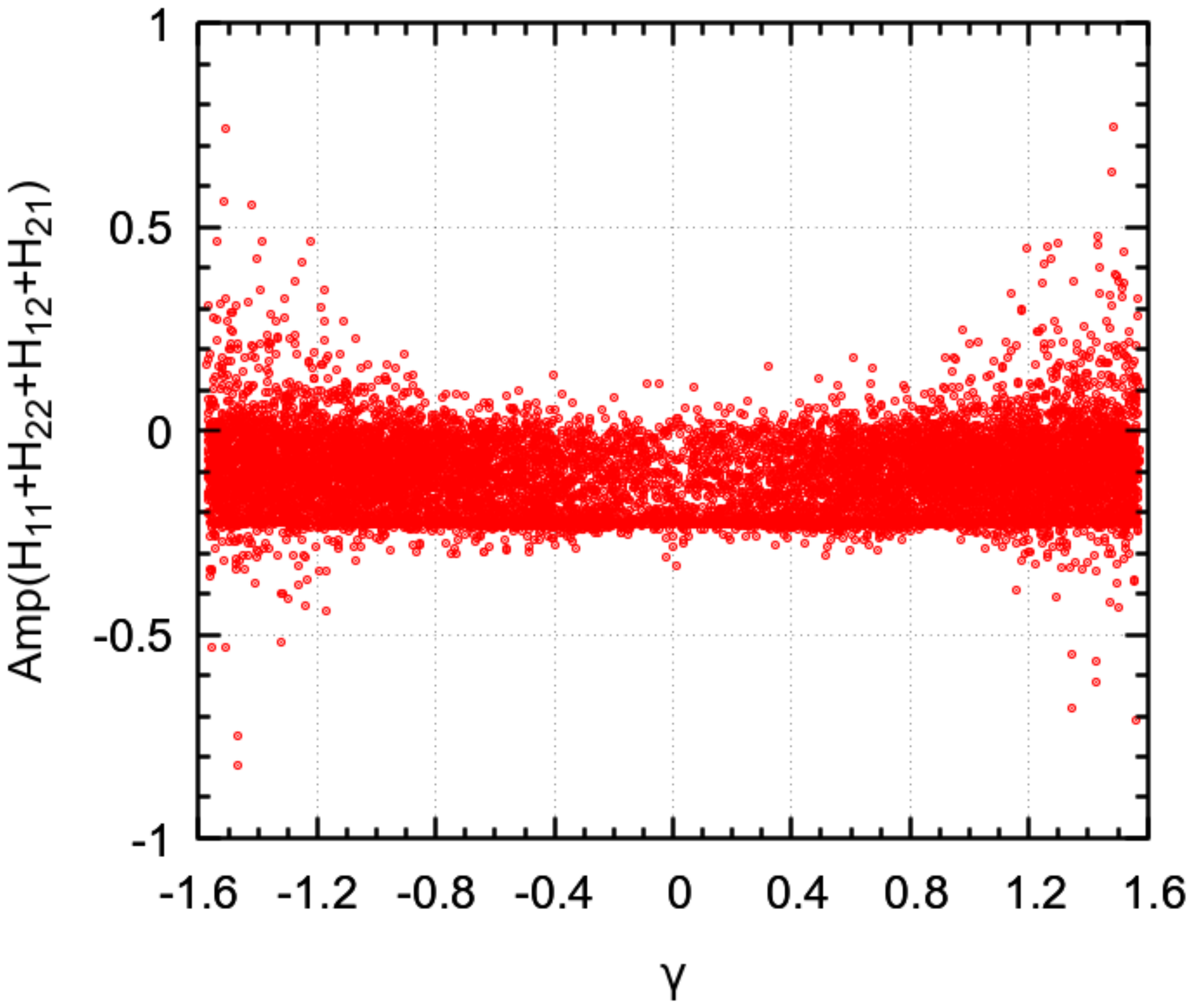}
  \end{tabular}
  \caption{Results for the charged scalar amplitudes contribution to
    $h\to Z \gamma$. On the left panel the sum of the product of
    couplings and on the right panel the complete result.}
  \label{fig:5}
\end{figure}

However this nice result will not help us in using the decay $h\to Z
\gamma$ to identify the novel coupling $Z H_1^+ H_2^-$ appearing in the
Zee model. The problem is that once we sum all contributions
we loose the dependence on $\gamma$. This can be seen on
fig.~\ref{fig:5} both for the products of the couplings in the left
panel, and for the final result for the charged scalar contribution to $h\to Z
\gamma$.

In conclusion, although the contribution of the charged scalars can
have both signs and also be zero, the dependence on $\gamma$ and
therefore on the mixing parameters $\mu_4$ is hidden. In fact we can
have the same behaviour of the charged scalar amplitudes in other
models like the 3HDM \cite{Boto:2021}.

\section{\label{sec:newdecay}Decays of the Charged Higgs}

\subsection{The decay $H_2^+ \to H_1^+ + Z$}

\begin{figure}[htb]
  \centering
  \begin{tabular}{cc}
    \includegraphics[width=0.45\textwidth]{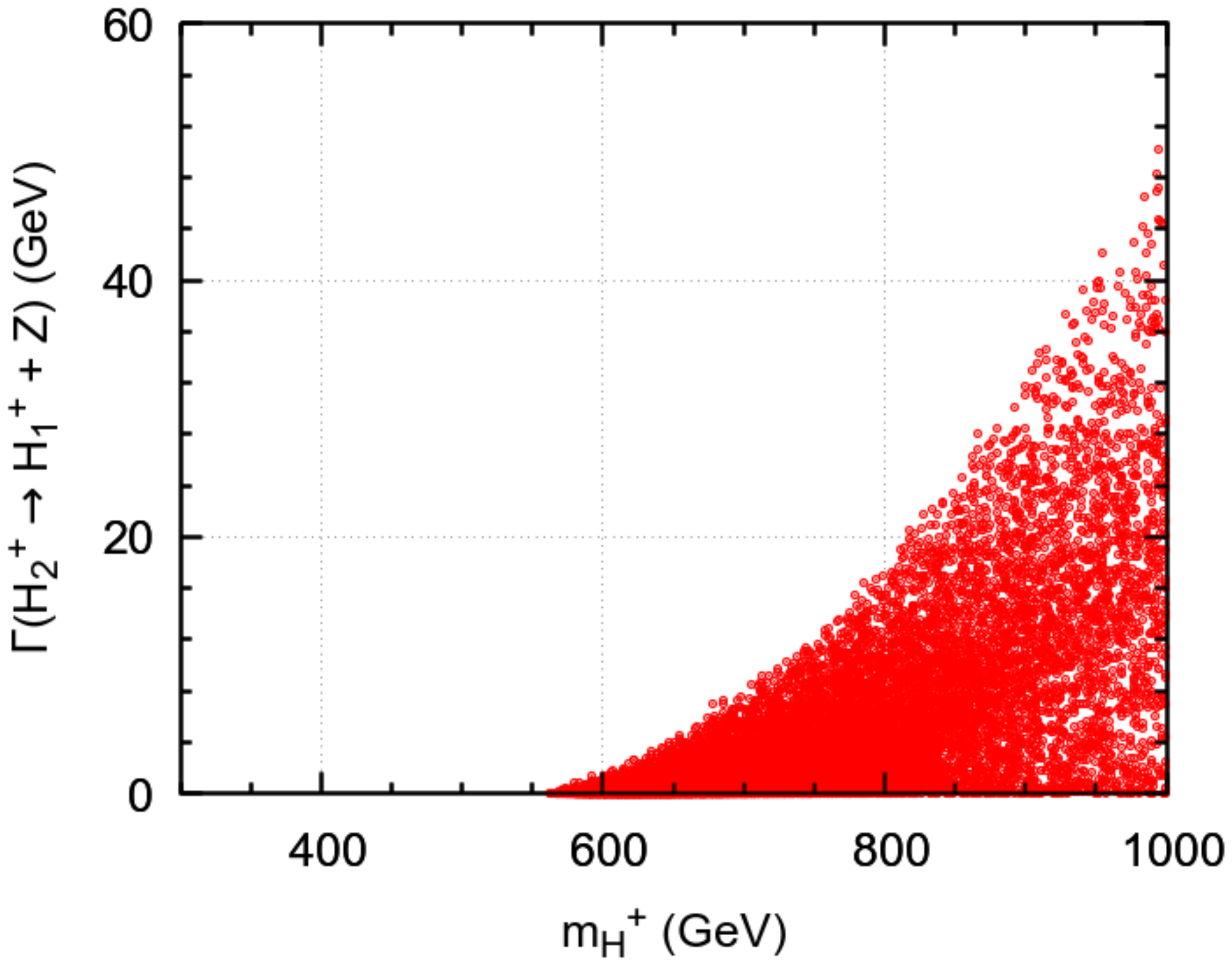}
&
    \includegraphics[width=0.45\textwidth]{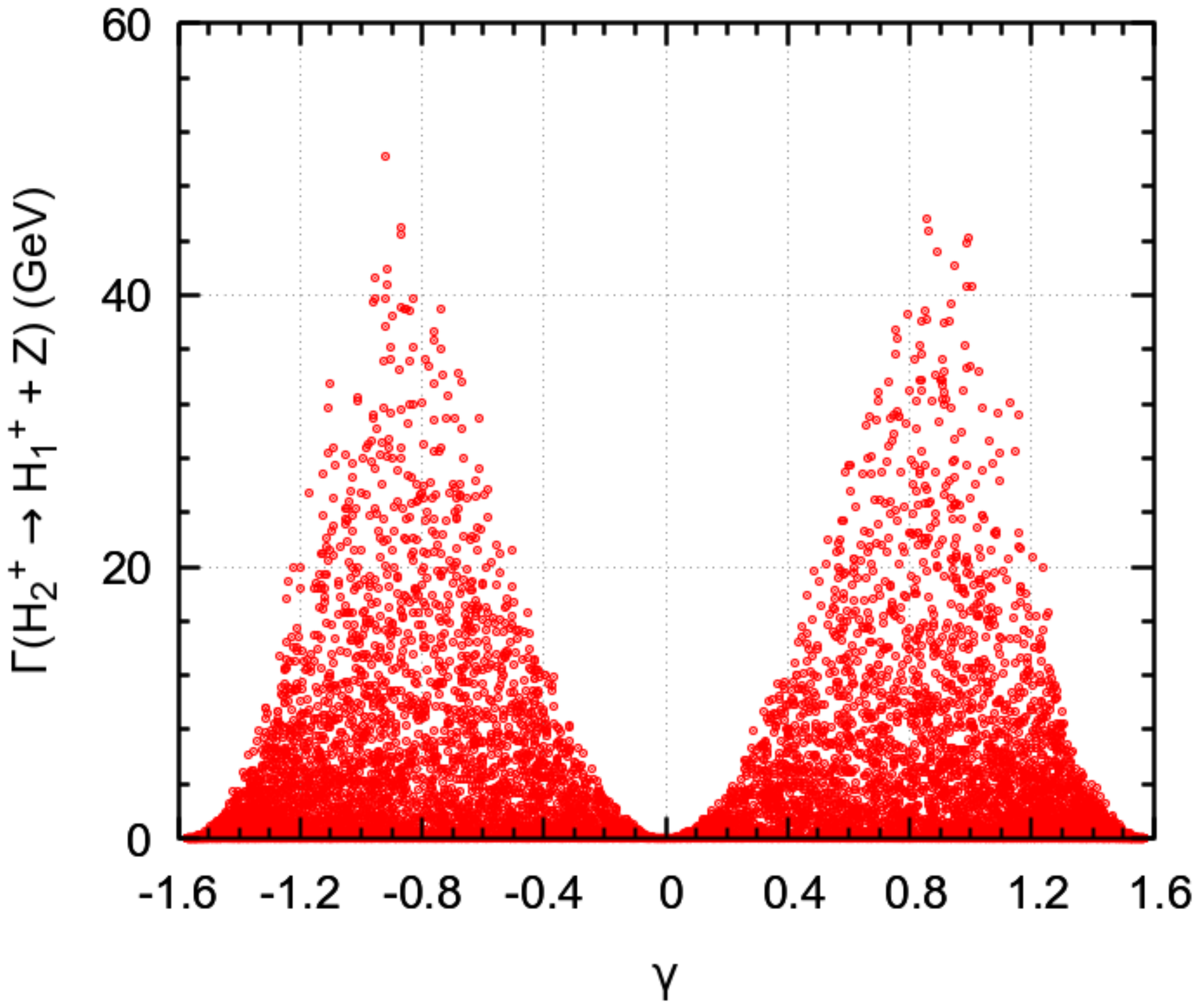}
  \end{tabular}
  \caption{ Decay with $H_2^+ \to H_1^+ + Z$.
    On the left panel the dependence on the mass
    of the decaying charged Higgs and on the right the dependence on $\gamma$.}
    \label{fig:6}
\end{figure}

If we want to have a unique signal for this model it would be the
decay of one charged Higgs in another one plus a $Z$ boson.
This is only possible if $\gamma \not= 0$.
We have checked that this can indeed occur,
as shown in fig.~\ref{fig:6}.
All points shown satisfy all the constraints discussed in
section~\ref{sec:constraints}. 
We see clearly that, as expected, one has to be away from $\gamma=0$ to have a
sizable decay width.

\subsubsection{Decays of the heavier $H_{2}^\pm$}

Depending on the masses the following decays are among the most important,
\begin{align}
H_{2}^\pm \to H_{1}^\pm + Z\, , &&
H_{2}^+ \to  t + \overline{b}\, ,&&
H_{2}^\pm \to H_{1}^\pm + h_i \, ,\\
H_{2}^\pm \to W^\pm + h_i\, , &&
H_{2}^+ \to  \nu_\tau + \tau^+ \, .
\end{align}
The first decay is unique to this type of models and not present in
NHDM. It requires a mixing between the charged Higgs from the doublets
with the charged Higgs from the singlets. The expression for the width
is
\begin{equation}
  \label{eq:9}
  \Gamma( H_{2}^\pm \to H_{1}^\pm + Z)=
\frac{g^2}{64\pi m_{H^+_j}^3 M_W^2}\, g_{\rm HpjHmkZ}[2,1]^2\,      
      \lambda(m_{H^+_2}^2,m_{H^+_1}^2,M_Z^2)^3,
\end{equation}
where the Källen function is given by
\begin{equation}
  \label{eq:10}
  \lambda(x^2,y^2,z^2)= \sqrt{x^4+y^4+z^4-2 x^2 y^2 -2 x^2 z^2 -2 y^2 z^2}.
\end{equation}
For the other decays we have
\begin{align}
  \label{eq:11}
  \Gamma(H_{2}^+ \to  t + \overline{b})=&
  \frac{3 g^2}{32\pi M_W^2} m_{H_2^+} \lambda(m_{H_2^+},m_t^2,m_b^2)
\left[(1-x_t-x_b) (Y_2^2 x_t+X_2^2 x_b)-
       4 x_t x_b X_2 Y_2 \right],
\end{align}
where
\begin{equation}
  \label{eq:12}
  x_t=\frac{m_t^2}{m_{H_2^+}}, \quad x_b=\frac{m_b^2}{m_{H_2^+}},
\end{equation}
and $X_k,Y_k$ are given in eq.~(\ref{eq:7}).
For the decay into the other charged Higgs and one neutral Higgs boson
we have,
\begin{align}
  \label{eq:13}
  \Gamma(H_{2}^\pm \to H_{1}^\pm + h_i) =&
  \frac{g_{\rm hjHpiHmk}[i,2,1]^2}{16\pi m_{H^+_2}^3}
\lambda(m_{H_2^+},m_{H_1^+},m_{h_i}^2).
\end{align}
The decay into one W and one neutral Higgs boson is similar to the
decay into the charged Higgs and Z. We obtain
\begin{align}
  \label{eq:14}
  \Gamma(H_{2}^\pm \to W^\pm + h_i) =&
  \frac{g^2}{64\pi m_{H^+_2}^3 M_W^2}\, g_{\rm hjHpkWm}[i,2]^2\,      
      \lambda(m_{H^+_2}^2,M_W^2,m_{h_i}^2)^3.
\end{align}
Finally the decay in the third family leptons (the others are
negligible) is given by
\begin{align}
  \label{eq:15}
  \Gamma(H_{2}^+ \to  \nu_\tau + \tau^+)=&
  \frac{g^2}{32\pi M_W^2}\, Z_2^2\, m_\tau^2\, m_{H^+_2}
  \left[1 - \frac{m_\tau^2}{m_{H^+_2}^2} \right]^2.
\end{align}

\subsubsection{Decays of the lighter $H_{1}^\pm$}

Except for the decays into another charged Higgs, that are not allowed
because we assume that $m_{H_1^+}< m_{H_2^+}$, the decays are
similar to those of the heavier charged scalar. If kinematically available,
the expressions for the decays can be easily obtained from the above
with index $2\to 1$. All the couplings needed
are given in appendix~\ref{app:A} and were obtained with the
help of the software \texttt{FeynMaster}\cite{Fontes:2019wqh}.

\section{\label{sec:benchmark}Benchmark points for the Zee model}

\subsection{Looking for a distinctive signature}

As we have discussed before, the Zee model provides an example of the
non-vanishing coupling between two different charged Higgs and the Z
boson. For instance, this cannot happen in any NHDM, even with a large
N. So we want to see if there is a signal of this coupling.

As we explained in section~\ref{sec:loopdecays}, the first idea was to look at the impact
on the BR($h_{125} \to Z \gamma$).
But it turns out that the effect of the extra diagrams is not
quantitatively different from the effect of a second charged scalar
coupling only diagonally to the $Z$ boson,
as occurs for instance in the 3HDM,
where there are two charged Higgs bosons, but no $Z H_1^+ H_2^-$
coupling \cite{Boto:2021}.
So, although there is an effect, for instance the
contribution of summing over all the charged Higgs diagrams can
vanish,
this is not an effect specific to the $Z H_1^+ H_2^-$ coupling.
So we turn to a distinctive decay:
\begin{equation}
  \label{eq:22}
  H_2^+ \to H_1^+ + Z,\quad \text{and}\quad  H_1^+ \to t + \overline{b}
\end{equation}
This decay has a very clear signature and should be searched for at the LHC.

\subsection{Benchmark Point $P_1$}

As the model has many independent parameters, if we try to plot the
various branching ratios of the $H_1^+$ or $H_2^+$ instead of
obtaining something similar to the famous plot \cite{Djouadi:2005gi}
of the SM Higgs boson BR's as a function of its mass
(when this mass was yet not known), we would get a
figure with all the points superimposed and no lines.
So, to have a better visualization we fix most of the parameters and show that
indeed the branching ratios for the processes in eq.~(\ref{eq:22}) can
be important, or even dominant. This leads us to the choice of
benchmark points. In choosing these benchmark points for the Zee model
we take in account all the theoretical and experimental constraints on
the model.

\begin{figure}[!htb]
  \centering
  \begin{tabular}{cc}
  \includegraphics[width=0.48\textwidth]{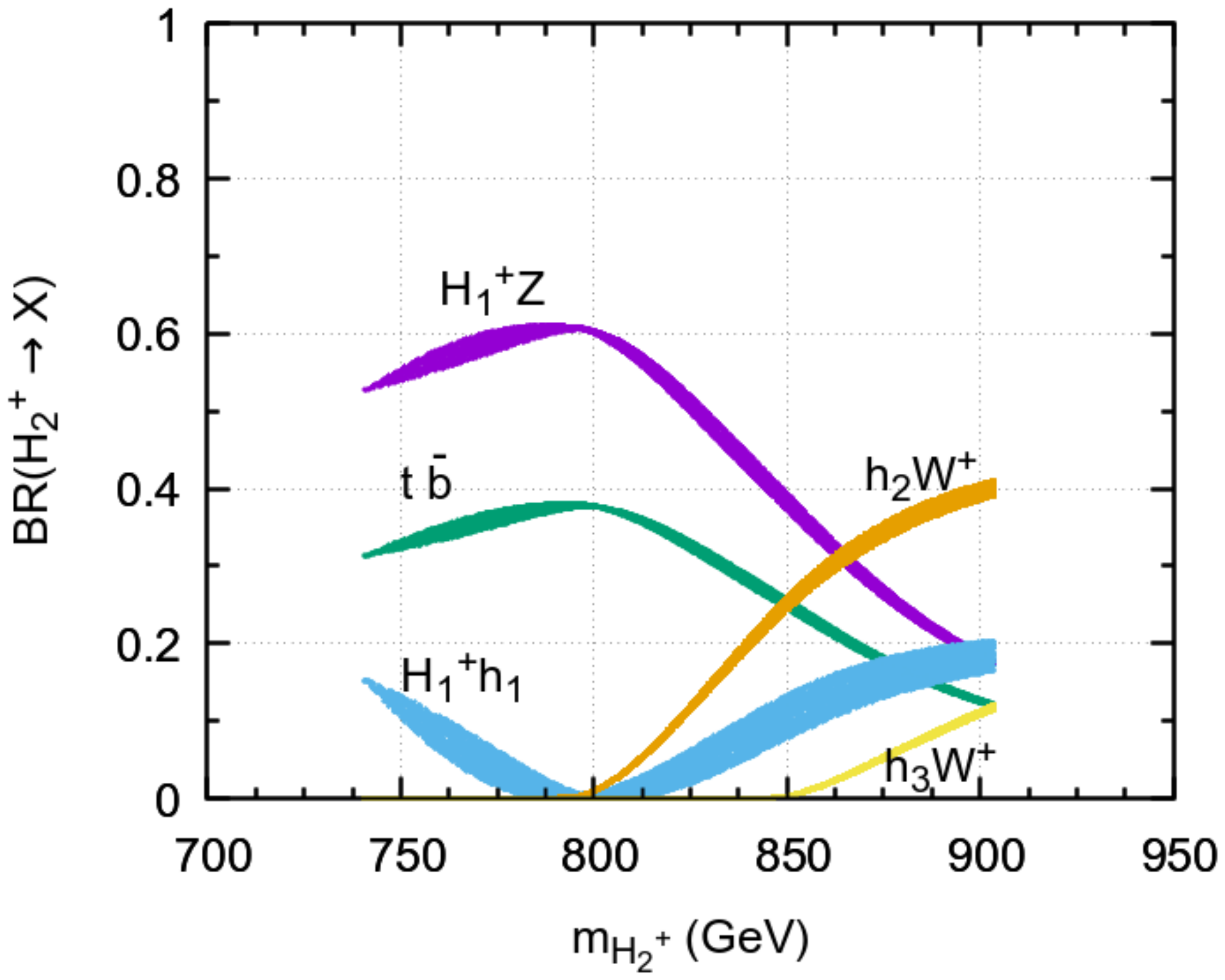} 
    &
    \includegraphics[width=0.48\textwidth]{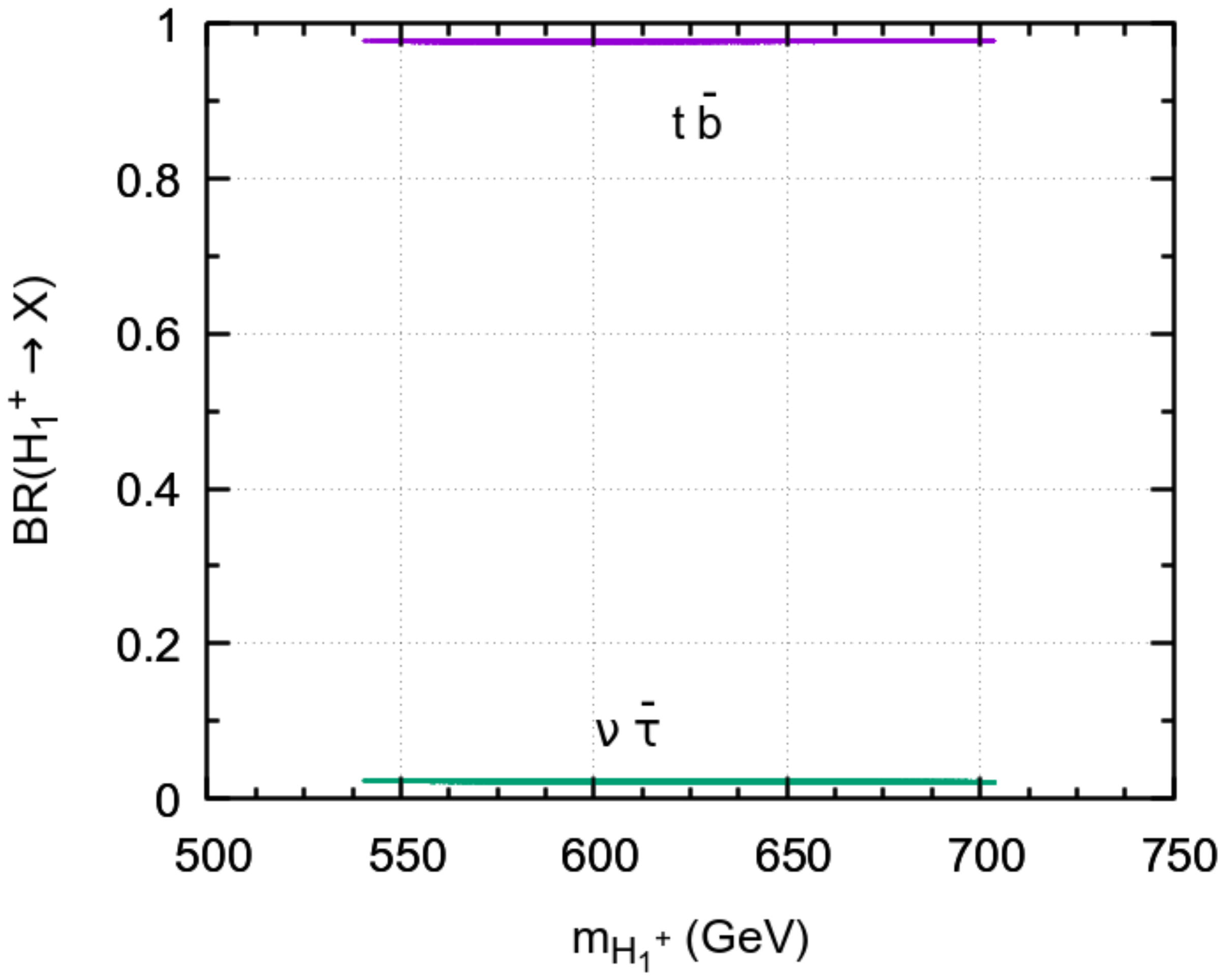}
   \end{tabular}
  \caption{Dominant BR's for $H_2^+$ (left panel) and  $H_1^+$ (right
    panel) for benchmark point $P_1$.}
  \label{fig:10}  
\end{figure}

For the first benchmark point, $P_1$, we choose a situation when both
masses are above\footnote{The starting point satisfied
  eq.~(\ref{eq:16}), but as we vary the masses some points
  are slightly below that limit.} the limit of eq.~(\ref{eq:16}). It is defined by the
following parameters,
\begin{subequations}   
\begin{align}
  \label{eq:24}
  &m_{h_1}=125\, \text{GeV} && m_{h_2}=714.98\, \text{GeV}&&
  m_{h_3}=767.42\, \text{GeV}  \\
&m_{H_1^+}= m_{H_2^+} -200\, \text{GeV} && \alpha=1.391   && \gamma=0.894\\
&m_{12}^2=8.828\times 10^{4}\, \text{GeV}^2 && \lambda_c= 0.4363  &&
k_{_1}=0.4633\\
&k_{_2}=0.4633&&  k_{_{12}}=5.427\times 10^{-2}&&
\end{align}
\end{subequations}   
The situation is shown in fig.~\ref{fig:10}.
We see that our signal decay has the largest branching ratio, while
$H_1^+$ decays almost 100\% into $t + \overline{b}$. This should provide
clear signatures at the LHC. A detailed analysis, with background
studies, should of course be done. The width of the bands comes from the
variation of $\tan\beta$ (at the percent level, because the good
points have $\alpha \simeq \beta$). All the points pass all the constraints,
including that of eq.~(\ref{eq:17}).

\subsection{Benchmark Point $P_2$}

One could argue that $P_1$ will lead to a situation where the
constraint  of eq.~(\ref{eq:17}) was verified, as we took the masses
to satisfy the bound of eq.~(\ref{eq:16}). Therefore we want to show
another benchmark point that would be excluded by
eq.~(\ref{eq:16}). That is, we do not exclude points \textit{a priori},
but for each point we evaluate the BR($B\to X_s\gamma$) to see if it
passes the bounds in eq.~(\ref{eq:17}). 

\begin{figure}[htb]
  \centering
  \begin{tabular}{cc}
  \includegraphics[width=0.48\textwidth]{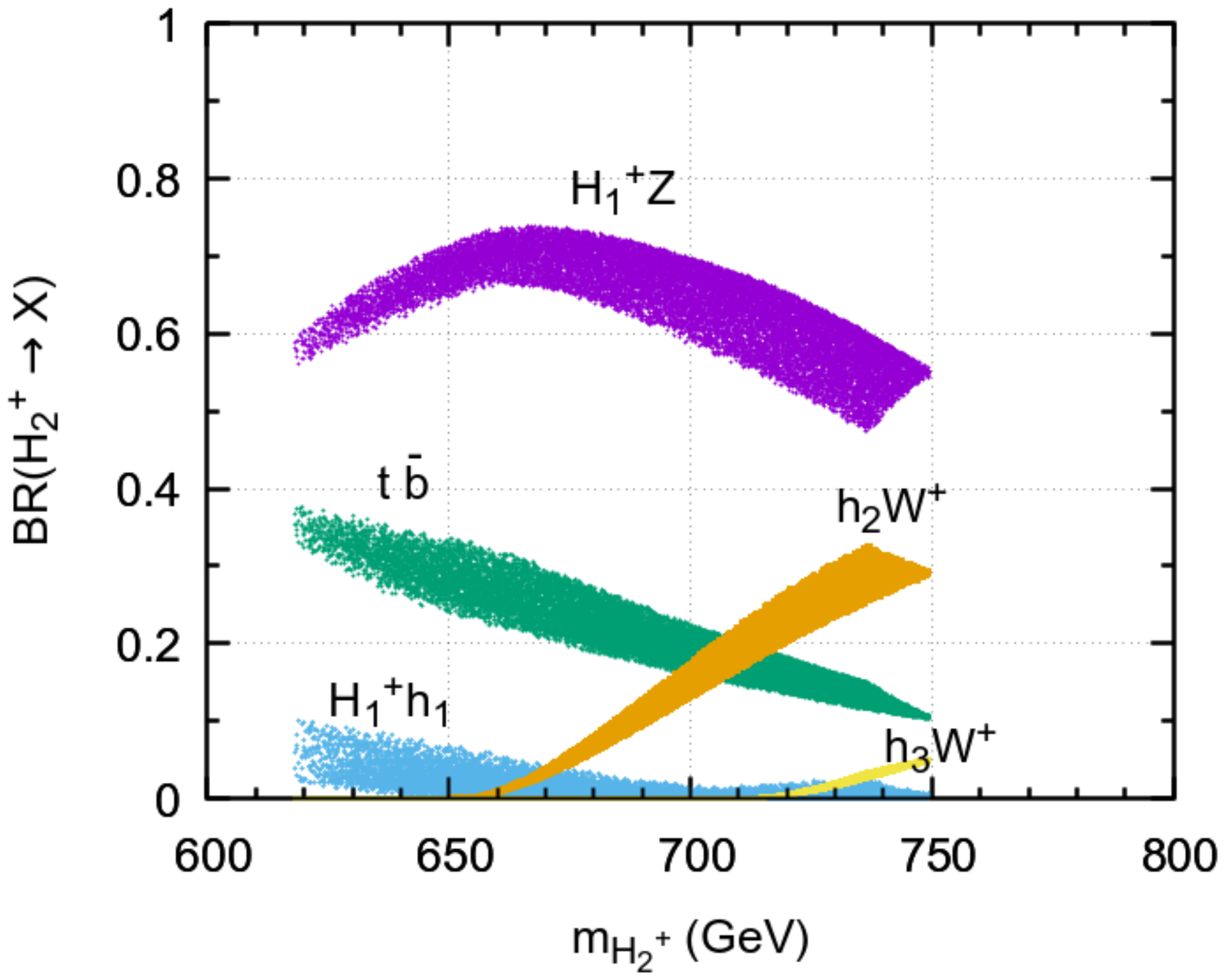} 
    &
    \includegraphics[width=0.48\textwidth]{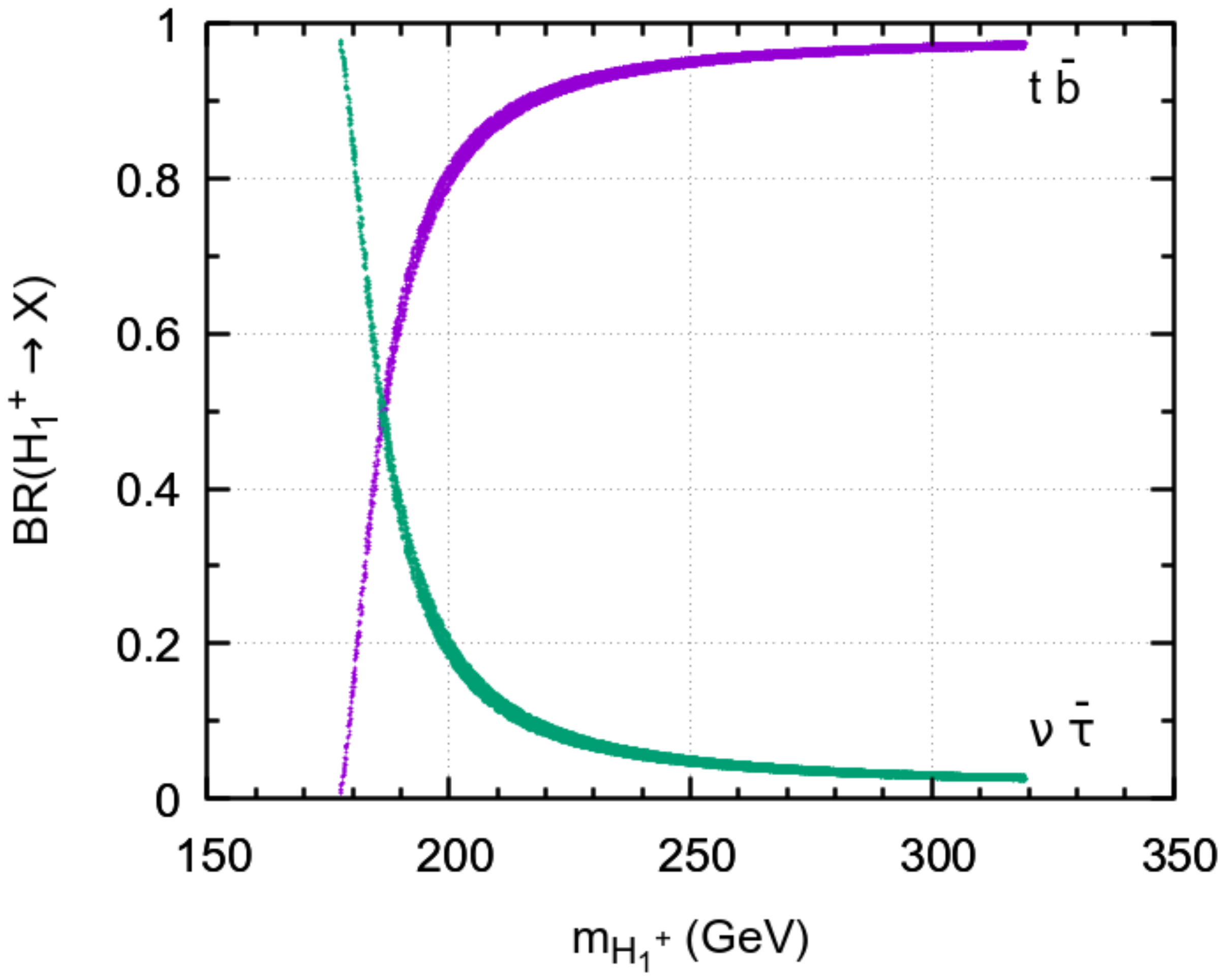} 
   \end{tabular}
  \caption{Dominant BR's for $H_2^+$ (left panel) and  $H_1^+$ (right
    panel) for benchmark point $P_2$.}
  \label{fig:11}  
\end{figure}

For the second benchmark point $P_2$ we therefore choose a situation
where the lowest charged Higgs mass is below that limit.
It is defined by the following parameters
\begin{subequations}
\begin{align}
  \label{eq:24}
  &m_{h_1}=125\, \text{GeV} && m_{h_2}= 580.7\, \text{GeV}
  && m_{h_3}= 633.7\, \text{GeV}\\
  &m_{H_1^+}, m_{H_2^+}\text{GeV},\ \text{scanned as shown} && \alpha= 1.398
  && \gamma=1.089\\
  &m_{12}^2=5.77\times 10^{4}\, \text{GeV}^2 && \lambda_c= 4.473
  && k_{_1}=1.082\\
&k_{_2}=3.98  \times 10^{-3} &&  k_{_{12}}=-1.266 \times 10^{-3} &&
\end{align}
\end{subequations}
The situation is shown in fig.~\ref{fig:11}.
We see that our signal decay has the largest branching ratio, while
$H_1^+$ decays almost 100\% into $t + \overline{b}$. This should be
clear signatures at the LHC, although background studies should be
done.  The width of the bands comes from the variation of
$\tan\beta$ and $m_{H_1^+}, m_{H_2^+}$ which were varied
independently. All the points pass all the constraints, including that
of eq.~(\ref{eq:17}).

\subsection{Benchmark Point $P_3$}

\begin{figure}[htb]
  \centering
  \begin{tabular}{cc}
  \includegraphics[width=0.48\textwidth]{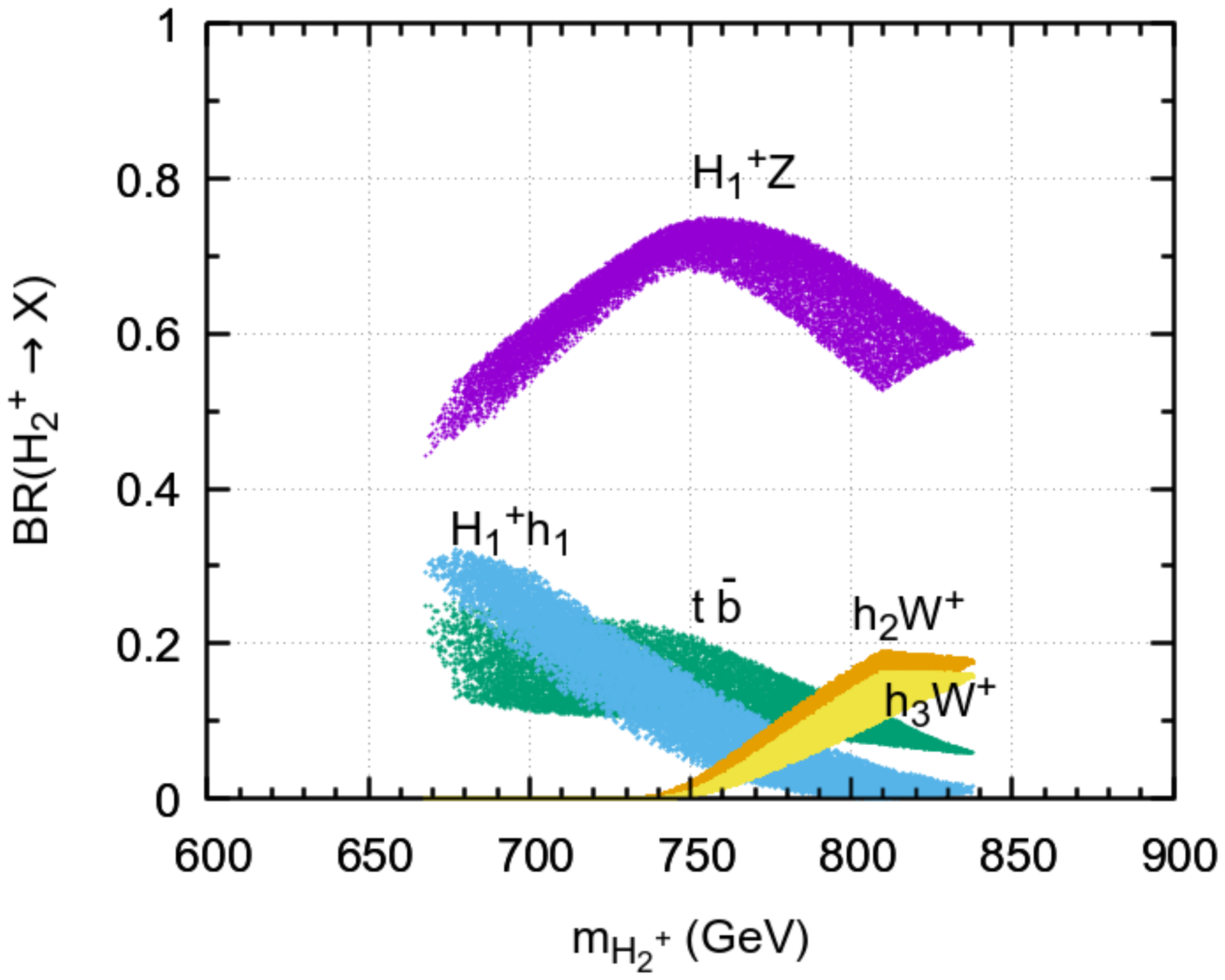} 
    &
    \includegraphics[width=0.48\textwidth]{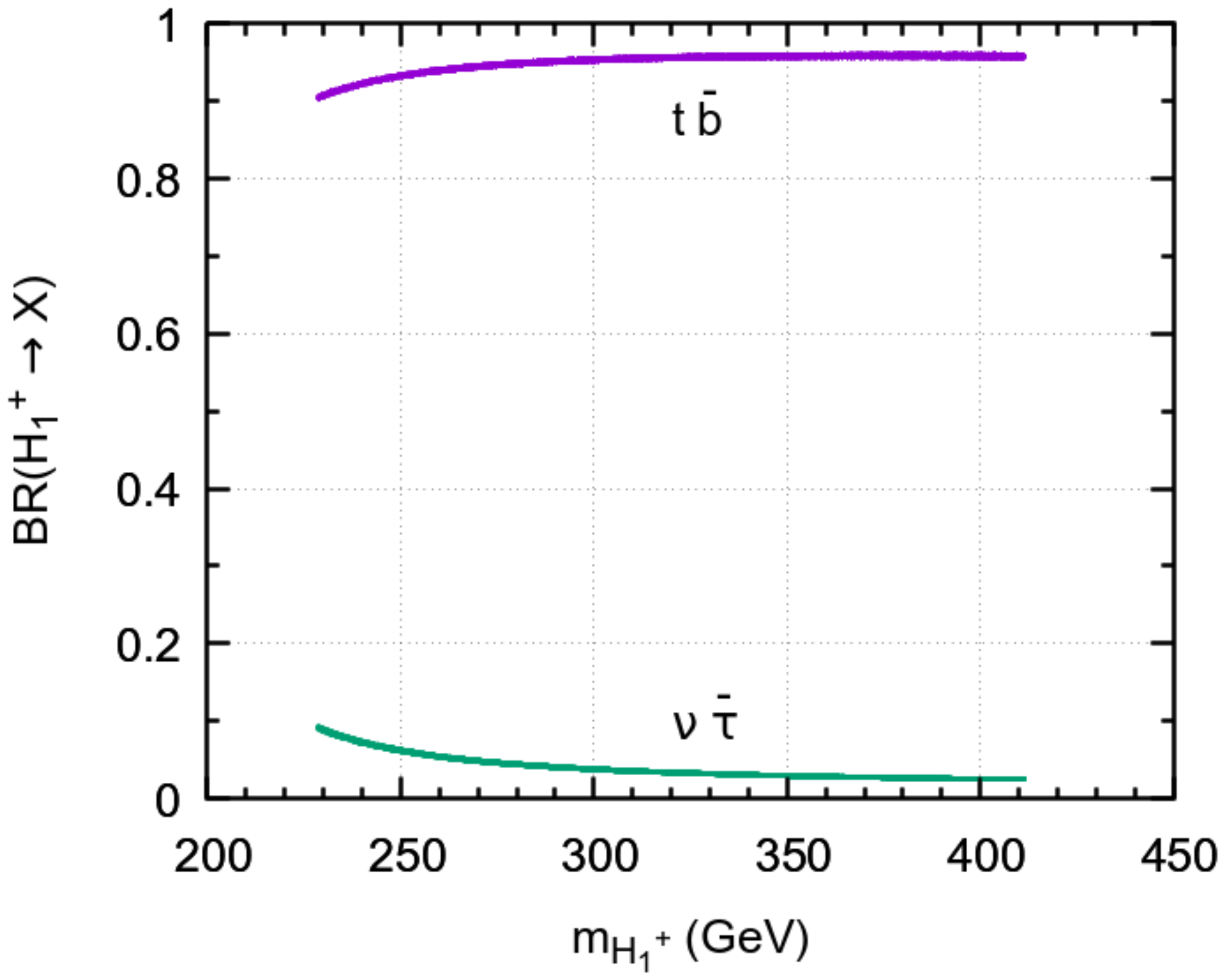} 
   \end{tabular}
  \caption{Dominant BR's for $H_2^+$ (left panel) and  $H_1^+$ (right
    panel) for benchmark point $P_3$.}
  \label{fig:12}  
\end{figure}

We have a large set of benchmark points that illustrate our signal,
the decay $H_2^+ \to H_1^+ + Z$. We just give another example, our
benchmark point $P_3$.
It is defined by the following parameters,
\begin{subequations}
\begin{align}
  \label{eq:24}
  &m_{h_1}=125\, \text{GeV} && m_{h_2}= 728.3\, \text{GeV}
  && m_{h_3}= 720.5\, \text{GeV}\\
  &m_{H_1^+},m_{H_2^+}\text{GeV},\ \text{scanned as shown}&& \alpha= 1.401
  && \gamma=-1.145\\
  &m_{12}^2= 9.48\times 10^{4}\, \text{GeV}^2 && \lambda_c= 2.67  \times 10^{-2}
  && k_{_1}=7.149 \\
&k_{_2}=1.425 \times 10^{-2} &&  k_{_{12}}=1.29 \times 10^{-2}&&
\end{align}
\end{subequations}
The situation is shown in fig.~\ref{fig:12}.
Again we see that our signal decay has the largest branching ratio, while
$H_1^+$ decays almost 100\% into $t + \overline{b}$.
The width of the bands comes from the variation of
$\tan\beta$ and $m_{H_1^+}, m_{H_2^+}$ which were varied
independently. All the points pass all the constraints, including that
of eq.~(\ref{eq:17}).

\subsection{Benchmark Point $P_4$}

\begin{figure}[htb]
  \centering
  \begin{tabular}{cc}
  \includegraphics[width=0.48\textwidth]{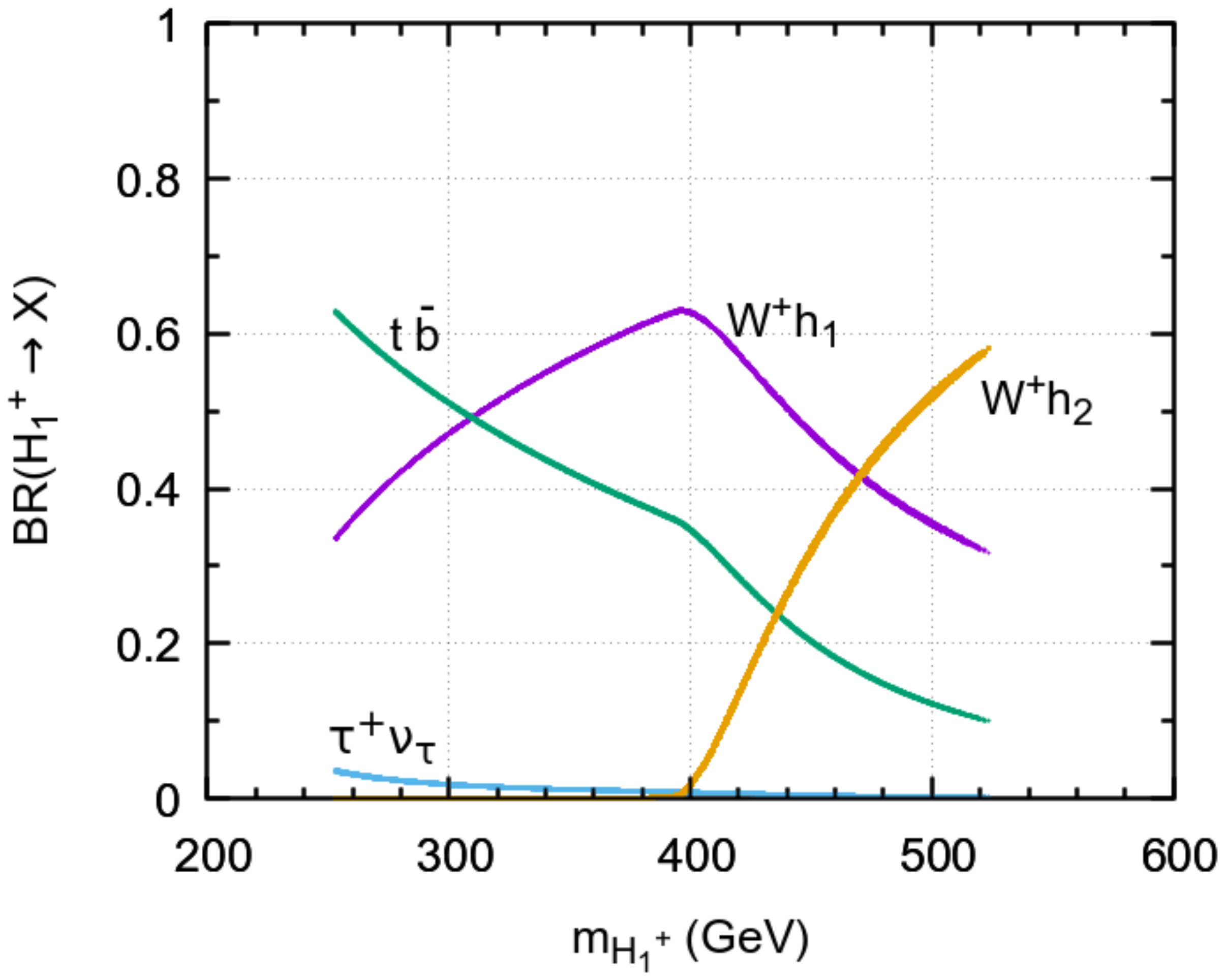} 
    &
    \includegraphics[width=0.48\textwidth]{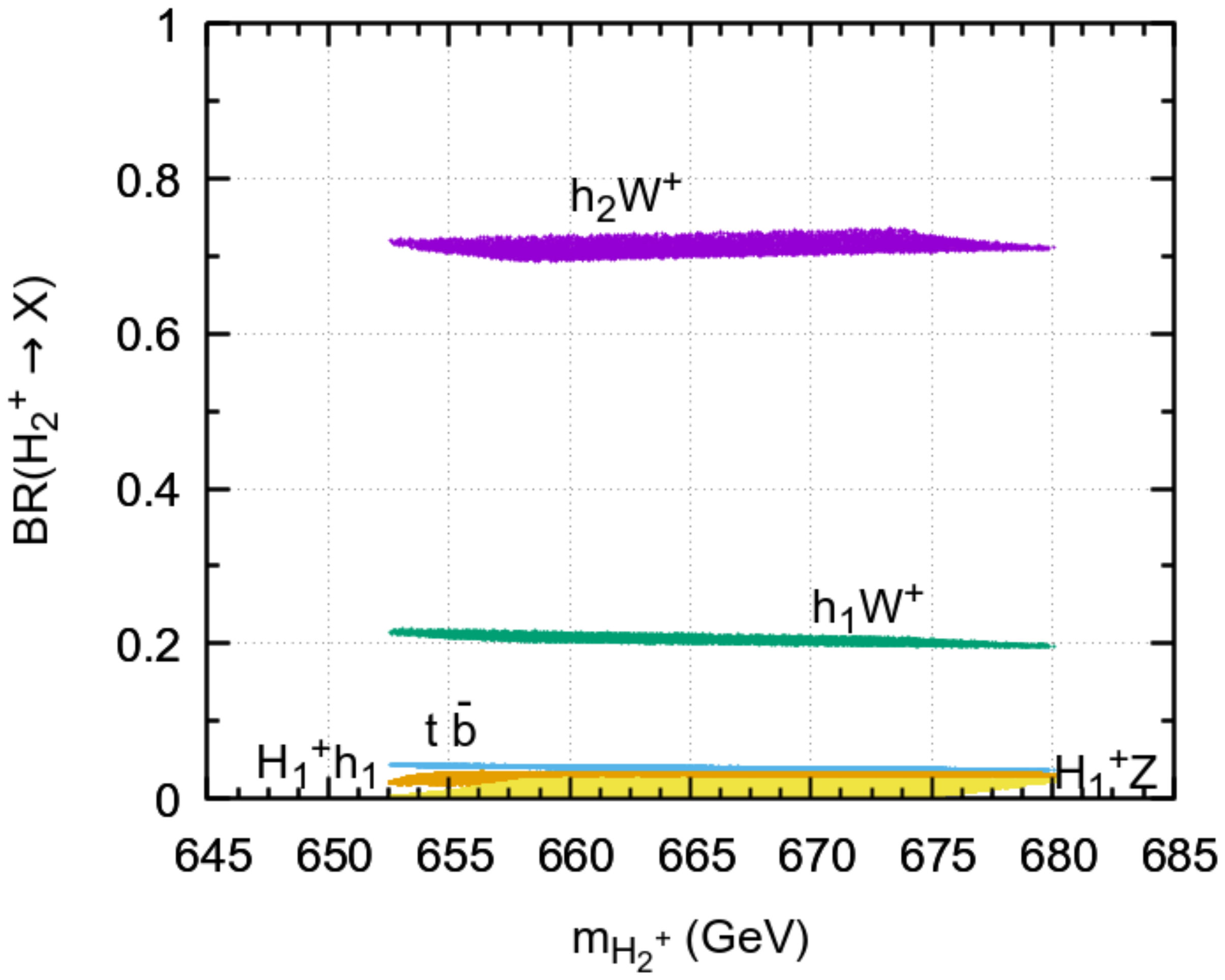} 
   \end{tabular}
  \caption{Dominant BR's for $H_1^+$ (left panel) and  $H_2^+$ (right
    panel) for benchmark point $P_4$.}
  \label{fig:7}  
\end{figure}

It has been pointed out recently \cite{Bahl:2021str},
that there are some
decay channels for the charged Higgs that have not been investigated
at LHC. One of them is the decay $H_1^+ \to W^+ + h_1$. We looked in our
data sample for points where the BR$(H_1^+ \to W^+ + h_1)$ could be
large. For our model, after passing through
the HiggsBounds 5, there are not many points of the general scan that
have a large BR$(H_1^+ \to W^+ + h_1)$. We took one of these which is our
benchmark point $P_4$.
It is defined by the following parameters,
\begin{subequations}
\begin{align}
  \label{eq:24}
  &m_{h_1}=125\, \text{GeV} && m_{h_2}= 314.9\, \text{GeV}
  && m_{h_3}= 651.3\, \text{GeV}\\
  &m_{H_1^+}, m_{H_2^+}\, \text{GeV}, \text{scanned as shown} && \alpha= -1.402
  && \gamma=-1.421\\
  &m_{12}^2= 1.85\times 10^{4}\, \text{GeV}^2 && \lambda_c= 2.00 \times 10^{-2}
  && k_{_1}=1.422 \times 10^{-2}\\
&k_{_2}=0.432 &&  k_{_{12}}=-9.597 \times 10^{-3} &&
\end{align}
\end{subequations}
The situation is shown in fig.~\ref{fig:7}.
We see that, in our model, both BR$(H_1^+ \to W^+ + h_1)$ and
BR$(H_1^+ \to W^+ + h_2)$ can be sizable.
In this case, the BR$(H_2^+ \to H_1^+ + Z)$ is
very small, around 2\%. However the BR$(H_2^+ \to W^+ + h_1)$ and
BR$(H_2^+ \to W^+ + h_2)$ can also be large, making this an
interesting benchmark point.  The width of the bands comes from the
variation of $\tan\beta$, $m_{H_1^+}$, and $m_{H_2^+}$,
which were varied
independently. All the points pass all the constraints, including that
of eq.~(\ref{eq:17}).

\subsection{Production cross-sections and Experimental bounds}

\begin{figure}[htb]
  \centering
  \begin{tabular}{cc}
    \includegraphics[width=0.48\textwidth]{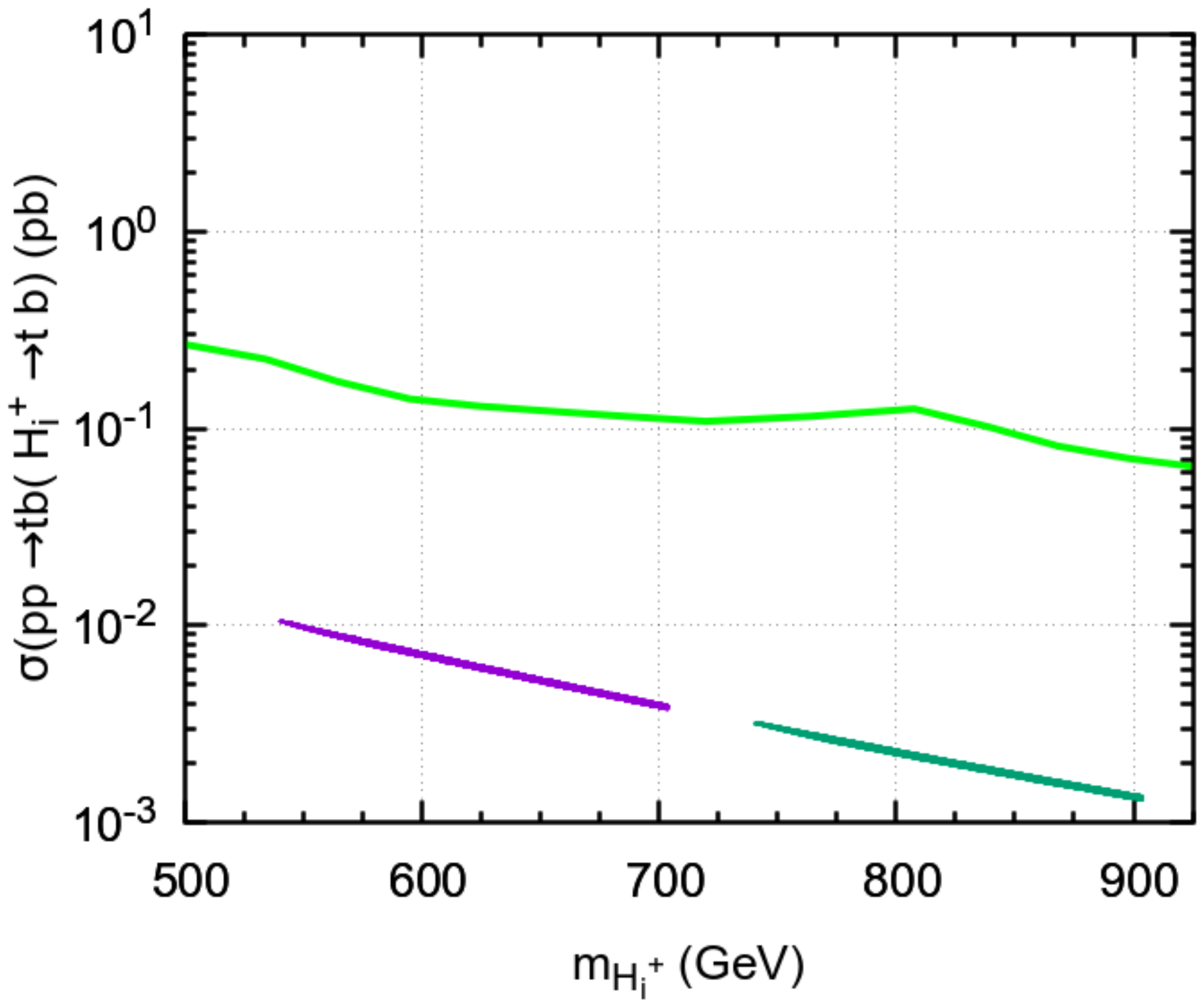}
&
    \includegraphics[width=0.48\textwidth]{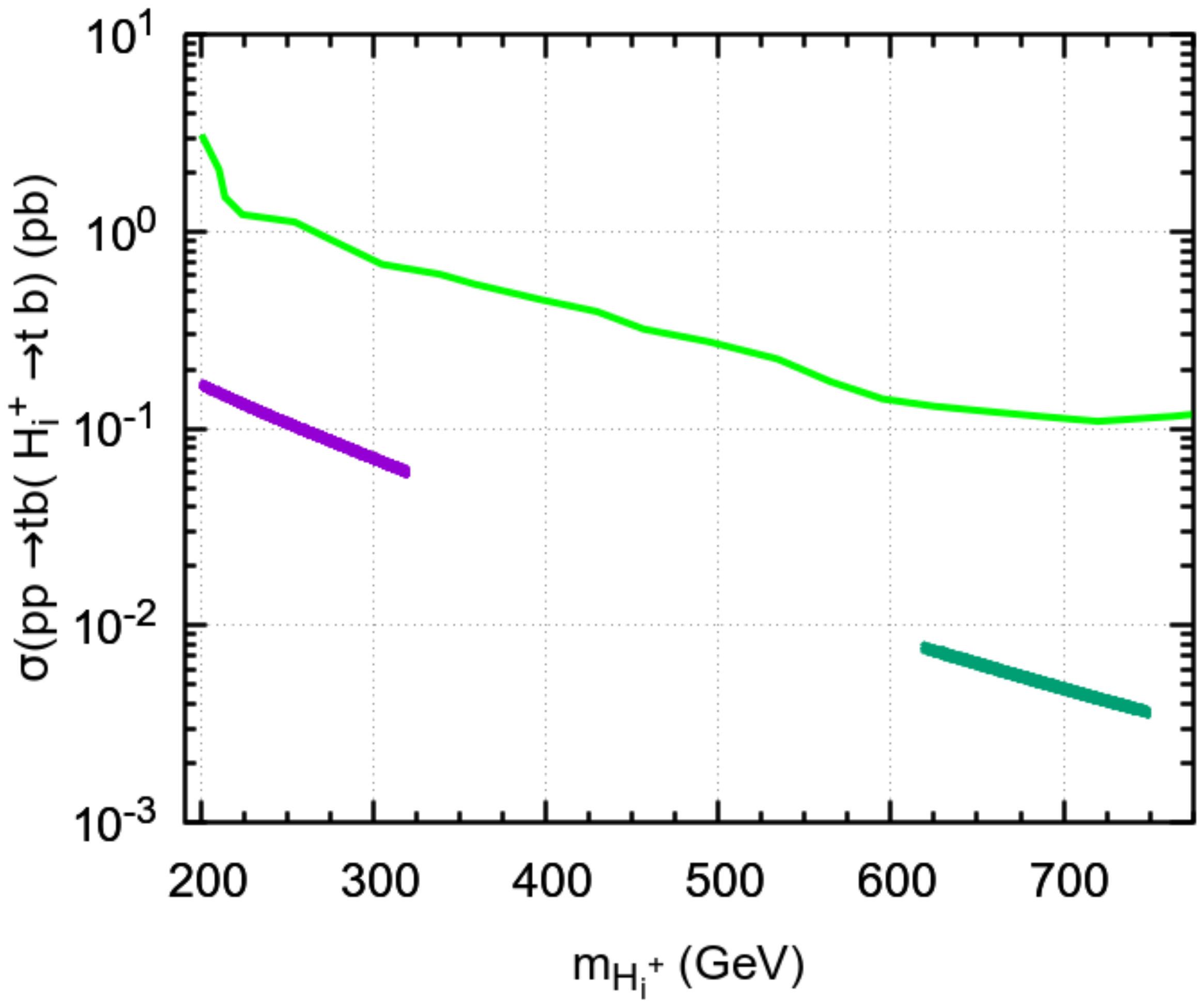}
  \end{tabular}
  \caption{$\sigma( pp \to t b H^+)\times BR(H^+\to t b)$ versus the
    charged Higgs mass for benchmark points $P_1$ (left panel) and
    $P_2$ (right panel). We took BR($H^+\to t b$)$=1$.
    The green line is the
  current LHC limit.}
  \label{fig:8}
\end{figure}

\begin{figure}[htb]
  \centering
  \begin{tabular}{cc}
    \includegraphics[width=0.48\textwidth]{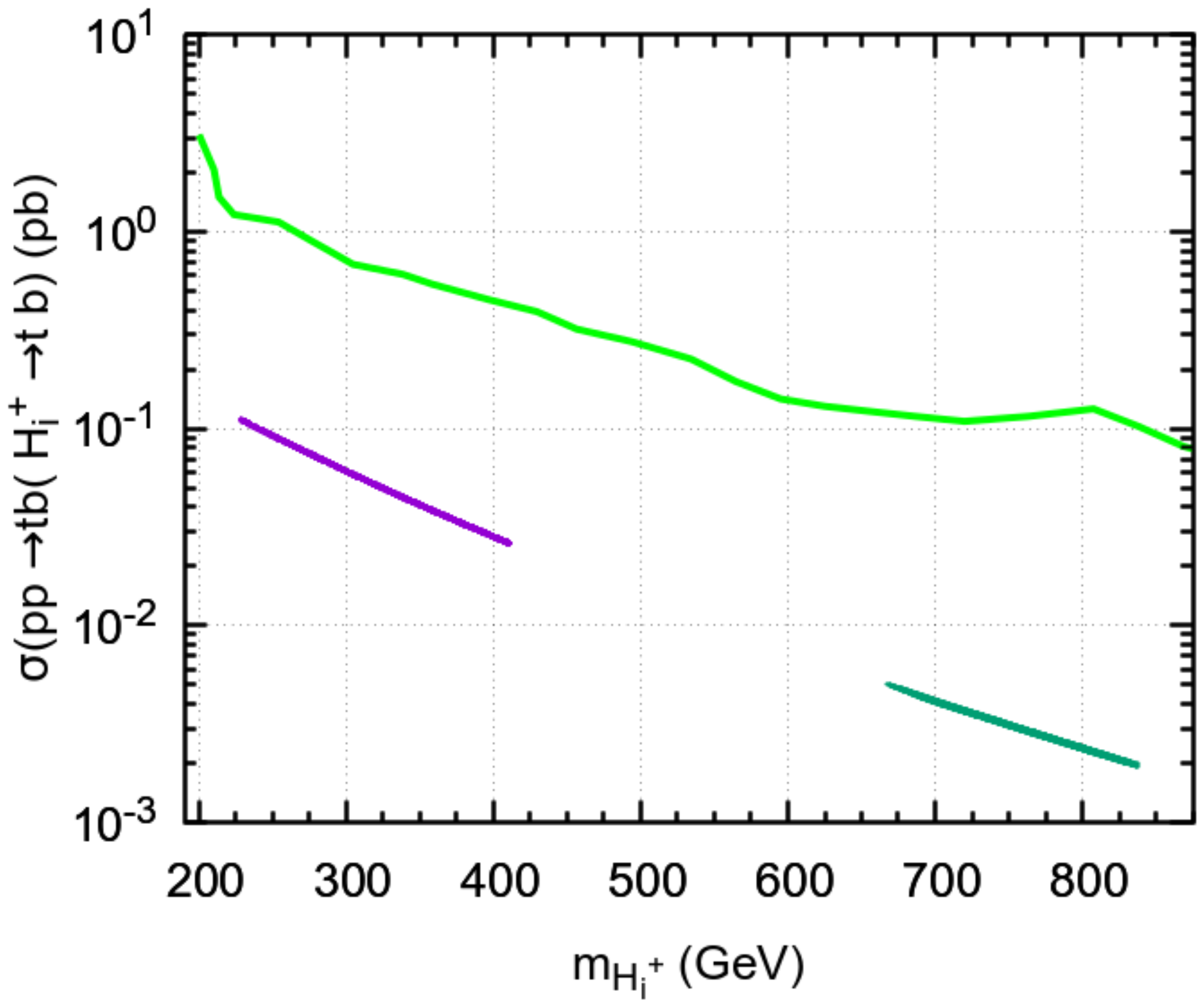}
&
    \includegraphics[width=0.48\textwidth]{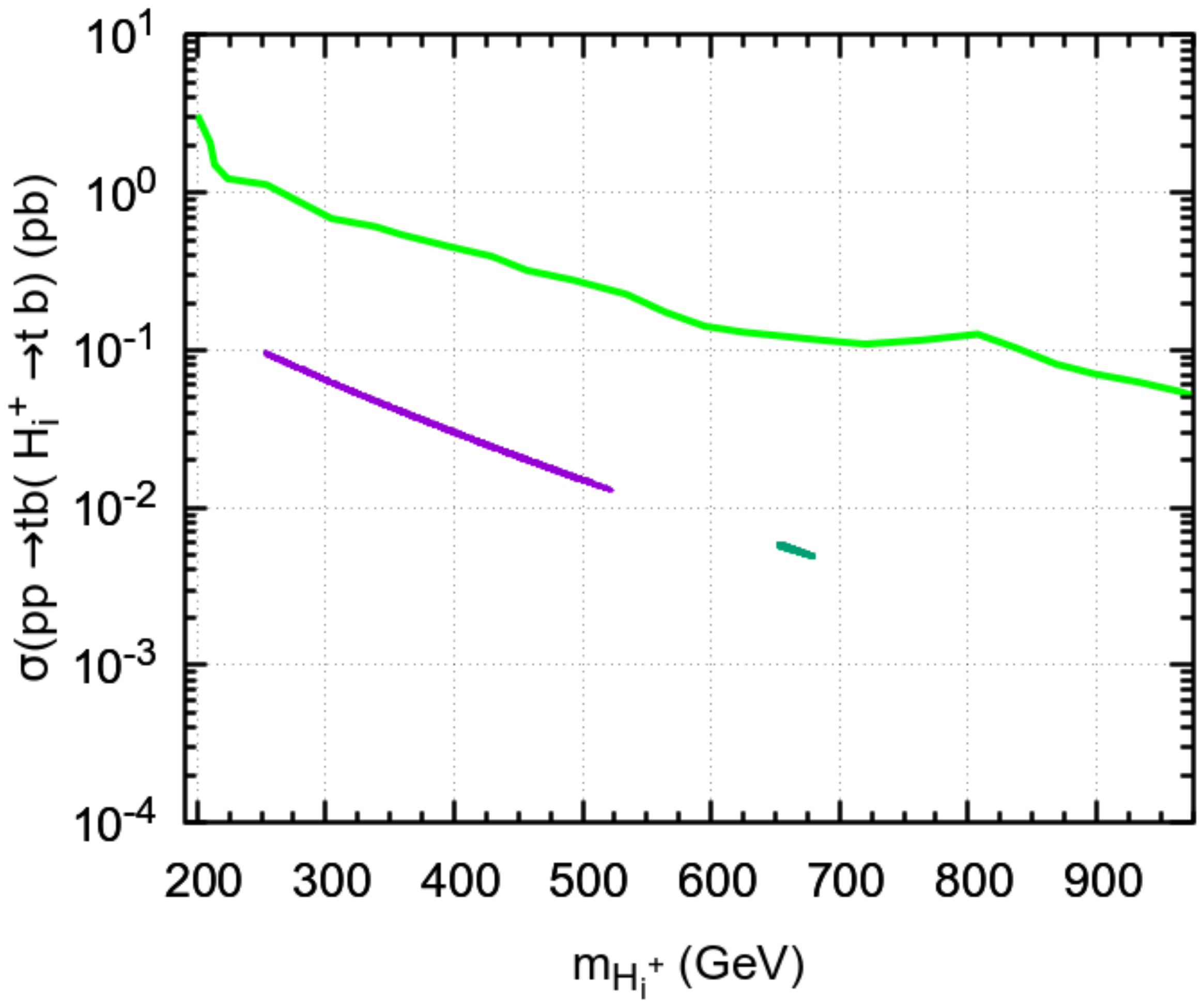}
  \end{tabular}
  \caption{$\sigma( pp \to t b H^+)\times BR(H^+\to t b)$ versus the
    charged Higgs mass for benchmark points $P_1$ (left panel) and
    $P_2$ (right panel). We took BR($H^+\to t b$)$=1$.
    The green line is the
  current LHC limit.}
  \label{fig:9}
\end{figure}

One can ask if a charged Higgs boson with a large
BR($H^+ \to t \overline{b}$)
is not in contradiction with experimental bounds from the
LHC. Although we have checked
 all the points with HiggsBounds 5.9.0 \cite{Bechtle:2020pkv},
it is perhaps helpful to show it explicitly for our benchmark
points. The results are shown in fig.~\ref{fig:8} and fig.~\ref{fig:9}.
We used the values for the production cross-section
$\sigma( pp \to t b H^+)$ from
ref.~\cite{Degrande:2015xnm,Degrande:2016hyf}.
To see if the points are
allowed we considered the worst case scenario where BR($H^+\to t b$)=1
(although for our benchmark points this is only true for the lightest
charged Higgs boson); see fig.~\ref{fig:4} and fig.~\ref{fig:5}.
The green line is the current experimental bound
from ATLAS \cite{ATLAS:2020jqj} as discussed in
ref.~\cite{Bahl:2021str}. So we conclude that all our benchmark points
are consistent with the latest LHC data.

\section{\label{sec:conclusions}Conclusions}

A singular feature of models with multiple scalar doublets and charged singlets
is the presence of off-diagonal $Z H_1^\pm H_2^\mp$ couplings.
We have studied this feature in detail, using the scalar sector of
the Zee model as an example.

Some formulae are presented in a form useful for generic models
with any number of doublet and singlet scalars.
We use in our scans all known theoretical constraints,
including a careful analysis of the BFB conditions,
the exclusion of lower-lying CB vacua, and the
unitarity conditions derived here for this model.

We show that $Z H_1^\pm H_2^\mp$ couplings appear in
$h \rightarrow Z \gamma$ and $B \rightarrow X_s \gamma$,
but that there they do not impose features beyond those
already present in generic 3HDM (where such off-diagonal couplings
are not present).

We stress the importance of looking experimentally for
$H_2^+ \rightarrow H_1^+ Z$ decays and propose interesting benchmark
points.
We also found in our model interesting values for the decays
recently proposed in \cite{Bahl:2021str}.
We found that there are regions of parameter space
consistent with large branching ratios for
$H_1^+ \rightarrow W^+ h_{1,2}$ or $H_2^+ \rightarrow W^+ h_{1,2}$.
But, in those cases, we found no case where simultaneously
BR($H_2^+ \rightarrow H_1^+ Z$) was large.
We strongly urge a search for $H_2^+ \rightarrow H_1^+ Z$ decays.

\section*{Acknowledgments}
We are very grateful to C. Greub for detailed
explanations on his Refs.~\cite{Borzumati:1998tg,Borzumati:1998nx}.
JPS is grateful to Z. Ligeti for discussions.
This work is supported in part by the Portuguese
Funda\c{c}\~{a}o para a Ci\^{e}ncia
e Tecnologia\/ (FCT) under Contracts
CERN/FIS-PAR/0008/2019,
PTDC/FIS-PAR/29436/2017,
UIDB/00777/2020,
and UIDP/00777/2020;
these projects are partially funded through POCTI (FEDER),
COMPETE,
QREN,
and the EU.

%\newpage
\appendix

\section{\label{app:A}Couplings of the charged Higgs}

\subsection{Couplings to the Z boson}
\label{sec:couplings}

We define the coupling as

\begin{equation}
  \label{eq:1}
  [H^+_j,H^-_k,Z]:= -i\ \frac{g}{2 c_W}  (p_{H_j^+} -p_{H_k^-})^\mu
  g_{\rm HpjHmkZ}[j,k],
\end{equation}
where all particles are entering the vertex and
\begin{subequations}
\begin{align}
  g_{\rm HpjHmkZ}[1,1]=& \frac{1}{2} \left(c_W^2-3 s_W^2+\cos(2
    \gamma)\right) \label{eq:3a},  \\
  g_{\rm HpjHmkZ}[1,2]=& - \frac{1}{2} \sin(2 \gamma)
  \label{eq:3b},\\
  g_{\rm HpjHmkZ}[2,1]=&  -\frac{1}{2} \sin(2 \gamma)
  \label{eq:3c},\\
  g_{\rm HpjHmkZ}[2,2]=& \frac{1}{2}\left(c_W^2-3 s_W^2-\cos(2
    \gamma)\right), \label{eq:3d}
\end{align}
\end{subequations}
Notice that when the mixing angle $\gamma$ vanishes the singlet
decouples from the doublet and there is no $[H^+_j,H^-_k,Z]$ vertex
for $j\not =k$. 

\subsection{Couplings to the W boson}

For CP even neutral Higgs bosons ($j=1,2$) we define the coupling as
\begin{equation}
\label{eq:4}
  [h_j,H^+_k,W^-]: - i\, \frac{g}{2} (p_{H^+_k} -p_{h_j})^\mu g_{\rm hjHpkWm}[j,k], 
\end{equation}
where all particles are entering the vertex. For CP odd neutral Higgs
boson ($j=3$) we define
\begin{equation}
  \label{eq:5}
  [h_3,H^+_k,W^-]:= \frac{g}{2} (p_{h_3} - p_{H^+_k})^\mu g_{\rm hjHpkWm}[3,k],
\end{equation}
where
\begin{subequations}
\begin{align}
  \label{eq:2}
     g_{\rm hjHpkWM}[1,1]=& \cos(\gamma) \sin(\alpha - \beta)),\\
     g_{\rm hjHpkWM}[1,2]=& - \sin(\alpha - \beta) \sin(\gamma),\\
     g_{\rm hjHpkWM}[2,1]=& \cos(\alpha - \beta) \cos(\gamma),\\
     g_{\rm hjHpkWM}[2,2]=& - \cos(\alpha - \beta) \sin(\gamma),\\
     g_{\rm hjHpkWM}[3,1]=& \cos(\gamma),\\
     g_{\rm hjHpkWM}[3,2]=& - \sin(\gamma),
\end{align}
\end{subequations}

\subsection{Couplings to quarks and leptons}

The interactions of charged Higgs bosons with quarks are given by
the following Lagrangian 
\begin{equation}
  \label{eq:6}
  \mathcal{L}= \frac{g}{\sqrt{2} } \left[
 \frac{m_{d_j}}{M_W} X_k  \overline{u}_i V_{ij} P_R d_j +
 \frac{m_{u_i}}{M_W} X_k  \overline{u}_i V_{ij} Y_k P_L d_j +
 \frac{m_{l}}{M_W} Z_k  \overline{\nu}_l  P_R e_l 
\right]
 H_k^+ + \text{h.c.} 
\end{equation}
where $k=1,2$ and we have used the conventions of Borzumati and
Greub \cite{Borzumati:1998tg}, extended in ref.~\cite{Akeroyd:2020nfj},
which is convenient for the BR($B\to X_s\gamma$) calculation. We get
\begin{subequations}
\begin{align}
  \label{eq:7}
  X_1=& \tan\beta \cos\gamma, &&Y_1=\cot\beta \cos\gamma,
  &&Z_1=  \tan\beta \cos\gamma, \\
  X_2=& -\tan\beta \sin\gamma, && Y_2=-\cot\beta \sin\gamma,
  &&Z_2=-\tan\beta \sin\gamma,
\end{align}
\end{subequations}

\subsection{Couplings to neutral Higgs}

Finally the couplings to the neutral Higgs are given by the Lagrangian
\begin{equation}
  \label{eq:8}
  \mathcal{L}= H^+_i H^-_k h_j g_{\rm hjHpiHmk}[j,i,k],
\end{equation}
where $g_{\rm hjHpiHmk}$ are long expressions that we do not reproduce
here. Note however that $g_{\rm hjHpiHmk}(3,i,k)=0$.

\section{\label{sec:hdecays}The decays $h \rightarrow \gamma  \gamma$ 
and $h \rightarrow Z \gamma$}

These decays were calculated for the 2HDM to one loop approximation
in \cite{Fontes:2014xva}.
Since most terms in the Lagrangian of our model only differ
by multiplicative constants, our results will only change by some factors.
We adapt from \cite{Fontes:2014xva} for the next results.
The major difference occurs in $h \rightarrow Z \gamma$,
where the presence of the $Z H_1^\pm H_2^\mp$ coupling allows
for the new diagrams in fig.~\ref{fig:HZA}.

\subsection{\label{subsec:fL}Fermion Loops}

The fermion loops are easily obtained plugging the couplings of eq.~\eqref{eq:f} in the results of \cite{Fontes:2014xva}:
\begin{align}
    X_F^{\gamma\gamma}&=-\sum_fN_c^f2a_f^2Q_f^2\tau_f[1+(1-\tau_f)f(\tau_f)]\, ,\nonumber\\
    Y_F^{\gamma\gamma}&=-\sum_fN_c^f2b_f^2Q_f^2\tau_ff(\tau_f)\, ,\nonumber\\
    X_F^{Z\gamma}&=-\sum_fN_c^f\frac{4a_f^2g_V^fQ_fm_f^2}{s_Wc_W}\Bigg[\frac{2M_Z^2}{(m_h^2-M_Z^2)^2}\left[B_0(m_h^2,m_f^2,m_f^2)-B_0(M_Z^2,m_f^2,m_f^2)\right]\\
    &+\frac{1}{m_h^2-M_Z^2}\left[(4m_f^2-m_h^2+M_Z^2)C_0(M_Z^2,0,m_h^2,m_f^2,m_f^2,m_f^2)+2\right]\Bigg]\, ,\nonumber\\
    Y_F^{Z\gamma}&=-\sum_fN_c^f\frac{4b_f^2g_V^fQ_fm_f^2}{s_Wc_W}C_0(M_Z^2,0,m_h^2,m_f^2,m_f^2,m_f^2)\, ,\nonumber
\end{align}
where $N_c^f$ is 3 for quarks and 1 for leptons, $Q_f$ is the fermion charge, $g_V^f$ is the fermion's vector coupling to the $Z$ boson and the sums run over all fermions $f$.
The function appearing is defined as
\begin{equation}
f(\tau)=
-\frac{2 m_f^2}{\tau_f} C_0(0,0,m_h^2,m_f^2,m_f^2,m_f^2)
=
\left\{
\begin{array}{ll}
\left[\sin^{-1} \left(\sqrt{1/\tau}\right)\right]^2,
&\ \  \text{if}\ \tau\ge 1\ \\[+2mm]
-\frac{1}{4}
\left[ \displaystyle
\ln\left(\frac{1+\sqrt{1-\tau}}{1-\sqrt{1-\tau}}\right)
-i \pi \right]^2 ,
&\ \  \text{if}\ \tau<1\, 
\end{array}
\right.\, ,
\label{eq:9}
\end{equation}
while $B_0$ and $C_0$ are the Passarino-Veltman functions.
\subsection{\label{subsec:KL}Charged gauge boson loops}

The only change in these loops comes from the $hVV$ vertex, which is multiplied by a factor $\text{Re}(\omega^\dagger V)^\beta$, and so is the loop. Using the notation of \cite{Fontes:2014xva}, we have
\begin{align}
    X_W^{\gamma\gamma}&=\text{Re}(\omega^\dagger V)^\beta\left[2+3\tau_W+3\tau_W(2-\tau_W)f(\tau_W)\right]\, ,\nonumber\\
    X_W^{Z\gamma}&=\frac{\text{Re}(\omega^\dagger V)^\beta}{\tan\theta_W}I_W\, ,
\end{align}
where,
\begin{align}
    \omega_a&=v_a/v,\;\;\;\tau_W=\frac{4M_W^2}{m_h^2}\, ,\nonumber\\
    I_W&=\frac{1}{(m_h^2-M_Z^2)^2}\left[m_h^2(1-\tan^2\theta_W)-2M_W^2(-5+\tan^2\theta_W)\right]M_Z^2\Delta B_0\, ,\nonumber\\
    &+\frac{1}{m_h^2-M_Z^2}[m_h^2(1-\tan^2\theta_W)-2M_W^2(-5+\tan^2\theta_W)\, ,\\
    &+2M_W^2\left[(-5+\tan^2\theta_W)(m_h^2-2M_W^2)-2M_Z^2(-3+\tan^2\theta_W)]C_0\right]M_Z^2\Delta B_0\, ,\nonumber\\
    \Delta B_0&=B_0(m_h^2,M_W^2,M_W^2)-B_0(M_Z^2,M_W^2,M_W^2)\, ,\nonumber\\
    C_0&=C_0(M_Z^2,0,m_h^2,M_W^2,M_W^2,M_W^2)\, .\nonumber
\end{align}

\subsection{\label{subsec:VL}Charged Scalar Loops}

For the decay to $\gamma\gamma$, the loops are the same as the one presented in \cite{Fontes:2014xva} with the cubic scalar vertex replaced by the ones we defined in eq.~\eqref{eq:h}. Besides this replacement, we only need to sum over the charged scalars, obtaining
\begin{equation}
    X_H^{\gamma\gamma}=-\sum_\alpha\frac{\lambda^{2\alpha\alpha}v^2}{2m_{\pm\alpha}^2}\tau_{\pm\alpha}[1-\tau_{\pm\alpha}f(\tau_{\pm\alpha})]\, ,
\end{equation}
where $\tau_{\pm\alpha}=4m_h^2/m_{\pm\alpha}$.
Regarding the decay to $Z\gamma$, we can allow two different scalars to
run within the same loop, as seen in fig.~\ref{fig:HZA}.
This generalizes the result in \cite{Fontes:2014xva}. We obtain
\begin{align}
  X_H^{Z\gamma}&=-\sum_{\alpha_1\alpha_2}
  \frac{(2s_W^2\delta^{\alpha_1\alpha_2}-(U^\dagger
    U)^{\alpha_2\alpha_1})}{\sin\theta_W\cos\theta_W}
  \frac{g^{2\alpha_1\alpha_2}}{m_h^2-M_Z^2}
  \Bigg[\frac{M_Z^2}{m_h^2-M_Z^2}\left(B_0(m_h^2,m_{\pm\alpha_1}^2,m_{\pm\alpha_2}^2)
\right.\nonumber\\
&\hskip 10mm
\left. -B_0(M_Z^2,m_{\pm\alpha_1}^2,m_{\pm\alpha_2}^2)\right)
  +1+m_{\pm\alpha_1}^2
  C_0(M_Z^2,0,m_h^2,m_{\pm\alpha_1}^2,m_{\pm\alpha_1}^2,m_{\pm\alpha_2}^2)\nonumber\\
  &\hskip 10mm
  +m_{\pm\alpha_2}^2
  C_0(M_Z^2,0,m_h^2,m_{\pm\alpha_2}^2,m_{\pm\alpha_2}^2,m_{\pm\alpha_1}^2)\Bigg]\, .
\end{align}
If there were no cubic terms in eq.~\eqref{eq:pot} ($\mu_4^{abi}=0$),
then $(U^\dagger U)^{\alpha_2\alpha_1} = \delta^{\alpha_2\alpha_1}$
and there would be no diagrams involving simultaneously two different
charged scalars.

\subsection{\label{subsec:WL}Final widths for loop decays}

The final widths are given by
\begin{align}
    \Gamma(h\rightarrow\gamma\gamma)&=\frac{G_F\alpha^2m_h^2}{128\sqrt{2}\pi^3}\left(|X_F^{\gamma\gamma}+X_W^{\gamma\gamma}+X_H^{\gamma\gamma}|^2+|Y_F^{\gamma\gamma}|^2\right)\, ,\nonumber\\
    \Gamma(h\rightarrow Z\gamma)&=\frac{G_F\alpha^2m_h^2}{64\sqrt{2}\pi^3}\left(1-\frac{M_Z^2}{m_h^2}\right)^3\left(|X_F^{Z\gamma}+X_W^{Z\gamma}+X_H^{Z\gamma}|^2+|Y_F^{Z\gamma}|^2\right)\, .
\end{align}

\section{\label{sec:PertUni} Perturbative unitarity}

We write here the scattering matrices for the various $(Q,Y)$
combinations and list all the eigenvalues at the end.
This is presented here for the first time.
We follow the notation of \cite{Bento:2017eti}

\subsection{$Q=2,Y=1$}

For the combination of states $S^{++}_\alpha$ in eq.~(\ref{equnit:8a})
we have 
\begin{equation}
  \label{eq:36}
  M^{++}_2 =
  \begin{bmatrix}
 \lambda_1 & 0 & 0 & \lambda_5 & 0 & 0 \\
 0 & \lambda_3+\lambda_4 & 0 & 0 & 0 & 0 \\
 0 & 0 & k_{1} & 0 & -k_{12} & 0 \\
 \lambda_5 & 0 & 0 & \lambda_2 & 0 & 0 \\
 0 & 0 & -k_{12} & 0 & k_{2} & 0 \\
 0 & 0 & 0 & 0 & 0 & 2 \lambda_c \\
  \end{bmatrix} \, .
\end{equation}

\subsection{$Q=1,Y=1$}

For the combination of states $S^{+}_\alpha$ in eq.~(\ref{equnit:8b})
we have 
\begin{equation}
  \label{eq:36}
  M^{+}_2 =
  \begin{bmatrix}
 \lambda_1 & 0 & 0 & \lambda_5 & 0 & 0 \\
 0 & \lambda_3 & \lambda_4 & 0 & 0 & 0 \\
 0 & \lambda_4 & \lambda_3 & 0 & 0 & 0 \\
 \lambda_5 & 0 & 0 & \lambda_2 & 0 & 0 \\
 0 & 0 & 0 & 0 & k_{1} & -k_{12} \\
 0 & 0 & 0 & 0 & -k_{12} & k_{2} \\
  \end{bmatrix}\, .
\end{equation}

\subsection{$Q=1,Y=0$}

For the combination of states $T^{+}_\alpha$ in eq.~(\ref{equnit:8c})
we have 
\begin{equation}
  \label{eq:36}
  M^{+}_0 =
  \begin{bmatrix}
 \lambda_1 & 0 & 0 & \lambda_4 & 0 & 0 \\
 0 & \lambda_3 & \lambda_5 & 0 & 0 & 0 \\
 0 & \lambda_5 & \lambda_3 & 0 & 0 & 0 \\
 \lambda_4 & 0 & 0 & \lambda_2 & 0 & 0 \\
 0 & 0 & 0 & 0 & k_{1} & -k_{12} \\
 0 & 0 & 0 & 0 & -k_{12} & k_{2} \\
  \end{bmatrix}\, .
\end{equation}

\subsection{$Q=0,Y=1$}

For the combination of states $S^{0}_\alpha$ in eq.~(\ref{equnit:8d})
we have 
\begin{equation}
  \label{eq:36}
  M^{0}_2 =
  \begin{bmatrix}
 0 & 0 & 0 \\
 0 & 0 & 0 \\
 0 & 0 & 0 \\
  \end{bmatrix}\, .
\end{equation}

\subsection{$Q=0,Y=0$}

\setcounter{MaxMatrixCols}{13}
For the combination of states $T^{0}_\alpha$ in eq.~(\ref{equnit:8e})
we have 
\begin{equation}
  \label{eq:36}
  M^{0}_0 =
  \begin{bmatrix}
 2 \lambda_1 & 0 & 0 & 0 & \lambda_{34} & 0 & 0 & 0 &
   k_{1} & \lambda_1 & 0 & 0 & \lambda_3 \\
 0 & 2 \lambda_5 & 0 & \lambda_{34} & 0 & 0 & 0 & 0 &
   -k_{12} & 0 & \lambda_5 & \lambda_4 & 0 \\
 0 & 0 & 0 & 0 & 0 & 0 & k_{1} & -k_{12} & 0 & 0 & 0 & 0 & 0 \\
 0 & \lambda_{34} & 0 & 2 \lambda_5 & 0 & 0 & 0 & 0 &
   -k_{12} & 0 & \lambda_4 & \lambda_5 & 0 \\
 \lambda_{34} & 0 & 0 & 0 & 2 \lambda_2 & 0 & 0 & 0 &
   k_{2} & \lambda_3 & 0 & 0 & \lambda_2 \\
 0 & 0 & 0 & 0 & 0 & 0 & -k_{12} & k_{2} & 0 & 0 & 0 & 0 & 0 \\
 0 & 0 & k_{1} & 0 & 0 & -k_{12} & 0 & 0 & 0 & 0 & 0 & 0 & 0 \\
 0 & 0 & -k_{12} & 0 & 0 & k_{2} & 0 & 0 & 0 & 0 & 0 & 0 & 0 \\
 k_{1} & -k_{12} & 0 & -k_{12} & k_{2} & 0 & 0 & 0 & 4
 \lambda_c & k_{1} & -k_{12} & -k_{12} & k_{2} \\ 
 \lambda_1 & 0 & 0 & 0 & \lambda_3 & 0 & 0 & 0 & k_{1} & 2
   \lambda_1 & 0 & 0 & \lambda_{34} \\
 0 & \lambda_5 & 0 & \lambda_4 & 0 & 0 & 0 & 0 & -k_{12} & 0 & 2
   \lambda_5 & \lambda_{34} & 0 \\
 0 & \lambda_4 & 0 & \lambda_5 & 0 & 0 & 0 & 0 & -k_{12} & 0 &
   \lambda_{34} & 2 \lambda_5 & 0 \\
 \lambda_3 & 0 & 0 & 0 & \lambda_2 & 0 & 0 & 0 & k_{2} &
   \lambda_{34} & 0 & 0 & 2 \lambda_2 \\
  \end{bmatrix}\, ,
\end{equation}
where for simplicity we have defined
\begin{equation}
  \label{eq:37}
  \lambda_{34}\equiv \lambda_3+\lambda_4 \, .
\end{equation}

\subsection{The independent eigenvalues}

We can obtain easily the eigenvalues for all the matrices except for
$M^{0}_0$ in eq.~(\ref{eq:36}) for which we have to solve numerically
a fourth order polynomial. The list of independent eigenvalues is,
\begin{subequations}
\begin{align}
  \Lambda_1=&\frac{1}{2} \left(-\sqrt{(k_{1})^2-2 k_{1} k_{2}+4
     (k_{12})^2+(k_{2})^2}+k_{1}+k_{2}\right)\, ,\\[+2mm]
  \Lambda_2=&\frac{1}{2}
   \left(\sqrt{(k_{1})^2-2 k_{1} k_{2}+4
      (k_{12})^2+(k_{2})^2}+k_{1}+k_{2}\right)\, ,\\[+2mm]
   \Lambda_3=&\lambda_3+\lambda_4\\[+2mm]
   \Lambda_4=&\frac{1}{2} \left(-\sqrt{\lambda_1^2-2
       \lambda_1 \lambda_2+\lambda_2^2
       +4 \lambda_5^2}+\lambda_1+\lambda_2\right)\, ,\\[+2mm]
\Lambda_5=&\frac{1}{2}
\left(\sqrt{\lambda_1^2-2 \lambda_1
    \lambda_2+\lambda_2^2+4 \lambda_5^2}
  +\lambda_1 +\lambda_2\right)\, ,\\[+2mm]
\Lambda_6=&2 \lambda_c\, ,\\[+2mm]
\Lambda_7=&\lambda_3-\lambda_4\, ,\\[+2mm]
\Lambda_8=&\frac{1}{2} 
\left(-\sqrt{\lambda_1^2-2 \lambda_1
    \lambda_2+\lambda_2^2+4 \lambda_4^2}
  +\lambda_1+\lambda_2\right)\, ,\\[+2mm]
\Lambda_9=&\frac{1}{2} \left(\sqrt{\lambda_1^2-2 \lambda_1
    \lambda_2+\lambda_2^2+4 \lambda_4^2}
  +\lambda_1+\lambda_2\right)\, ,\\[+2mm]
\Lambda_{10}=&\lambda_3-\lambda_5\, ,\\[+2mm]
\Lambda_{11}=&\lambda_3+\lambda_5\, ,\\[+2mm]
\Lambda_{12}=&\frac{1}{2} \left(-\sqrt{(k_{1})^2-2 k_{1} k_{2}+4
   (k_{12})^2+(k_{2})^2}-k_{1}-k_{2}\right)\, ,\\[+2mm]
\Lambda_{13}=&\frac{1}{2}
   \left(\sqrt{(k_{1})^2-2 k_{1} k_{2}+4
      (k_{12})^2+(k_{2})^2}-k_{1}-k_{2}\right)\, ,\\[+2mm]
   \Lambda_{14}=&\lambda_5-\lambda_3\, ,\\[+2mm]
   \Lambda_{15}=&-\lambda_3-2 \lambda_4+3 \lambda_5 \, .
 \end{align}
 \end{subequations}
The remaining eigenvalues, $\Lambda_{16}-\Lambda_{19}$, are the roots of the polynomial of fourth
degree
\begin{equation}
  \label{eq:38}
  c_0+ c_1\, \eta + c_2\, \eta^2+ c_3\, \eta^2+ c_4\, \eta^4=0\, ,
\end{equation}
where
\begin{subequations}
\begin{align}
  c_0=&6 (k_{1})^2 \lambda_2 \lambda_3+12 (k_{1})^2 \lambda_2 \lambda_4+18 (k_{1})^2
   \lambda_2 \lambda_5-8 k_{1} k_{2} \lambda_3^2-20 k_{1} k_{2}
   \lambda_3 \lambda_4\nonumber\\
   &-24 k_{1} k_{2} \lambda_3 \lambda_5-8 k_{1} k_{2}
   \lambda_4^2-12 k_{1} k_{2} \lambda_4 \lambda_5+36(k_{12})^2 \lambda_1
   \lambda_2\nonumber\\
   &-16(k_{12})^2 \lambda_3^2-16(k_{12})^2 \lambda_3 \lambda_4-4
  (k_{12})^2 \lambda_4^2+6 (k_{2})^2 \lambda_1 \lambda_3+12
  (k_{2})^2 \lambda_1 
  \lambda_4\nonumber\\
  &+18 (k_{2})^2 \lambda_1 \lambda_5-36 \lambda_1 \lambda_2 \lambda_3
   \lambda_c-72 \lambda_1 \lambda_2 \lambda_4 \lambda_c-108 \lambda_1 \lambda_2
   \lambda_5 \lambda_c+48 \lambda_3^2 \lambda_4 \lambda_c\nonumber\\
   &+48 \lambda_3^2 \lambda_5 
   \lambda_c+16 \lambda_3^3 \lambda_c+36 \lambda_3 \lambda_4^2
   \lambda_c+48 \lambda_3 
   \lambda_4 \lambda_5 \lambda_c+12 \lambda_4^2 \lambda_5 \lambda_c+8 \lambda_4^3
   \lambda_c   \, ,\\[+2mm]
  c_1=&-6 (k_{1})^2 \lambda_2-2 (k_{1})^2 \lambda_3-4
  (k_{1})^2 \lambda_4-6 (k_{1})^2   \lambda_5
  +8 k_{1} k_{2} \lambda_3+4 k_{1} k_{2}
  \lambda_4\nonumber\\
  &-12 (k_{12})^2 \lambda_1-12(k_{12})^2 \lambda_2-6 (k_{2})^2 \lambda_1-2
   (k_{2})^2 \lambda_3-4 (k_{2})^2 \lambda_4-6 (k_{2})^2
   \lambda_5\nonumber\\
   &+9 \lambda_1  \lambda_2 \lambda_3+18 \lambda_1 \lambda_2
   \lambda_4+27 \lambda_1 \lambda_2 
   \lambda_5+36 \lambda_1 \lambda_2 \lambda_c+12 \lambda_1 \lambda_3 \lambda_c+24
   \lambda_1 \lambda_4 \lambda_c\nonumber\\
   &+36 \lambda_1 \lambda_5 \lambda_c+12 \lambda_2
   \lambda_3 \lambda_c+24 \lambda_2 \lambda_4 \lambda_c+36 \lambda_2 \lambda_5
   \lambda_c-12 \lambda_3^2 \lambda_4-12 \lambda_3^2 \lambda_5-16 \lambda_3^2
   \lambda_c\nonumber\\
   &-4 \lambda_3^3-9 \lambda_3 \lambda_4^2-12 \lambda_3 \lambda_4 \lambda_5-16
   \lambda_3 \lambda_4 \lambda_c-3 \lambda_4^2 \lambda_5-4 \lambda_4^2
   \lambda_c-2    \lambda_4^3 \, ,
  \\[+2mm]
  c_2=&2 (k_{1})^2+4(k_{12})^2+2 (k_{2})^2-9 \lambda_1
  \lambda_2-3 \lambda_1 
  \lambda_3-6 \lambda_1 \lambda_4-9 \lambda_1 \lambda_5
  -12 \lambda_1 \lambda_c\nonumber\\
  &-3 \lambda_2 \lambda_3-6 \lambda_2 \lambda_4-9 \lambda_2
  \lambda_5-12 \lambda_2  
   \lambda_c+4 \lambda_3^2+4 \lambda_3 \lambda_4-4 \lambda_3
   \lambda_c+\lambda_4^2\nonumber\\
   &-8   \lambda_4 \lambda_c-12 \lambda_5 \lambda_c \, ,
  \\[+2mm]
  c_3=&3 \lambda_1+3 \lambda_2+\lambda_3+2 \lambda_4+3 \lambda_5
  +4 \lambda_c   \, ,\\[+2mm]
  c_4=& -1 \, .
\end{align}
\end{subequations}

\providecommand{\href}[2]{#2}\begingroup\raggedright\endgroup

%\bibliographystyle{jhep}
%\bibliography{romao-ref}

\end{document}